\documentclass[]{aa}
\usepackage{natbib}
\usepackage{graphicx}
\usepackage{txfonts}
\usepackage{psfig}

\newcommand{\resp}{resp{.}}
\newcommand{\ie}{i{.}e{.}~}

\newcommand{\eq}{Eq{.}~}                 
\newcommand{\eqs}{Eqs{.}~}
\newcommand{\fg}{Fig{.}~}
\newcommand{\fgs}{Figs{.}~}
\newcommand{\sct}{Sect{.}~}
\newcommand{\scts}{Sects{.}~}

\newcommand{\kms}{~km~s$^{-1}$}
\newcommand{\hmpc}{~$h^{-1}$~Mpc}

\newcommand{\hmpcc}{~$h^{-3}$~Mpc$^3$}
\newcommand{\phiunit}{~$h^3$~Mpc$^{-3}$~mag$^{-1}$}
\newcommand{\etal}{{\it et\thinspace al.}\ }

\begin{document}

\title{Spatial clustering in the ESO-Sculptor Survey: two-point
correlation functions by galaxy type at redshifts 0.1 -- 0.5
\thanks{Based on observations collected at the European Southern
Observatory (ESO), La Silla, Chile.}  }

\titlerunning{ESO-Sculptor clustering by galaxy type at $z\simeq0.1-0.5$}
\authorrunning{de Lapparent and Slezak}

\author{
 V. de Lapparent \inst{1}
 \and E. Slezak \inst{2}}

\offprints{V. de Lapparent}

\institute{
 Institut d'Astrophysique de Paris, UMR7095 CNRS, Univ. Pierre et Marie Curie, 
 98 bis Boulevard Arago, F-75014 Paris, France\\
 \email{lapparen@iap.fr}
\and
 Observatoire de la C\^ote d'Azur, BP 4229, F-06304 Nice Cedex 4, France\\
 \email{Eric.Slezak@oca.eu}
}

\date{Received 23 January 2007 / Accepted 4 June 2007}

\abstract{
{Galaxy clustering shows segregation effects with galaxy type, color and
luminosity, which bring clues on the relationship with the underlying
density field.}
{We explore these effects among the populations of giant and dwarf
galaxies detected in the ESO-Sculptor survey.}
{We calculate the spatial two-point auto and cross-correlation
functions for the 765 galaxies with $R_\mathrm{c}\le21.5$ and $0.1\le
z\le 0.51$ and for subsets by spectral type and luminosity.}
{At separation of $0.3$\hmpc, pairs of early-type galaxies dominate
the clustering over all the other types of pairs.  At intermediate
scales, $0.3-5$\hmpc, mixed pairs of dwarf and giant galaxies
contribute equally as pairs of giant galaxies, whereas the latter
dominate at $\simeq10$\hmpc. Moreover, the correlation functions per
galaxy type display the expected transition between the 1-halo and
2-halo regimes in the scenario of hierarchical merging of dark matter
halos. The 1-halo component of the early-type galaxies largely outdoes
that for the late spiral galaxies, and that for the dwarf galaxies is
intermediate between both. In contrast, the 2-halo component of the
early-type galaxies and late spiral galaxies are comparable, whereas
that for the dwarf galaxies is consistent with null clustering. }
{We link the clustering segregation of the early-type and late spiral
galaxies to their spatial distribution within the underlying dark
matter halos. The early-type galaxies are preferentially located near
the centers of the most massive halos, whereas late spiral galaxies
tend to occupy their outskirts or the centers of less massive
halos. This appears to be independent of luminosity for the early-type
galaxies, whereas faint late spiral galaxies might reside in less
dense regions than their bright analogs. The present analysis also
unveils unprecedented results on the contribution from dwarf galaxies:
at the scale at which they significantly cluster inside the halos
($\le0.3$\hmpc), they are poorly mixed with the late spiral galaxies,
and appear preferentially as satellites of early-type galaxies.}


}

\maketitle


\section{Introduction                       \label{intro}}

The two-point correlation function is a fundamental statistic for
characterizing the galaxy distribution. It partly quantifies the
visual impression of clustering provided by the redshift maps, and
subsequently allows one to perform direct comparison with the
theoretical predictions. One of the issues is to determine how galaxies
trace the underlying mass distribution, and whether and how this is
related to their internal properties. This in turn can provide crucial
information on how galaxies have formed and evolved until now.

Here we use the ESO-Sculptor Survey (hereafter ESS;
\citealt{lapparent03}) to statistically characterize the large-scale
clustering of galaxies at $z\la0.5$, and to examine its dependence on
galaxy type. The ESS provides a nearly complete
redshift survey of galaxies at $z\la0.5$ over a contiguous area of the
sky \citep{bellanger95a}, supplemented by CCD-based photometry
\citep{arnouts97} and a template-free spectral classification
\citep{galaz98}. In agreement with the other existing redshift surveys
to smaller or similar distances
\citep{lapparent86,shectman96,small97a,colless01,zehavi02}, the ESS
redshift map reveals a highly structured cell-like distribution out to
$z\sim0.5$ in which numerous sharp walls or filaments alternate with
regions devoid of galaxies on a typical scale of $\sim 25$\hmpc\
\citep{bellanger95b}.  The deep pencil-beam geometry of the survey is
characterized by a long line-of-sight of $1300$\hmpc, and a transverse
extent of $\sim11$\hmpc\ at $z=0.3$, corresponding to $\sim3$
correlation lengths (quoted scales are in comoving coordinates). The
ESS therefore provides a sufficiently large sample for performing a
useful two-point correlation analysis.

Using the ESS spectral classification and the corresponding luminosity
functions per galaxy class \citep{lapparent03}, we examine the
variations in the ESS clustering as a function of galaxy type. Various
surveys have detected the stronger clustering of early-type/red
galaxies over late-type/blue galaxies at redshifts $z\la0.1$
\citep{loveday99,giuricin01,norberg02,magliocchetti03,zehavi05}, and
at higher redshifts $z\ga0.5$
\citep{shepherd01,phleps03,coil04,meneux06}.  Further details were
obtained by \citet{li06} from low redshift galaxies, whose analysis
shows that the observed clustering differences between red and blue
galaxies, namely a higher amplitude and steeper slope than for blue
galaxies, are largest at small scales and for low mass galaxies; the
authors also measure the same clustering segregation effects when
considering galaxy age as traced by the $4000$\AA\ break strength,
instead of galaxy color. To examine in further details the
relationship between galaxy type and clustering, we propose here a new
approach based on the separation of the giant and dwarf galaxies. We
also measure the cross-correlation of the various samples, which
provides complementary clues on the relative distribution of the
different galaxy types.

The paper is organized as follows. Section \ref{data} reviews the
characteristics of the ESS galaxy redshift survey, defines the
sub-samples used in the present analysis, and describes the luminosity
functions used for calculating the selection functions. In
\sct\ref{errors}, we evaluate the various sources of random and
systematic errors. Results on the redshift space auto-correlation
function $\xi(s)$ for the full ESS sample and the sub-samples by
galaxy types are given in \sct\ref{xis}.  The correlation as a
function of projected separation $w(r_\mathrm{p})$ for the various ESS
samples are described in \sct\ref{wrp}, along with the
cross-correlation functions between the different galaxy types. The
auto and cross-correlation functions are then interpreted in
\sct\ref{halo} in terms of the occupation of the dark matter halos by
the different galaxy types. In \sct\ref{comp}, we compare our results
on $w(r_\mathrm{p})$ to those from the other existing redshift
surveys. Finally, \sct\ref{summary} summarizes our conclusions and
\sct\ref{discussion} discusses them in view of other existing results
on galaxy clustering. In the Appendix, we describe the estimators
which we use for calculating the two-point correlation functions, and
we address the issues of the weighting scheme, the normalization of
the correlation function, and the estimation of the mean density.

Throughout the present analysis, we assume a flat Universe with, at
the present epoch, a scale parameter $H_0 = 100 h$\kms~Mpc$^{-1}$, a
matter density $\Omega_\mathrm{m}=0.3$ and a cosmological energy
density $\Omega_\Lambda=0.7$
\citep{riess98,perlmutter99,phillips01,tonry03}.  All absolute
absolute magnitude are defined modulo $+5\log h$.


\section{The ESO-Sculptor Survey              \label{data}}

The ESS covers a rectangular area of 0.37~deg$^2$ defined as a thin
strip of $1.53^{\circ}\times0.24^{\circ}$ near the south Galactic pole
($b_{II}\sim -83^{\circ}$) and centered at
$0^\mathrm{h}22.5^\mathrm{m},-30.1^{\circ}$ (2000) in the Sculptor
constellation. The observations were performed using the New
Technology Telescope (NTT) and the 3.6 m telescope at the European
Southern Observatory (ESO).  The photometric catalogue is based on CCD
multicolor imaging in the Johnson-Cousins $BVR_\mathrm{c}$ system, and
contains nearly 13~000 galaxies to $V\simeq24$ \citep{arnouts97}.  The
spectroscopic survey provides flux-calibrated spectra and redshifts
(with an rms ``external'' uncertainty $\sigma(z)\sim0.00055$) for
$\sim600$ galaxies with $R_\mathrm{c}\le20.5$, within a slightly
smaller field of $\sim 0.25$ deg$^2$=$1.02^\circ\times 0.24^\circ$
\citep{bellanger95a}. The $R_\mathrm{c}\le20.5$ sample has a 92\%
redshift completeness and its median redshift and effective depth are
$z\simeq$~0.3 and $z\sim$~0.5, respectively.  Additional redshifts for
$\sim 250$ galaxies with $20.5< R_\mathrm{c} \le 21.5$ were also
measured in the same area, leading to a 52\% redshift completeness to
this fainter limit (see \citealp{lapparent03} for details).

The ESS spectroscopic catalogue was also used to perform a
template-free spectral classification based on a principal component
analysis of the flux-calibrated spectra, which yields a well-defined
sequence parameterized continuously using 2 indices denoted $\delta$
and $\theta$ \citep{galaz98,lapparent03}: $\delta$ measures the shape
of the continuum, hence the relative contribution from red and blue
stellar populations, whereas the departures of $\theta$ from the
sequence measure the strength of the nebular emission lines, hence the
current star formation rate. Comparison with the Kennicutt templates
\citep{kennicutt92} shows that the $\delta-\theta$ sequence is
strongly related to the Hubble morphological type \citep[see
also][]{folkes96,bromley98,baldi01}, and provides a better estimate of
the galaxy Hubble type than any color information \citep[as used for
example in][]{lilly95,lin99}.

The ESS catalogue is also complemented by precise type-dependent
K-corrections derived from the joint use of the spectral
classification and the PEGASE spectrophotometric models
\citep{fioc97}. These in turn yield absolute magnitudes in the
rest-frame filter bands (Johnson-Cousins $BVR_\mathrm{c}$), from which
\citet{lapparent03} have derived detailed luminosity functions as a
function of spectral type. Here, we make use of these various
parameters and characteristics of the ESS survey, to measure the
two-point correlation functions.


\subsection{The galaxy samples                \label{samples}}

The ESS spectroscopic sample was selected in $R_\mathrm{c}$ magnitude,
and is thus most complete in this band, whereas the $V$ and $B$
samples suffer color-related selection effects at faint magnitudes
\citep[see][]{lapparent03}. The present analysis is therefore based on
the $R_\mathrm{c}\le21.5$ redshift sample, and all the quoted absolute
magnitudes are in this band.  In order to use only galaxies with a
redshift value unaffected by peculiar motions within the local group,
we omit in all samples the few ESS galaxies with $z\le0.1$.  We also
reject distant galaxies with $z\ge0.51$: at these redshifts, the
selection function becomes of the order of $10$\% and decreases
steeply, which causes a very sparse sampling of the large-scale
structures.  The bounding redshifts $0.1$ and $0.51$ correspond to
comoving distances
\begin{equation}
\begin{array}{ll}
r_{\min}&=284.33\;h^{-1}~\mathrm{Mpc}\\
r_{\max}&=1338.04\;h^{-1}~\mathrm{Mpc},
\label{eq:rlim}
\end{array}
\end{equation}
respectively.  Within these redshift boundaries, the ESS contains 765
galaxies with a reliable redshift, which lie in the $R_\mathrm{c}$
absolute magnitude interval $-23 \leq M +5\log h \leq -16$.

To examine the varying clustering properties of the ESS with galaxy
type, we define various sub-samples by galaxy type and absolute
luminosity. We first consider the 3 spectral classes defined by
\citet{lapparent03}: early-type with $\delta\leq-5$, intermediate-type
with $-5<\delta\leq 3$, and late-type with $3<\delta$. As shown in
\citet{lapparent03}, projection of the \citet{kennicutt92} templates
onto the ESS classification space indicates that these 3 classes
approximately correspond to the following mixes of giant morphological
types: E + S0 + Sa in the early class; Sb + Sc in the
intermediate class; and Sc + Sd/Sm in the late class.

Furthermore, the analysis of the ESS luminosity functions suggests
that the ESS intermediate-type and late-type classes also contain
dwarf morphological populations \citep{lapparent03}: (1) gas-poor
dwarf galaxies which are classified as intermediate spectral type due
to their intermediate color, and most likely include dwarf elliptical
(dE) and dwarf lenticular (dS0) galaxies, together with their
nucleated analogs \citep{grant05}; here, these objects are altogether
denoted dE; (2) dwarf irregular (dI) galaxies, which are detected in
the late spectral class due to their richer gas content, hence blue
colors.

In the hierarchical scenario of galaxy formation, variations in the
clustering properties of high mass and low mass galaxies are predicted
\citep{pierce01,yang04}, which lead to the expectation of variations
in the clustering properties of giant and dwarf galaxies.  To separate
the giant and dwarf galaxy populations which are mixed within the
intermediate-type and late-type ESS spectral classes, we take
advantage of their respective relative contribution at the bright and
faint ends of the corresponding luminosity functions (see \fg 11 in
\citealt{lapparent03} or \fg 1 of \citealt{lapparent04}): we apply an
$R_\mathrm{c}$ absolute luminosity cut at $M(R_\mathrm{c})=-19.3$ for
the intermediate-type galaxies, and at $M(R_\mathrm{c})=-19.57$ for
the late-type galaxies.  We then merge the bright, \resp\ faint,
populations of both spectral classes, to built 2 samples which are
expected to include essentially:
\begin{itemize}
\item late spiral galaxies: Sb + Sc + Sd/Sm;
\item dwarf galaxies: dE + dI.
\end{itemize}

To examine the dependence of the two-point correlation function on the
absolute luminosity, we further divide the giant galaxy classes into
``bright'' and ``faint'' sub-samples defined by the median
$R_\mathrm{c}$ absolute magnitude of the sample. Moreover, because the
ESS exhibits a marked over-dense region in the interval $0.41\le
z<0.44$ which causes a strengthening of the clustering signal (see
\sct\ref{xis_overdens}), we also consider the sub-samples by galaxy
type after removal of all galaxies in that particular redshift
interval.  At last, for further testing the sensitivity of the
correlation function to cosmic variance, we also exclude the
under-dense region defined by $0.34\le z\le 0.39$ (see
\sct\ref{xis_ls}). The characteristics of these various sub-samples
are listed in Table~\ref{T1}.

\begin{table*}[t]
\caption{Definition of the ESO-Sculptor survey sub-samples used for calculation
of the two-point correlation function.}
\label{T1}
\begin{center}
\begin{tabular}{lccr@{$\:\;<\;\;$}c@{$\;\;\leq\;\:$}rrc}
\hline
\hline
Sub-sample & Redshift range & $R_\mathrm{c}$ absolute magnitude range & \multicolumn{3}{c}
{\ Spectral range} & \multicolumn{1}{c}{$N_\mathrm{d}$} & $<\delta>$ \\
(1) & (2) & (3) & \multicolumn{3}{c}{(4)} & \multicolumn{1}{c}{(5)} & (6)  \\
\hline
\multicolumn{2}{l}{\bf all galaxies:}    & & & &   \\
-- with the over-density                & $]0.1;0.51]$             & $-23.0 \leq M-5\log h\leq -16.0$         & $-20$ & $\delta$ & $20$ & 765 & $-0.53$  \\
-- without the over-density             & $]0.1;0.51]-[0.41;0.44]$ & $-23.0 \leq M-5\log h\leq -16.0$         & $-20$ & $\delta$ & $20$ & 654 & $-0.09$  \\
-- without the under-density            & $]0.1;0.51]-[0.34;0.39]$ & $-23.0 \leq M-5\log h\leq -16.0$         & $-20$ & $\delta$ & $20$ & 709 & $-0.44$  \\
\hline
\multicolumn{2}{l}{\bf early-type:}     & & & &   \\
-- with the over-density                & $]0.1;0.51]$             & $-23.0 \leq M-5\log h\leq -16.0$         & $-20$ & $\delta$ & $-5$ & 274 & $-8.46$  \\
$\;\;\bullet$ bright early-type         & $]0.1;0.51]$             & $-23.0 \leq M-5\log h\leq -21.14$        & $-20$ & $\delta$ & $-5$ & 137 & $-8.63$  \\
$\;\;\bullet$ faint early-type          & $]0.1;0.51]$             & $-21.14\leq M-5\log h\leq -16.0$         & $-20$ & $\delta$ & $-5$ & 137 & $-8.29$  \\
-- without the over-density             & $]0.1;0.51]-[0.41;0.44]$ & $-23.0 \leq M-5\log h\leq -16.0$         & $-20$ & $\delta$ & $-5$ & 218 & $-8.43$  \\
$\;\;\bullet$ bright early-type         & $]0.1;0.51]-[0.41;0.44]$ & $-23.0 \leq M-5\log h\leq -21.14$        & $-20$ & $\delta$ & $-5$ &  97 & $-8.45$  \\
$\;\;\bullet$ faint early-type          & $]0.1;0.51]-[0.41;0.44]$ & $-21.14\leq M-5\log h\leq -16.0$         & $-20$ & $\delta$ & $-5$ & 121 & $-8.41$  \\
-- without the under-density            & $]0.1;0.51]-[0.34;0.39]$ & $-23.0 \leq M-5\log h\leq -16.0$         & $-20$ & $\delta$ & $-5$ & 245 & $-8.45$  \\
\hline
\multicolumn{2}{l}{\bf intermediate-type:} & & & &   \\
-- with the over-density                & $]0.1;0.51]$             & $-23.0 \leq M-5\log h\leq -16.0$         &  $-5$ & $\delta$ &  $3$ & 240 & $-1.04$  \\
$\;\;\bullet$ bright intermediate-type  & $]0.1;0.51]$             & $-23.0 \leq M-5\log h\leq -19.30$        &  $-5$ & $\delta$ &  $3$ & 207 & $-1.15$  \\
-- without the over-density             & $]0.1;0.51]-[0.41;0.44]$ & $-23.0 \leq M-5\log h\leq -16.0$         &  $-5$ & $\delta$ &  $3$ & 206 & $-1.03$  \\
$\;\;\bullet$ bright intermediate-type  & $]0.1;0.51]-[0.41;0.44]$ & $-23.0 \leq M-5\log h\leq -19.30$        &  $-5$ & $\delta$ &  $3$ & 173 & $-1.16$  \\
-- without the under-density            & $]0.1;0.51]-[0.34;0.39]$ & $-23.0 \leq M-5\log h\leq -16.0$         &  $-5$ & $\delta$ &  $3$ & 231 & $-1.02$  \\
\hline
\multicolumn{2}{l}{\bf late-type:} & &  & &   \\
-- with the over-density                & $]0.1;0.51]$             & $-23.0 \leq M-5\log h\leq -19.30/-19.57$ &   $3$ & $\delta$ & $20$ & 251 & $+8.61$  \\
$\;\;\bullet$ bright late-type          & $]0.1;0.51]$             & $-23.0 \leq M-5\log h\leq -19.57$        &   $3$ & $\delta$ & $20$ & 125 & $+7.53$  \\
-- without the over-density             & $]0.1;0.51]-[0.41;0.44]$ & $-23.0 \leq M-5\log h\leq -16.0$         &   $3$ & $\delta$ & $20$ & 230 & $+8.66$  \\
$\;\;\bullet$ bright late-type          & $]0.1;0.51]-[0.41;0.44]$ & $-23.0 \leq M-5\log h\leq -19.57$        &   $3$ & $\delta$ & $20$ & 106 & $+7.53$  \\
-- without the under-density            & $]0.1;0.51]-[0.34;0.39]$ & $-23.0 \leq M-5\log h\leq -16.0$         &   $3$ & $\delta$ & $20$ & 233 & $+8.55$  \\
\hline
\multicolumn{2}{l}{\bf late spiral galaxies:} & &  & &   \\
-- with the over-density                & $]0.1;0.51]$             & $-23.0 \leq M-5\log h\leq -19.30/-19.57$ &  $-5$ & $\delta$ & $20$ & 332 & $+2.11$  \\
$\;\;\bullet$ bright late spiral        & $]0.1;0.51]$             & $-23.0 \leq M-5\log h\leq -20.56$        &  $-5$ & $\delta$ & $20$ & 166 & $+0.65$  \\
$\;\;\bullet$ faint  late spiral        & $]0.1;0.51]$             & $-20.56\leq M-5\log h\leq -19.30/-19.57$ &  $-5$ & $\delta$ & $20$ & 166 & $+3.59$  \\
-- without the over-density             & $]0.1;0.51]-[0.41;0.44]$ & $-23.0 \leq M-5\log h\leq -19.30/-19.57$ &  $-5$ & $\delta$ & $20$ & 279 & $+2.14$  \\
$\;\;\bullet$ bright late spiral        & $]0.1;0.51]-[0.41;0.44]$ & $-23.0 \leq M-5\log h\leq -20.56$        &  $-5$ & $\delta$ & $20$ & 127 & $+0.65$  \\
$\;\;\bullet$ faint  late spiral        & $]0.1;0.51]-[0.41;0.44]$ & $-20.56\leq M-5\log h\leq -19.30/-19.57$ &  $-5$ & $\delta$ & $20$ & 152 & $+3.39$  \\
\hline
\multicolumn{2}{l}{\bf dwarf galaxies:} & & & &   \\
-- with the over-density                & $]0.1;0.51]$             & $-19.30/-19.57 \leq M-5\log h\leq -16.0$ &  $-5$ & $\delta$ & $20$ & 159 & $+7.59$  \\
$\;\;\bullet$ faint intermediate-type   & $]0.1;0.51]$             & $-19.30\leq M-5\log h\leq -16.0$         &  $-5$ & $\delta$ &  $3$ &  33 & $-0.37$  \\
$\;\;\bullet$ faint late-type           & $]0.1;0.51]$             & $-19.57\leq M-5\log h\leq -16.0$         &   $3$ & $\delta$ & $20$ & 126 & $+9.68$  \\
\hline
\end{tabular}
\smallskip
\\
\end{center}
\begin{list}{}{}
\item[\underline{Definition of columns:}]
\item(1) identification of the sub-sample; (2) redshift interval; (3) interval of absolute magnitude; 
(4) interval of spectral index $\delta$; (5) number of galaxies in the sub-sample; 
(6) average spectral type $\delta$ for the sub-sample.
\end{list}
\end{table*}


\subsection{The luminosity functions          \label{lf}}

\begin{table*}
\caption{Parameters of the Gaussian and Schechter components of the
  composite luminosity functions fitted to the ESO-Sculptor spectral
  classes at $R_\mathrm{c}\le21.5$.}
\label{T2}
\begin{center}
\begin{tabular}{lcccc}
\hline
\hline
\multicolumn{2}{l}{Luminosity function parameters}             & early-type                    & intermediate-type                 & late -type                       \\
\hline
\multicolumn{2}{l}{\emph{Gaussian component}:} & \\
                         & Morphol. content    & E + S0 + Sa                   & Sb + Sc                 & Sc + Sd/Sm                  \\
                         & $M_0-5\log h$       & $-20.87\pm0.23$               & $-20.27\pm0.21$         & $-19.16\pm0.29$             \\
                         & $\sigma$            & $0.84\pm0.24$ / $1.37\pm0.36$ & $0.91\pm0.18$           & $0.97\pm0.13$               \\
                         & $\phi_0$            & $0.00333$                     & $0.00326$               & $0.00194[1+3.51\;(z-0.15)]$ \\
\hline
\multicolumn{2}{l}{\emph{Schechter component}:} & \\
                         & Morphol. content    &                               & dE                    & dI                          \\
                         & $M^*-5\log h$       &                               & $-19.28\pm0.37$         & $-18.12\pm0.22$             \\
                         & $\alpha$            &                               & $-1.53\pm0.33$          & $-0.30$                     \\
                         & $\phi^*$            &                               & $0.00426$               & $0.02106[1+3.51\:(z-0.15)]$ \\
  
\hline
\end{tabular}
\smallskip
\\
\end{center}
\begin{list}{}{}
\item[\underline{Notes:}]
\item[--] For the early-type luminosity function, the 2 listed values
of $\sigma$ are $\sigma_a$ / $\sigma_b$ (see
\eq \ref{eq:lf-gauss-2wing}).
\item[--] The amplitudes $\phi_0$ and $\phi^*$ are in units of\phiunit.
\end{list}
\end{table*}

For the luminosity function $\phi(M)$, we use the composite luminosity
functions proposed by \citet{lapparent03}: a two-wing Gaussian
function for the early-type galaxies, and the sum of a Gaussian
function and a \citet{schechter76} function for the intermediate-type
and late-type galaxies. These composite fits are motivated by their
better adjustment to the ESS luminosity functions than pure Schechter
functions, and by their good agreement with the luminosity functions
per galaxy type measured locally (\citealt{jerjen97b}; see
\citealt{lapparent03} for details). The composite fits of the ESS
luminosity functions also confirm the morphological content of the ESS
spectral classes in terms of giant galaxies and dwarf galaxies.  

Table~\ref{T2} lists the various giant and dwarf components of the
luminosity functions for the 3 spectral classes and the corresponding
parameters. The two-wing Gaussian luminosity function for the
early-type galaxies is parameterized as
\begin{equation}
\label{eq:lf-gauss-2wing}
\begin{array}{ll}
\phi(M)\; \mathrm{d}M & = \phi_0 e^{-( M_0 - M ) ^2 / 2 \sigma_a^2}\; \mathrm{d}M \; {\rm for}\; M\le M_0\\
           & = \phi_0 e^{-( M_0 - M ) ^2 / 2 \sigma_b^2}\; \mathrm{d}M \; {\rm for}\; M\ge M_0\\
\end{array}
\end{equation}
where $M_0$ is the peak magnitude, and $\sigma_a$ and $\sigma_b$ are
the dispersion values for the 2 wings. The Gaussian component of the
intermediate-type and late-type luminosity functions is parameterized
as
\begin{equation}
\label{eq:lf-gauss}
\phi(M)\; \mathrm{d}M = \phi_0 e^{-( M_0 - M ) ^2 / 2 \sigma^2}\; \mathrm{d}M,
\end{equation}
where $M_0$ and $\sigma$ are the peak and rms dispersion
respectively.  The \citet{schechter76} component of the
intermediate-type and late-type luminosity functions is parameterized
as
\begin{equation}
\begin{array}{ll}
\label{eq:lf-schec-mag}
\phi(M)\; \mathrm{d}M & = 0.4 \ln 10\; \phi^* e^{-X} X^{\alpha+1}\; \mathrm{d}M \\
{\rm with}& \\
X & \equiv {L\over L^*} = 10^{\;0.4\,(M^* - M)} \\
\end{array}
\end{equation}
where $M^*$ is the characteristic magnitude, and $\alpha+1$ the
``faint-end slope''.  For the late spiral and dwarf sub-samples, we
use the bright and faint parts \resp~of the luminosity functions by
spectral type (see Table~\ref{T1}).

The parameters listed in Table~\ref{T2} are those derived by
\citet{lapparent04} from the $R_\mathrm{c}\le21.5$ sample.  For the
early-type and intermediate-type classes, the listed values of
$\phi_0$ are those listed as $\Phi_1(0.51)$ in Table 2 of
\citet{lapparent04}, derived using an ``equal pair'' weighting for the
mean density estimator (see \eqs\ref{eq:mean-dens} and \ref{eq:w1});
they thus yield a total expected number of galaxies over the redshift
interval $0.1-0.51$ for a homogeneous distribution which is equal to
the observed number of galaxies in each spectral class.  For the
late-type class, we use (and list here in Table~\ref{T2}) the
parameterization $\Phi_1(z)$ provided in Table 2 of
\citet{lapparent04}, which reflects the marked evolution of the
amplitudes $\phi_0$ and $\phi^*$ for the ESS late-type galaxies. This
evolution could also be due to pure luminosity evolution, but the
available data do not allow us to discriminate between the 2 effects
\citep{lapparent04}. Whatever the nature of this evolution, the
parameterized amplitude evolution listed in Table~\ref{T2} does allow
us to estimate the selection function (\sct\ref{selfunc}) required for
calculating correlation functions.

For the intermediate-type and late-type class, the values of $\phi^*$
are derived from the values of $\phi_0$ using the ratios
\begin{eqnarray}
{\phi_0\over0.4\ln10\;\phi^*}=&0.83 &\quad\mathrm{for}\quad\mathrm{intermediate-type}\\ 
{\phi_0\over0.4\ln10\;\phi^*}=&0.10 &\quad\mathrm{for}\quad\mathrm{late-type}
\end{eqnarray}
also provided by \citet[][ see their Table 1]{lapparent04}.  

\begin{figure}
\resizebox{\hsize}{!}{\includegraphics{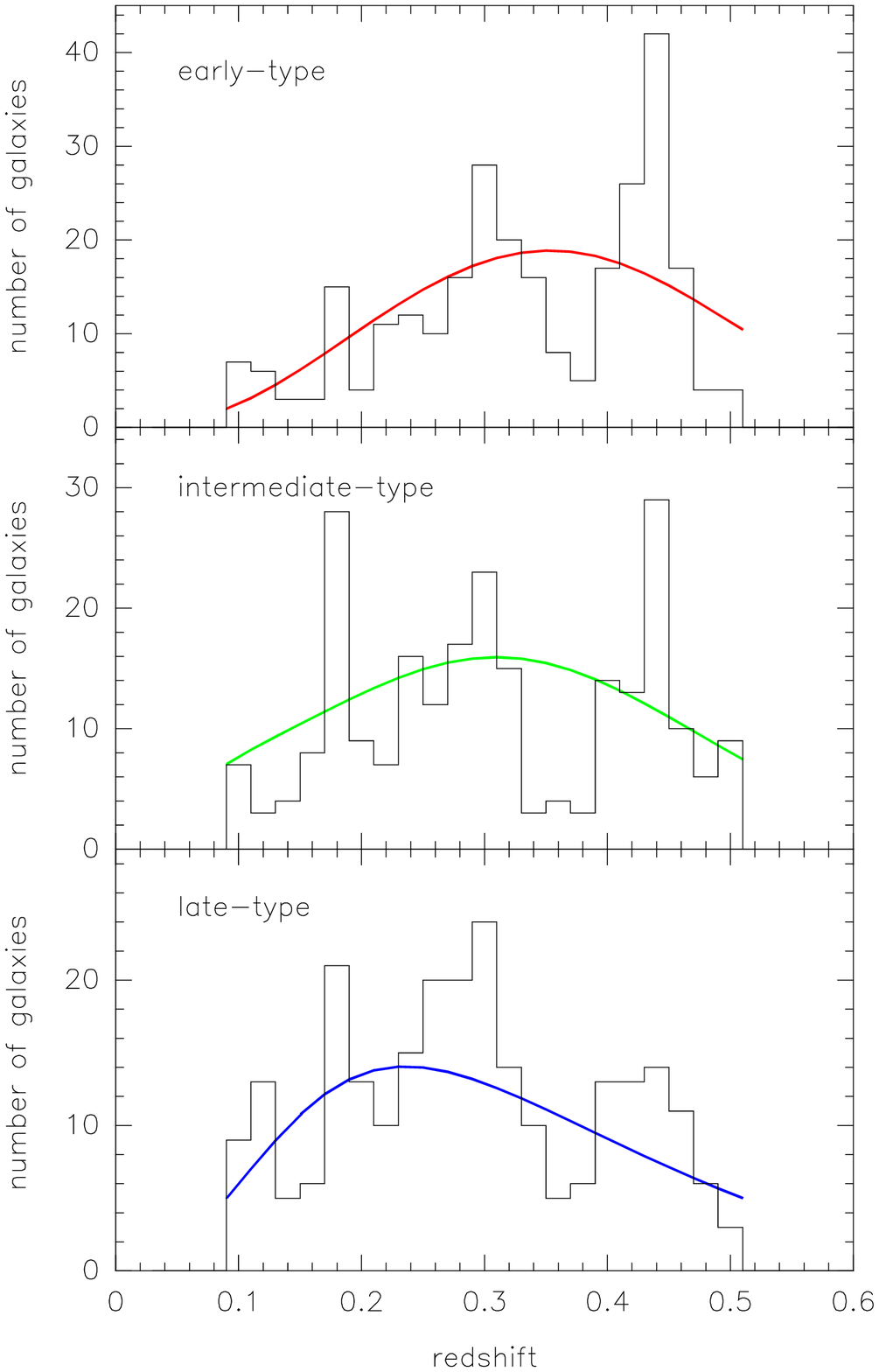}}
\caption{Redshift histograms for the early-type (top panel),
intermediate-type (middle panel), and late-type (bottom panel)
spectral classes of the ESO-Sculptor survey, using a redshift bin of
$0.02$. The solid curves show the expected distributions for a
homogeneous sample given the luminosity functions defined in
Table~\ref{T2}, the magnitude limits and the angular coverage of the
survey.}
\label{histo-type}
\end{figure}
\begin{figure}
\resizebox{\hsize}{!}{\includegraphics{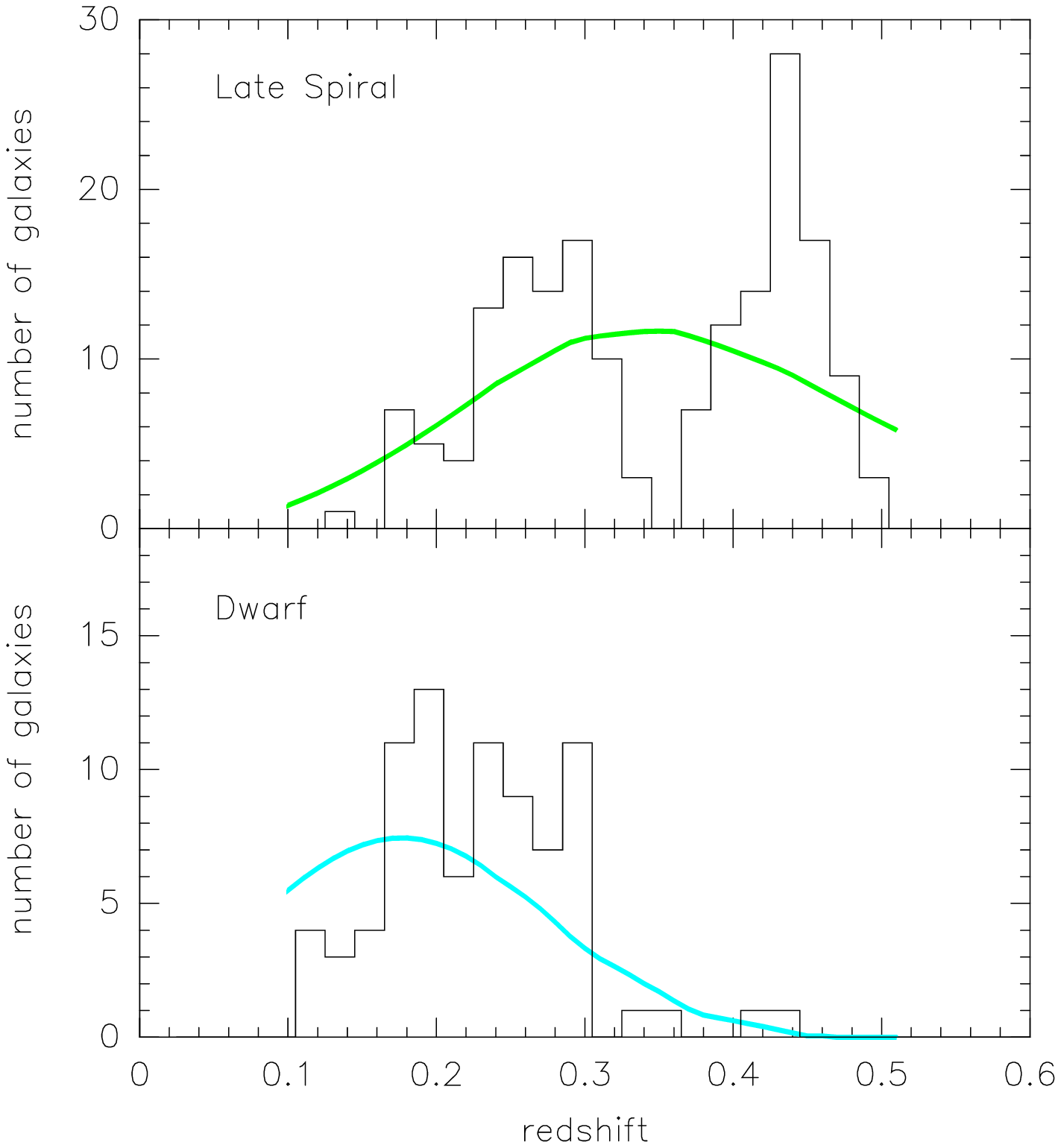}}
\caption{Redshift histograms for the late spiral (top panel) and dwarf
(bottom panel) sub-samples of the ESO-Sculptor survey (see
\fg\ref{histo-type} for details).}
\label{histo-il}
\end{figure}

These various luminosity functions allow us to derive the
corresponding selection functions affecting the ESS sub-samples (see
\sct\ref{selfunc}). These yield the expected redshift distribution for
homogeneous samples, which are shown in \fgs\ref{histo-type} and
\ref{histo-il} and are compared with the ESS observed distributions
for each spectral class and galaxy type.  To calculate the expected
redshift distribution, we also use for each sub-sample the
$K$-correction function $K(z,\delta)$ provided by \citet{lapparent04}
at the mean value of the spectral-type $\delta$ listed in the
Table~\ref{T1}. Note that comparison of the observed and expected
distributions for the 3 spectral classes shown in \fg\ref{histo-type}
confirms the validity of the luminosity functions and amplitudes
listed in Table~\ref{T2}, and in particular validates the
parameterization of the amplitude evolution for the late-type
galaxies.


\section{Computing the correlation function and its uncertainties    \label{errors}}

The formalism for derivation of the comoving distances $r$ from the
redshifts, for calculating the three different estimators of the
redshift-space and projected correlation function (Davis-Peebles DP,
Landy-Szalay LS, and Hamilton H), the various selection functions and
the corresponding three weighting schemes (J3, ``equal volume'' EV,
and ``equal pair'' EP), the normalization, and the mean density are 
defined in the Appendix.  Because the variance in the estimates of
$\xi(s)$ depends on the chosen statistical weights, it is useful to
apply the three weighting schemes to each of the 3 estimators of the
correlation function, and to cross-check the results; this is done in
the \sct\ref{xis}. Here, as a preliminary, we describe and estimate
the various uncertainties at play in the correlation function
measurements.


\subsection{Statistical noise in $\xi(s)$  \label{errors_xis}}

From the experimental point of view, statistical error bars in
$\xi(s)$ would be best determined from the ensemble error by splitting
the survey into independent regions containing approximately equal
numbers of galaxies and taking the standard deviation of the
correlation function estimates calculated in these
sub-areas. Unfortunately, the size of the ESS redshift sample is too
small to allow for such an approach. Here we consider both the Poisson
error
\begin{equation}
\sigma_\mathrm{Poiss}(s) = [\ 1+\xi(s)\ ]\ DD(s)^{-1/2};
\label{eq:sigma-poisson}
\end{equation}
and the bootstrap error $\sigma_\mathrm{boot}(s)$, which is derived as
the standard deviation in the estimates $\xi_{(k)}$ from 50 randomly
data sets of $N_\mathrm{d}$ points obtained from the original
$N_\mathrm{d}$ galaxies by re-sampling with replacement, without any
correction for possible systematic biases \citep{ling86}.  The Poisson
error is an underestimate of the true uncertainty for correlated data
but is appropriate in the regime of weak clustering, \ie at large
scales ($s\ga10$\hmpc). In contrast, the bootstrap error tends to
overestimate the true error in over-dense regions by roughly a factor
two, as shown by \citet{fisher94a}. In the following, we thus adopt
the bootstrap error at all scales $s$ as the statistical random
uncertainty in the measured values of $\xi(s)$ and
$\xi(r_\mathrm{p},\pi)$.


\subsection{Uncertainties from distances   \label{errors_z}}

A priori, derivation of distances for the ESS galaxies requires to
correct the ESS velocities for the motion within the Local Group
\citep{yahil77,courteau99}, the infall onto Virgo
\citep{ramella97,ekholm01}, and the cosmic microwave background dipole
\citep{smoot91}.  We estimate the impact of these velocity corrections
by calculating the difference in correction value for 4 imaginary
points located at the extreme 4 ``corners'' or the ESS survey
region. Each ``corner'' point is defined by the constant value of
either one of the 2 coordinates RA (J2000), Dec (J2000) among
\begin{equation}
\begin{array}{rlrl}
\mathrm{RA}_{\min}&=0^\mathrm{h}19.0^\mathrm{m}\quad
&\mathrm{RA}_{\max}&=0^\mathrm{h}23.5^\mathrm{m}\\
\mathrm{Dec}_{\min}&=-30^\circ 14^\prime\quad
&\mathrm{Dec}_{\max}&=-29^\circ 58^\prime,
\label{eq:corners}
\end{array}
\end{equation}
while the other coordinate takes the 2 possible values.
The largest resulting velocity differences are 2.5\kms~for
pairs of points in which only RA varies; 1.4\kms~for pairs of
points in which only Dec varies.  The largest velocity difference
of 3.7\kms~is obtained for the pair of points with coordinates
(RA$_{\min}$,Dec$_{\max})$ and (RA$_{\max}$,Dec$_{\min})$.
These values are negligeable compared to the ESS rms uncertainty in
the velocities of $\sigma(v)\sim165$\kms. 
We thus neglect the effects of the 3 mentioned systematic motions in
the estimation of distances.

The uncertainties in the ESS redshifts ($\sigma(z)\sim0.00055$) is
also a source of random error in the measurement of the correlation
function. To estimate its impact, we add to all galaxies in the ESS
early-type sub-sample an additional dispersion in the redshift $\delta
z$ defined by a Gaussian probability distribution centered at zero and
with an rms deviation $\sigma(z)\sim0.00055$. We repeat this
procedure 30 times, and measure an rms dispersion in
$\xi_\mathrm{LS}(s)$ at scales $1<s<15$\hmpc\ which is 2 to 3 times
smaller than the bootstrap uncertainty. Similar results are obtained
using the intermediate-type and late-type sub-samples. We thus neglect
this additional source of random error.


\subsection{Uncertainties from the luminosity function            \label{errors_lf}}

As far as systematic errors are concerned, a dominant contribution to
the correlation function may be the uncertainty in the selection
function (\sct\ref{selfunc}), which results from the uncertainties in
the luminosity function. As shown by \citet{peebles80}, the quadratic
H estimator (\eq\ref{eq:xi-H}) is more affected than the DP and LS
estimators (\eqs\ref{eq:xi-DP} and \ref{eq:xi-LS}; see
\sct\ref{estimators}). For power-law correlation functions
(\eq\ref{eq:xi-power}), such biases have more impact on the
correlation length $s_0$ than on the power-law index $\gamma$.  As
described in \sct\ref{lf} and Table~\ref{T2}, five luminosity functions
are involved in the computation according to the considered spectral
class.  In principle, any derivation of $s_0$ and $\gamma$ should be
repeated by varying the selection function parameters along the six
principal axes of its error ellipsoid, and the scatter among the
results added in quadrature to their statistical uncertainties. In
practise, we change the value of $M_0$ for the early-type and
intermediate-type luminosity functions, and the value of the Schechter
slope $\alpha$ for the intermediate-type and late-type luminosity
functions by plus or minus their rms uncertainty (listed in
Table~\ref{T2}). The resulting systematic shifts in
$\xi_\mathrm{LS}(s)$ for $1\le s\le5$\hmpc\ are $\pm4.2$\% and
$\pm8.6$\% when changing $M_0$ for the early-type and
intermediate-type samples \resp; and $^{+4.0\%}_{-3.5\%}$ and
$^{+9.0\%}_{5.3\%}$ when changing $\alpha$ for the intermediate-type
and late-type samples respectively; these various shifts are $\sim4$
to $10$ times smaller than the bootstrap errors for each sample.


\subsection{Uncertainties from the $J_3$ weighting              \label{errors_J3}}

Another source of uncertainty is the choice of the parameterization
for $J_3(s)$ in the case of a $J_3$ weighting scheme (see
\eq\ref{eq:w-J3}).  We estimate this uncertainty by comparing
$\xi_\mathrm{LS}(s)$ for the ESS early-type sub-sample with that using
the following $J_3$ weighting parameterization
\begin{equation}
\begin{array}{lll}
J_3(s)&=14.98\;s^{1.4}\ h^{-3}\ \mathrm{Mpc}^3 &\quad\mathrm{for}\quad s\le s_\mathrm{c},\\ 
J_3(s)&=1752\ h^{-3}\ \mathrm{Mpc}^3 &\quad\mathrm{for}\quad s>s_\mathrm{c}.
\end{array}
\label{eq:J3-power-2}
\end{equation}
(with $s_\mathrm{c}\equiv30$\hmpc). This other parameterization of
$J_3(s)$ is obtained from the power-law description of the correlation
function (\eq\ref{eq:xi-power}) with a $10$\% increase of $s_0$ above
the value given in \eq\ref{eq:xi-power-param}.  The relative change in
$\xi_\mathrm{LS}(s)$ is then $\la 1.1$~\% at all scales, which is one
to two orders of magnitude smaller than the bootstrap random
uncertainties.  Performing a similar test in which only the slope
$\gamma$ of the power-law parameterization is increased by $10$\% over
the value in \eq\ref{eq:xi-power-param} yields slightly larger
variations in $\xi(s)$ ($<2.2$~\%). Decreasing either $s_0$ or
$\gamma$ by similar amounts yields the corresponding opposite shifts in
$\xi(s)$. The systematic uncertainties in $J_3(s)$ therefore have a
very small impact on the error budget in $\xi(s)$.


\subsection{Cosmic bias                                         \label{cosmic_bias}}

The correlation function is also affected by a systematic bias called
``cosmic bias'', which is caused by using the observed density of
galaxies in the sample for normalization of the number of pairs (see
\sct\ref{norm}). This results in an implicit normalization to zero of
the integral of the correlation function over the survey
volume. However, in finite samples, the two-point correlation function
is positive out to scales of $\sim20$\hmpc, and the mean density of
galaxies estimated from a sample of comparable scale is an
over-estimate of the mean density of the Universe.  The corresponding
bias in the correlation function is then expected to be negative. In
that case, the cosmic bias can be corrected for by an additive
correction, called ``integral constraint''
\citep{ratcliffe96,brainerd95}. In contrast, for sample sizes of
$\sim100$\hmpc~ or larger, the presence of an under-density occupying
a large volume may instead lead to an under-estimation of the mean
density and a corresponding over-estimation of the correlation
function.

To estimate the cosmic bias for the ESS, we vary by $\pm4$\% (which
corresponds to $\sqrt{N}/N\simeq0.036$, where $N=765$ is the number of
galaxies in the full ESS sample, see Table~\ref{T1}) the amplitude
$\phi_0$ of the Gaussian component of the early-type,
intermediate-type and late-type luminosity functions, and the
amplitude $\phi^*$ of the Schechter components of the
intermediate-type and late-type luminosity functions (see
Table~\ref{T2}), and calculate the resulting $\xi(s)$ for the full ESS
sample.  The shift in $\xi(s)$ for the full ESS sample is $\la 4$\%
for $1\le s\le 9$\hmpc, which is 10 times smaller than the bootstrap
error; the shift then increases at larger scales, taking its largest
value at $s\simeq14-28$\hmpc, as it is comparable to the transverse
extent of the survey.

Nevertheless, we show in \sct\ref{xis_overdens} that a marked
over-dense region of the ESS causes a larger systematic shift in the
correlation function than the above estimated cosmic bias, and
furthermore, this shift is in the opposite direction (an excess
correlation).  We thus choose to neglect the standard ``cosmic bias'',
and instead, we evaluate the impact of this over-density onto the
various measured correlation functions in \scts\ref{xis_all},
\ref{xis_overdens}, \ref{xis_ls}, \ref{wrp_all} and \ref{wrp_type}. In
\sct\ref{wrp_overdens}, we further discuss the role of this structure
in terms of cosmic variance.


\section{The redshift-space correlation function $\xi(s)$                \label{xis}}


\subsection{General behavior of $\xi(s)$                                \label{xis_all}}

\begin{figure}
\resizebox{\hsize}{!}{\includegraphics{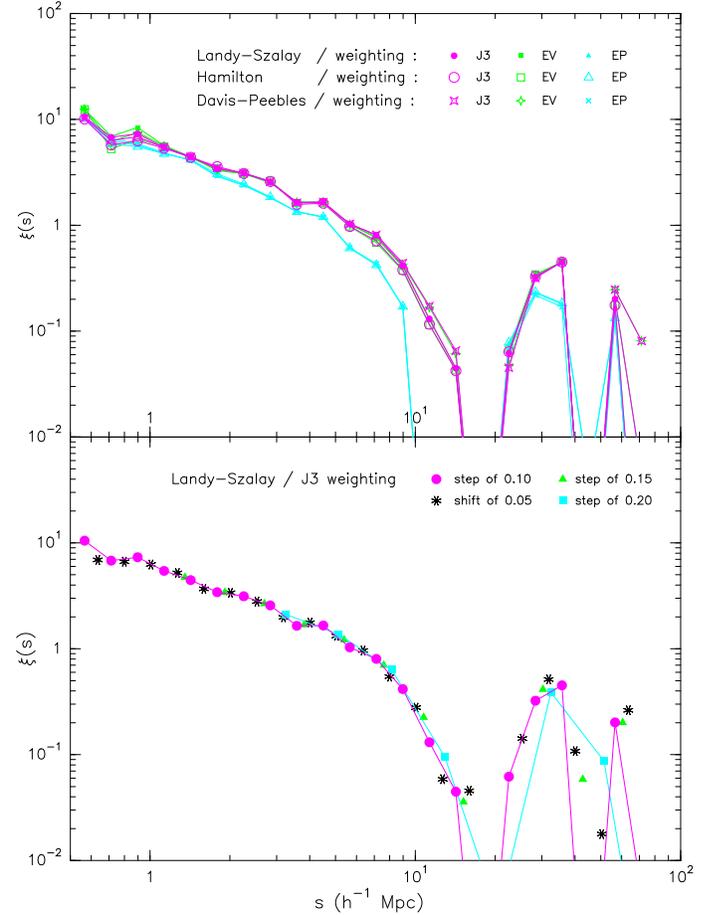}}
\caption{The redshift-space correlation function $\xi(s)$ for all
ESO-Sculptor galaxies with $0.10<z<0.51$. The top panel displays the
result obtained with a bin size of $\Delta\log(s)=0.10$ for the 3
estimators Landy-Szalay (filled symbols), Hamilton (open symbols),
Davis-Peebles (starred symbols), and the 3 weighting schemes J3
(magenta circles and diagonal stars), ``equal volume'' (denoted EV,
green squares and stars) and ``equal pair'' (denoted EP, cyan
triangles and cross). The bottom panel compares the Landy-Szalay
estimate of $\xi(s)$ with a $J3$-weighting for different values of the
bin size: $\Delta\log(s)=0.10$ (magenta circles), $\Delta\log(s)=0.15$
(green triangles), $\Delta\log(s)=0.20$ (cyan squares); the bin size
$\Delta\log(s)=0.10$ with a 0.05 shift in $\Delta\log(s)$ from the
origin is also plotted (black asterisks).  For sake of clarity, only
two sets of points are connected by a solid line.}
\label{ksi-all}
\end{figure}
\begin{figure}
\resizebox{\hsize}{!}{\includegraphics{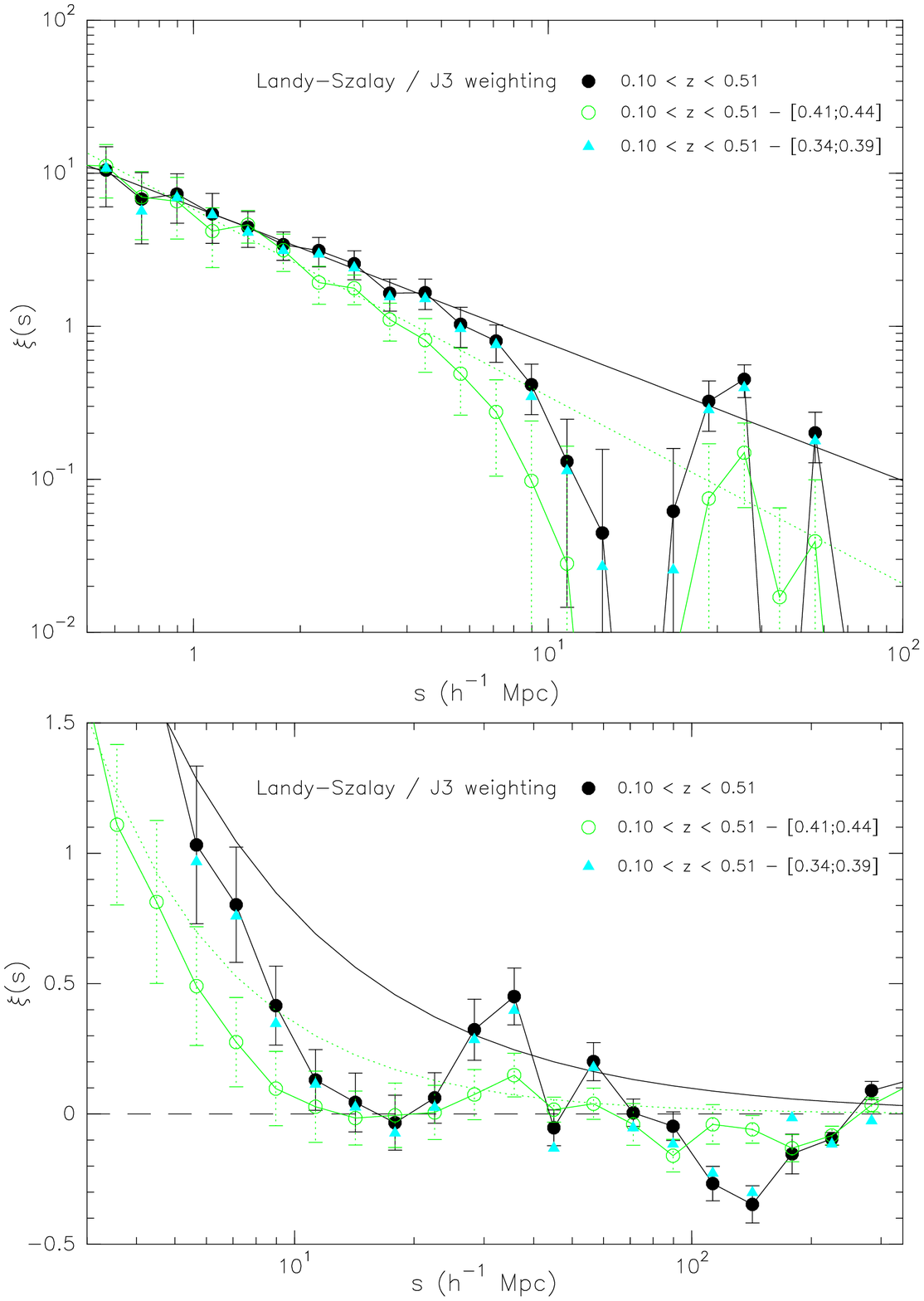}}
\caption{The redshift-space correlation function $\xi(s)$ for the
ESO-Sculptor galaxies, computed with the Landy-Szalay estimator using
the minimum-variance $J_3$-weighting scheme for (i) all galaxies with
$0.10<z<0.51$ (black filled circles); (ii) the sub-sample in which the
over-density in the interval $0.41<z<0.44$ is removed (green open
circles); (iii) the sub-sample in which the under-dense region in the
interval $0.34<z<0.39$ is removed (cyan triangles).  The bottom panel
shows the same curves in linear scale, restricted to $s>3$\hmpc. Both
panels display $\xi(s)$ with a bin size $\Delta \log(s)=0.10$, and
show the power-law model fitted to the full sample with $0.10<z<0.51$
in the interval $0.5<s<5.0$\hmpc\ (black solid line). The bootstrap
error bars for the sub-sample without the under-density at
$0.34<z<0.39$ are not shown, as they are similar to those for the full
sample.}
\label{ksi-wod}
\end{figure}

Top panel of \fg\ref{ksi-all} shows the 9 combinations of the 3
estimators H, LS and DP (see \sct\ref{estimators}) and the 3 weighting
schemes J3, EV and EP (see \sct\ref{weights}) applied to the full ESS
sample limited to $R_\mathrm{c}\le21.5$, $-23<M(R_\mathrm{c})-5\log
h<-16$ and $0.10<z<0.51$ (765 objects, see Table~\ref{T1}).
Comparison of the various estimators confirms that the $J3$ weighting
behaves as the EP weighting at small separation $s$, and as the EV
weighting at large $s$.  For a given weighting scheme and at scales
$s\ga0.2$\hmpc, the differences between the 3 estimators are
significantly smaller than the error bars in each estimate. The
bootstrap errors for the LS estimate of $\xi(s)$ are shown in
\fg\ref{ksi-wod}; the bootstrap errors in the DP and H estimates are
comparable to those in the LS estimate, but have a less stable
behavior with varying separation.  For this reason, and because the
$J3$ weighting is the minimum variance weighting and does not favor
nearby nor distant pairs, we only show and examine in the following
the LS estimate of $\xi(s)$ with $J3$ weighting; note, however, that
the results and conclusions of the article are unchanged using the
other estimators and weighting schemes.

The various pairs of objects in top panel of \fg\ref{ksi-all} are
counted in logarithmic bins of equal size with $\Delta\log s =0.1$.
In the following, we adopt this bin size, as it allows one to probe
the small scale regime $s\la1$\hmpc. Note that even if $\xi(s)$ for
the ESS is biased on these scales, $w(r_\mathrm{p})$ can successfully
be measured down to $s\simeq0.2$\hmpc\ (see \sct\ref{wrp}). We show in
bottom panel of \fg\ref{ksi-all} that the overall behavior of $\xi(s)$
for the ESS is kept unchanged when shifting the bin set by half the
bin size, or when using larger bin sizes with $\Delta\log s =0.15$ and
$\Delta\log s =0.2$; similar conclusions are drawn for the correlation
functions $\xi(r_\mathrm{p},\pi)$ and $w(r_\mathrm{p})$ considered in
the following sections.

In \fg\ref{ksi-wod}, we show the LS estimate of $\xi(s)$ with $J3$
weighting from the full ESS sample (labeled as ``$0.10<z<0.51$''),
along with its bootstrap errors; in top panel, $\xi(s)$ is in
logarithmic scale, and in bottom panel, in linear scale. The
redshift-space correlation function has the usual power-law behavior
at small scales, and a smooth roll-off at $s\la10$\hmpc.  The
adjustment of a power-law (see \eq\ref{eq:xi-power}) in the interval
$0.5<s< 5.0$\hmpc\ yields a correlation length and correlation slope
of
\begin{equation}
s_0=7.49 \pm 3.18\ h^{-1}{\rm Mpc,} \qquad \gamma=0.90 \pm 0.13,
\label{eq:xi-fit}
\end{equation}
\resp; this power-law fit is shown as a solid line in both panels of
\fg\ref{ksi-wod}.  

Top panel of \fg\ref{ksi-wod} shows that at larger scales, $s\sim
10$\hmpc, $\xi(s)$ breaks down from the power-law fit. Then at
$s\simeq15$\hmpc, the amplitude of the correlation function crosses
zero, as seen in bottom panel of \fg\ref{ksi-wod}. $\xi(s)$ rises up
again at scales $25\la s\la 40$\hmpc, and another peak occurs at
$50<s<60$\hmpc. This large-scale behavior is discussed in the next
sub-sections.

\begin{figure*}
\centerline{\psfig{figure=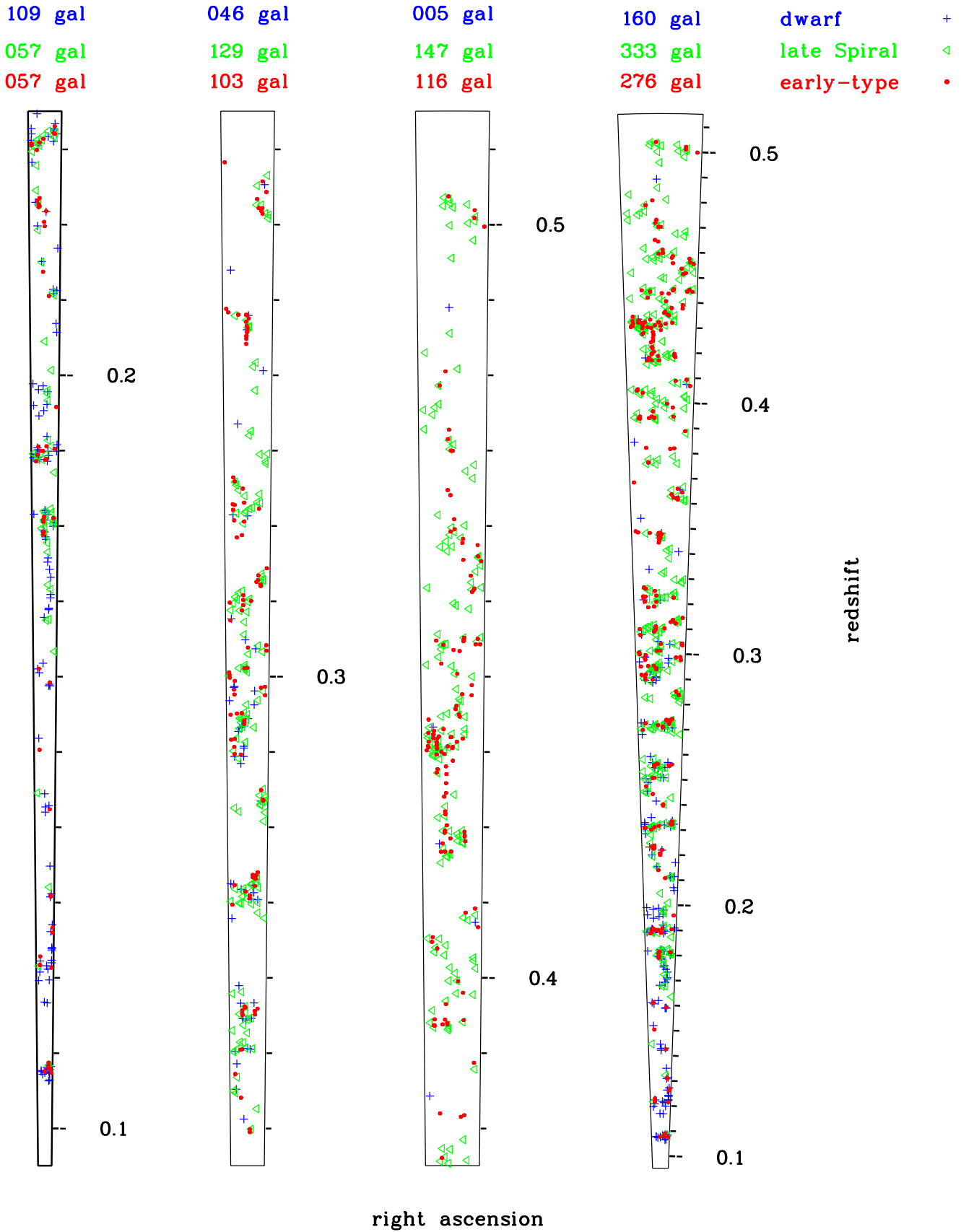}}
\caption{Redshift cone-diagrams for the ESO-Sculptor survey, truncated
into 3 redshift intervals (3 left cones). A total number of 769
galaxies with $R\leq 21.5$ and reliable redshift in the interval
$0.1\leq z\leq 0.51$ are plotted in the full cone (to the right),
which is stretched in angle by a factor 3.  Dwarf galaxies (blue
crosses) are plotted first, then the late spiral galaxies (green
triangles), and finally the early-type galaxies (red disks). These
graphs show that the survey intercepts many large-scale structures,
appearing as an alternation of voids and walls or filaments.  The
stronger clustering of the early-type galaxies over the late spiral
and dwarf galaxies is also visible.}
\label{z-cone}
\end{figure*}


\subsection{Impact of the over-density at $0.41<z<0.44$        \label{xis_overdens}}

Part of the deviations of $\xi(s)$ from null clustering at $s\ga
10$\hmpc\ (seen in bottom panel of \fg\ref{ksi-wod}) appear to be due
to the presence of a marked over-density in the redshift interval
$0.41< z<0.44$. This structure is clearly seen in the redshift-cones
of the survey shown in \fg\ref{z-cone}. It also appears in the
redshift histograms shown in \fg\ref{histo-type} as an integrated
excess by nearly a factor 2 over the expected number for a homogeneous
distribution (indicated by the solid lines) for the early-type and
late-type galaxies.

The ESS over-dense region at $0.41< z<0.44$ has an impact on the
redshift-space correlation function at both intermediate and large
scales. Top panel of \fg\ref{ksi-wod} shows that when removing the
over-density, $\xi(s)$ shifts to lower amplitudes in the interval
$0.5\le s\le 5$\hmpc, which results in a lower amplitude and steeper
slope for the power-law fit:
\begin{equation}
s_0=4.22 \pm 1.15\ h^{-1}{\rm Mpc,} \qquad \gamma=1.22 \pm 0.15,
\label{eq:xi-fit-wod}
\end{equation}
(measured for $0.5<s<5$\hmpc).  If the change in $\xi(s)$ was solely
due to the change in the normalizing density, a higher amplitude of
$\xi(s)$ would be expected when removing the galaxies in the $0.41<
z<0.44$ interval, as this over-dense region tends to artificially
increase the mean density of the sample, thus contributing to the
cosmic bias (see \sct\ref{cosmic_bias}). The decreasing amplitude of
$\xi(s)$ at $s\ga2$\hmpc\ implies that galaxies in the $0.41< z<0.44$
region have stronger clustering at medium and large scale than the
average for the rest of the survey. Indeed, detailed examination of
\fg\ref{z-cone} indicates that the large-scale structure at $0.41<
z<0.44$ is a dense collection of groups of galaxies extending over
$\Delta z\sim0.025$, that is $\sim60$\hmpc\ in comoving distance. In
\sct\ref{xis_type} below, we show that both the early-type and late
spiral galaxies contribute to the excess clustering in this region.

When the galaxies in the redshift interval $0.41< z<0.44$ are removed
from the ESS full sample, the deviations of $\xi(s)$ from zero at
scales $9\la s\la 150$\hmpc\ (see bottom panel of \fg\ref{ksi-wod})
are reduced to less than the bootstrap uncertainty for most points.
In particular, the second peak at $50<s<60$\hmpc\ becomes
insignificant. The peak at $25\la s\la 40$\hmpc\ is also significantly
reduced to a ``marginal'' detection, as $\xi(s)$ deviates by less than
twice the bootstrap error for $s\simeq35$\hmpc.  One possible
interpretation of the shift towards negative values of $\xi(s)$ at
$s\sim 90$\hmpc, when removing the over-density within $0.41< z<0.44$,
could be the artificial anti-correlation thus created between this
empty region and the foreground/background walls of galaxies: the
difference in comoving distance between $z=0.41$ and $z=0.44$ is
$\simeq72$\hmpc.

When removing a given redshift interval from an ESS sample, the
selection function of each of the 3 spectral-type samples is assigned
to zero in that interval, and no points are generated in that redshift
interval of the random distributions. To check that the procedure does
not introduces any bias, we also calculate $\xi(s)$ for the full ESS
sample with all galaxies in the redshift interval $0.34< z<0.39$
removed: this region of the ESS has a similar volume as the region
defined by $0.41< z<0.44$, and corresponds to an under-dense region in
the ESS early-type and intermediate-type redshift histograms
(\fg\ref{histo-type}). The resulting $\xi(s)$ with the $0.34< z<0.39$
interval removed is over-plotted in both panels of \fg\ref{ksi-wod},
and compared to that for the full ESS sample: it shows negligeable
changes at all scales. Although the $0.34< z<0.39$ region is
under-dense compared to the rest of the ESS, it has little impact on
the two-point correlation function at all scales.


\subsection{Large-scale power in $\xi(s)$                               \label{xis_ls}}

The excess in $\xi(s)$ at $25\la s\la 40$\hmpc, seen in bottom panel
of \fg\ref{ksi-wod}, might also be due in part to the pencil-beam
geometry of the ESS. The large-scale clustering of galaxies is
characterized by walls and filaments which delineate large voids
\citep{lapparent86,shectman96,small97a,colless01,zehavi02}. This
results in an alternation of voids and narrow portions of walls or
filaments intercepted at a wide range of angles with the line-of-sight
of the survey, as one then expects few walls, and even fewer
filaments, to be intercepted parallel to the line-of-sight, and thus
to appear as an extended over-density along the line-of-sight (the
over-density at $0.41< z<0.44$ may however be one of these rare
occurrences). Then, the particular line-of-sight and limited volume
sampled by the ESS may define a typical scale for the
wall/filament separation, which would appear as excess signal in the
two-point correlation.

\begin{figure}
\resizebox{\hsize}{!}{\includegraphics{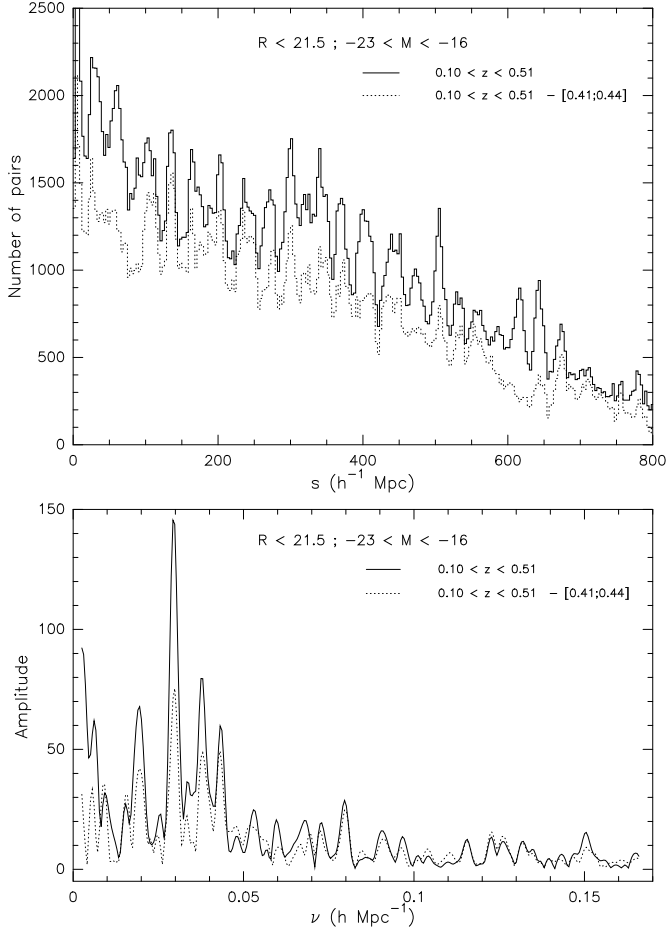}}
\caption{Pairs separations and the redshift-space correlation function
$\xi(s)$ at large scales. The top panel shows the histogram of the
data pair separations for all ESO-Sculptor galaxies with $0.10<z<0.51$. The
number of pairs decreases as the separation in comoving distance
increases, as expected in a pencil-beam survey. Superimposed on the
overall trend, numerous peaks are visible and define an apparently
regular pattern. The periodogram of the detrended signal is given in
the bottom panel. A period of 33.9\hmpc\ ($\nu=0.0295$) is 
evidenced. Also present are at least four other smaller peaks linked
to components with 26.4 and 23.3\hmpc\ periods and the related
multiples.}
\label{pair-sep}
\end{figure}

A pair separation analysis provides quantitative evidence for the
impact of the ESS pencil-beam geometry onto the detected large-scale
clustering.  Top panel of \fg\ref{pair-sep} shows the histogram of the
galaxy pair separations in comoving distance for the ESS full
sample. This distribution exhibits numerous peaks which seem to define
a preferred scale.  The bottom panel of \fg\ref{pair-sep} shows the
``periodogram'' obtained by a spectral density analysis of the pair
separation distribution from which the continuum has been subtracted.
A marked peak occurs at 34\hmpc\ ($\nu=0.0295$), which indicates an
increased probability of having pairs of ESS galaxies separated by
this scale. This in turn explains the excess signal in $\xi(s)$ at
$25\la s\la 40$\hmpc\ (bottom panel of \fg\ref{ksi-wod}). Note that
this scale corresponds to the mean interval picked up by the eye
between adjacent walls/filaments in the cone diagram of the ESS
(\fg\ref{z-cone}): 34\hmpc\ corresponds to $\Delta z=0.013$ at
$z=0.3$, that is 30\% larger than the tick mark separation. When the
over-dense region with $0.41< z<0.44$ is removed from the ESS, the
results of the pair separation analysis remain; the peak at 34\hmpc\
in the periodogram becomes however weaker, which confirms that the
over-dense region also contributes excess pairs in $\xi(s)$ at $25\la
s\la 40$\hmpc.

We emphasize that the result shown in \fg\ref{pair-sep} is \emph{not}
interpreted as evidence for periodicity in the galaxy distribution
(see \citealt{yoshida01}). It is symptomatic of the fact that the ESS
does not represent a fair sample of the galaxy distribution, due to
its limited volume: as a result, the alternation of voids and
walls/filaments along the line-of-sight, and the presence of one
over-density parallel to the line-of-sight, both leave an imprint in
the correlation function.


\subsection{$\xi(s)$ per galaxy type                \label{xis_type}}

\begin{figure}
\resizebox{\hsize}{!}{\includegraphics{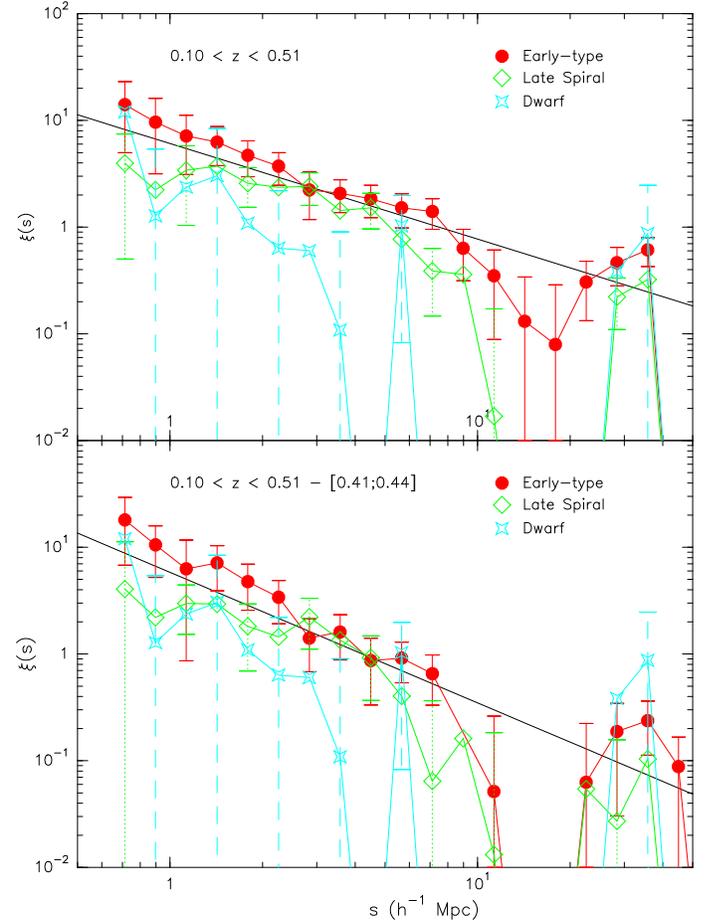}}
\caption{The redshift-space correlation function $\xi(s)$ for the ESO-Sculptor
sub-samples restricted to the 3 galaxy types: early-type (red
circles), late spiral (green diamonds), and dwarf galaxies (cyan open
stars). The top panel displays $\xi(s)$ for all samples with
$0.10<z<0.51$ while the bottom panel shows the results for the samples
in which galaxies within the over-density $0.41<z<0.44$ have been
excluded. The same correlation function for the dwarf galaxies (cyan
open stars) is shown in both panels as this sample contains no
galaxies with $z>0.41$ (see \fg\ref{histo-il}).  In each panel, the
continuous line indicates the power-law fit to $\xi(s)$ for the full
sample (see \fg\ref{ksi-wod}, and
\eqs\ref{eq:xi-fit}-\ref{eq:xi-fit-wod}). For clarity, in both panels,
the one-sigma error bars are only plotted for every other point of the
late spiral and dwarf galaxies.}
\label{ksi-type}
\end{figure}

Top panel of \fg\ref{ksi-type} shows $\xi(s)$ for the early-type, late
spiral and dwarf galaxies (see Table~\ref{T1} for details on each
sub-sample). For comparison, the power-law fit to $\xi(s)$ for the
full sample is also plotted.  This comparison shows that the
correlation functions for the early-type, fitted by
\begin{equation}
s_0=7.67\pm4.08\ h^{-1}{\rm Mpc,} \qquad \gamma=1.06\pm0.18
\label{eq:xi-fit-E}
\end{equation}
(measured for $0.7<s<7.5$\hmpc), has a similar power-law behavior as
for the full sample (see \eq\ref{eq:xi-fit} and \fg\ref{ksi-all}).
The late spiral galaxies present a lower amplitude, and flatter slope
of $\xi(s)$ at small scales than for the early-type galaxies: the
power-law fit is defined by
\begin{equation}
s_0=4.90\pm2.55\ h^{-1}{\rm Mpc,} \qquad \gamma=0.96\pm0.18
\label{eq:xi-fit-I}
\end{equation}
(measured for $0.7<s<9.0$\hmpc). In contrast, the dwarf galaxies show
evidence for a lower clustering amplitude, by a factor $\sim3.5$, and
a steeper slope than for the earlier galaxy types:
\begin{equation}
s_0=1.94\pm0.81\ h^{-1}{\rm Mpc,} \qquad \gamma=2.18\pm0.71
\label{eq:xi-fit-L}
\end{equation}
(for $0.7<s<4.0$\hmpc). The marked cut-off in $\xi(s)$ around $s=
14$\hmpc\ and the excess power at $s\sim30$\hmpc, visible for the 3
spectral classes is symptomatic of the geometry and limited volume of
the ESS survey, already discussed in the previous sub-section.

We emphasize that the different clustering for the late spiral and
dwarf galaxies in \fg\ref{ksi-type} strongly supports the fact that
both galaxy types are physically relevant classes for distinguishing
the different components of the galaxy density field.  Note also that
in \fg\ref{ksi-type} and in most of the following graphs, the
mentioned clustering differences among the various sub-samples are
often significant at the $2\sigma$ level at most, due to the limited
size of the ESS. We however take them at face value and derive an
interpretation in terms of type segregation in the two-point
clustering.

The clustering differences among the ESS galaxy types confirm the
visual impression from the ESS redshift cone (\fg\ref{z-cone}): the
early-type galaxies are concentrated within the densest regions,
corresponding to groups of galaxies, whereas the late spiral and dwarf
galaxies also populate the sparser regions of the density field. The
observed behavior of the $\xi(s)$ for the 3 ESS galaxy types is
related to the type-density relation, originally named
``morphology-density'' relation \citep{dressler80,postman84}. The
elliptical galaxies tend to populate the densest regions of the
Universe, namely clusters and groups of galaxies, and thus tend to
have a stronger two-point correlation function on scales of a few
\hmpc, corresponding to the extent of these concentrations. The
late-type spiral galaxies are much more weakly clustered and populate
the lowest density regions, whereas the early spiral galaxies have an
intermediate behavior.

Recently, using the Sloan Digital Sky Survey, \citet{blanton05} showed
that it is the present star formation rate which is directly related
to the local density, and the correlation with the morphology is a
consequence of the relationship between the present star formation
rate of a galaxy and its morphology. The ESS spectral classification
has the advantage to be tightly related to the present star formation
rate \citep{galaz98}, and the observed trends of $\xi(s)$ seen in
\fg\ref{ksi-type} are in agreement with the earlier-type galaxies
residing in higher density environments than the late spiral galaxies.

The correlation function for the dwarf galaxies points to a more
subtle effect, which is detected here for the first time.  At the
smallest scales, the amplitude is as high as that for the early-type
galaxies, and it steadily decreases at larger and larger scales: at
$0.9\le s\le1.3$\hmpc, $\xi(s)$ for the dwarf galaxies is comparable
to that for the late spiral galaxies, and it becomes a factor of
$\sim2-10$ times lower at $1.3\le s\le4.0$\hmpc.  This is consistent
with the dwarf galaxies populating the dense groups (small scale
behavior of $\xi(s)$), whereas on scales of $2-4$\hmpc, they appear
much more weakly clustered than both types of giant galaxies
(early-type and late spiral).

Bottom panel of \fg\ref{ksi-type} shows $\xi(s)$ for the ESS
early-type and late spiral galaxies when excluding all galaxies with
$0.41<z<0.44$.  For the dwarf galaxies, we plot the same curve as in
\fg\ref{wrp-lum-wod} as this sample contains only 2 galaxies with $z>0.41$
(see \fg\ref{histo-il}). The power-law fit to $\xi(s)$ for the full
sample without the over-density is also plotted
(\eq\ref{eq:xi-fit-wod}).  A power-law model is still a good fit to
both functions. For the early-type class, we measure
\begin{equation}
s_0=4.79\pm1.69\ h^{-1}{\rm Mpc,} \qquad \gamma=1.47\pm0.20
\label{eq:xi-fit-wod-E}
\end{equation}
in the interval $0.7<s<7.5$\hmpc. For the late spiral galaxies, we
obtain
\begin{equation}
s_0=3.81\pm2.41\ h^{-1}{\rm Mpc,} \qquad \gamma=0.96\pm0.25
\label{eq:xi-fit-wod-I}
\end{equation}
(for $0.7<s<9.0$\hmpc). Therefore, the effect of removing the
over-density onto the correlation functions for the early-type and
late spiral galaxies is a decrease in amplitude at scales
$s\ga3$\hmpc: a lower amplitude $s_0$ for both types, and a steeper
slope for the early-type galaxies. The fact that $\xi(s)$ for both the
early-type and late spiral galaxies are affected by removing the
over-density indicates that these two populations contribute to the
excess clustering in this region.

The differences in the spatial correlation function between different
galaxy types may actually be larger than those seen in $\xi(s)$,
because of the redshift distortions caused by the galaxy peculiar
velocities (see \scts\ref{proj}), which tend to erase the type
effects. At small scales ($s\sim1$\hmpc), the redshift distortions
caused by the random motions in dense regions tend to weaken the
amplitude increase of $\xi(s)$ for the early-type galaxies with
respect to the other types; although there are no rich clusters of
galaxies in the ESS, the survey contains contains numerous groups of
galaxies (seen as small ``fingers-of-god'' in \fg\ref{z-cone}) which
contribute to this effect at small scales. At larger scales
($s\ga5$\hmpc), the coherent bulk flows tend to increase $\xi(s)$ for
the late spiral galaxies, which dominate in the medium and low density
environments.  In order to free the type measurements from the
redshift distortion effect, we calculate in the following sections the
projected spatial correlation function $w(r_\mathrm{p})$.


\section{The projected spatial correlation function $w(r_\mathrm{p})$  \label{wrp}}


\subsection{General behavior of $w(r_\mathrm{p})$                          \label{wrp_all}}

\begin{figure}
\resizebox{\hsize}{!}{\includegraphics[angle=0]{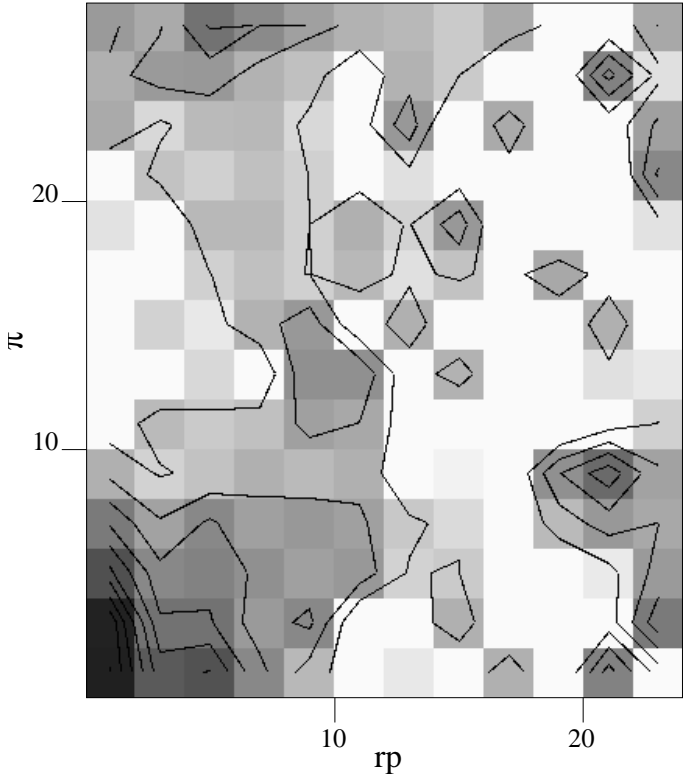}}
\caption{Line-of-sight/transverse correlation function
$\xi(r_\mathrm{p},\pi)$ for the ESO-Sculptor galaxies with
$0.10<z<0.51$, where $r_\mathrm{p}$ and $\pi$ are the separations
perpendicular and parallel to line-of-sight, both measured in unit of
\hmpc. The grey levels (from white to black) and the contours of
constant $\xi(r_\mathrm{p},\pi)$ are linearly spaced from
$\xi(r_\mathrm{p},\pi)=0.0$ to $\xi(r_\mathrm{p},\pi)=3.0$ with steps
of $\Delta\xi(r_\mathrm{p},\pi)=0.25$.}
\label{ksi-rppi}
\end{figure}
\begin{figure}
\resizebox{\hsize}{!}{\includegraphics{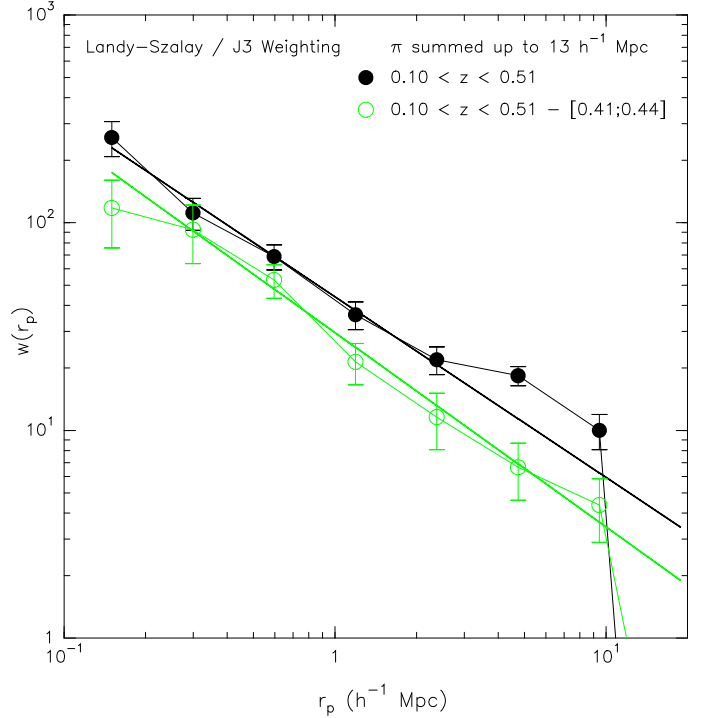}}
\caption{Projected correlation function $w(r_\mathrm{p})$ for the ESO-Sculptor 
sub-sample including all galaxies (black filled circles), and after removal
of the over-density in the interval $0.41<z<0.44$ (green empty circles).
The straight lines correspond to the best-fit power-laws whose parameters
are given in \eqs\ref{eq:wrp-fit} and \ref{eq:wrp-fit-wod}.}
\label{wrp-fit}
\end{figure}

To measure $w(r_\mathrm{p})$, one must first calculate the correlation
function $\xi(r_\mathrm{p},\pi)$ as a function of separation parallel
and perpendicular to the line-of-sight (see \sct\ref{proj}).
\fg\ref{ksi-rppi} shows $\xi(r_\mathrm{p},\pi)$ for all ESS galaxies
using the LS estimator (\eq\ref{eq:xi-LS}) with the minimum-variance
weights (\eq\ref{eq:w-J3}) and bin widths of 2\hmpc. The contours of
constant clustering amplitude are drawn as solid
lines. \fg\ref{ksi-rppi} exhibits a stretching along the line-of-sight
($\pi$ direction) for separations smaller than 4\hmpc, caused by the
peculiar velocities within the numerous groups of galaxies present in
the ESS (see \fg\ref{z-cone}).  The flattening of the contours of
$\xi(r_\mathrm{p},\pi)$ along the line-of-sight due to the coherent
infall of galaxies onto the over-dense regions, which was first
detected in the 2dFGRS by \citet{hawkins03}, is hardly seen here, due
to the limited angular extent of the ESS in right ascension, which
subtends $\sim 8$\hmpc\ at $z\sim0.2$.

Because $\xi(r_\mathrm{p},\pi)$ is a decreasing function of both
$r_\mathrm{p}$ and $\pi$, one can estimate the projected real-space 
correlation function $w(r_\mathrm{p})$ by an integral over $\pi$ (see
\eq\ref{eq:wrp}). In practise, the integral can only be performed to a
finite bound, which needs to be determined. To this end, we have
calculated the various functions $w(r_\mathrm{p})$ with integration
bounds of 6.31\hmpc, 12.59\hmpc, 25.12\hmpc, and 50.12\hmpc. The
12.59\hmpc\ integration bound appears as the smallest bound with
evidence for stabilization, and we thus adopt this value. The chosen
integration bound of $12.59$\hmpc\ also insures that the random noise
fluctuations visible at larger values of $\pi$ in \fg\ref{ksi-rppi}
are excluded, as we interpret these as symptomatic of the limited
sampling volume of the ESS (see \sct\ref{xis}).

The resulting projected real-space correlation function
$w(r_\mathrm{p})$ for the full ESS sample is shown in
\fg\ref{wrp-fit}.  In the interval $0.1<r_\mathrm{p}<3$\hmpc,
$w(r_\mathrm{p})$ is well fitted by a power-law with parameters
\begin{equation}
r_0= 5.25\pm1.82\ h^{-1}{\rm Mpc,} \qquad \gamma=1.87\pm0.07\ ,
\label{eq:wrp-fit}
\end{equation}
(the quoted uncertainties in $r_0$ and $\gamma$ ignore the correlation
between the various points of $w(r_\mathrm{p})$, and are therefore
underestimated).


\subsection{Nature of the over-density at $0.41<z<0.44$                 \label{wrp_overdens}}

We also plot in \fg\ref{wrp-fit} $w(r_\mathrm{p})$ for the ESS sample
without the over-density. The resulting correlation function can be
fitted by a power-law over a larger interval of $r_\mathrm{p}$ than
for the full ESS sample: for $0.2<r_\mathrm{p}<10$\hmpc, we obtain the
following best fit parameters
\begin{equation}
r_0= 3.50\pm1.21\ h^{-1}{\rm Mpc,} \qquad \gamma=1.93\pm0.09\;.
\label{eq:wrp-fit-wod}   
\end{equation}
The decrease in the amplitude $r_0$ of $w(r_\mathrm{p})$ from
\eq\ref{eq:wrp-fit} to \eq\ref{eq:wrp-fit-wod}, with a nearly constant
slope, indicates that the over-density at $0.41<z<0.44$ contributes
excess pairs of galaxies at all scales.  Note that $w(r_\mathrm{p})$
can be fitted by a power-law over a larger range of scales than
$\xi(s)$ (compare \fgs\ref{ksi-wod} and \ref{wrp-fit}), because the
latter function is affected by the peculiar velocities (but see
\sct\ref{halo} for discussion of the small deviations at
$r_\mathrm{p}\simeq0.15$\hmpc\ and $r_\mathrm{p}\simeq1.0$\hmpc\ in
\fg\ref{ksi-wod}).

Close examination of \fg\ref{wrp-fit} indicates that the over-density
in the $0.41<z<0.44$ redshift interval also has a differential effect
on $w(r_\mathrm{p})$ at both small and large scales, with an excess
clustering at $r_\mathrm{p}\la0.2$\hmpc\ and $3\la
r_\mathrm{p}\la10$\hmpc\ compared to the power-law fits.  The small
scale $0.2$\hmpc\ corresponds to the typical virial radius of galaxy
groups \citep{yang05d}, suggesting that the over-density may be due to
richer groups than in the rest of the survey. Besides, the
$3$\hmpc~intermediate scale is close to the minimum separation between
the galaxy groups \citep{yang05b}, which suggest that the groups
located within the over-density are more densely clustered than in the
rest of the ESS.

Calculation of $w(r_\mathrm{p})$ therefore clarifies the nature of the
over-density, and quantifies the visual impression that this redshift
interval contains richer groups and with a higher spatial contrast
than in the rest of the survey (see \fg\ref{z-cone}).  This region is
thus clearly peculiar. Because of its large size and strong excess in
clustering, it has a visible impact on the ESS clustering
measurements, which can be interpreted in terms of cosmic variance.
When this structure is excluded, the ESS clustering measurements are
in good agreement with the other measurements from larger redshift
surveys at $z\sim0$ and $z\sim0.5$ (see \sct\ref{comp}).  Moreover,
there is enough clustering signal in the ESS survey outside the
over-density for allowing us to measure the galaxy correlation
functions. Thanks to a detailed account of the various selection
effects related to the galaxy types and the luminosity functions, the
signal-to-noise in the ESS correlation measurements when excluding the
over-density is only slightly reduced. The gain in revealing the
1-halo and 2-halo components of dark matter halos (see \sct\ref{halo})
largely compensates for this slight loss in signal-to-noise.

In the following, we thus only consider the correlation functions
obtained when the over-density at $0.41<z<0.44$ is excluded, as these
best reflect the typical galaxy clustering.


\subsection{$w(r_\mathrm{p})$ per galaxy type                   \label{wrp_type}}

\begin{figure}
\resizebox{\hsize}{!}{\includegraphics{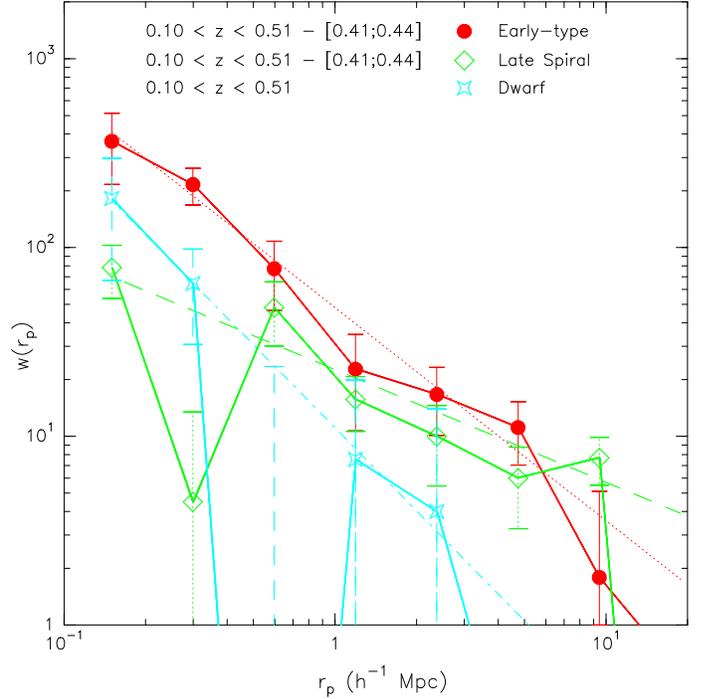}}
\caption{Projected correlation function $w(r_\mathrm{p})$ for the
ESO-Sculptor sub-samples per galaxy type: early-type galaxies (red
filled circles), late spiral galaxies (green open diamonds), and dwarf
galaxies (cyan open stars). For each galaxy type, the straight line
corresponds to the best-fit power-law whose parameters are given in
\eqs\ref{eq:wrp-early-fit} to \ref{eq:wrp-dwarf-fit}.  The
over-density in the interval $0.41<z<0.44$ has been removed from the
early-type and late spiral samples.}
\label{wrp-lum-wod}
\end{figure}

\fg\ref{wrp-lum-wod} shows the projected real-space correlation
function $w(r_\mathrm{p})$ for each of the 3 ESS galaxy types:
early-type, late spiral and dwarf galaxies. The relative behavior of
$w(r_\mathrm{p})$ for the late spiral galaxies and early-type galaxies
is somewhat similar to that seen in $\xi(s)$. The late spiral have a
significantly weaker correlation function than for the early-type
galaxies at small scales ($r_{\mathrm p}\le0.3$\hmpc). Then at larger
scale, $w(r_\mathrm{p})$ for the late spiral galaxies has a very
similar behavior to that for the early-type galaxies, with a
comparable slope, and a factor $\sim1.5-2.0$ lower amplitude.

\begin{figure}
\resizebox{\hsize}{!}{\includegraphics{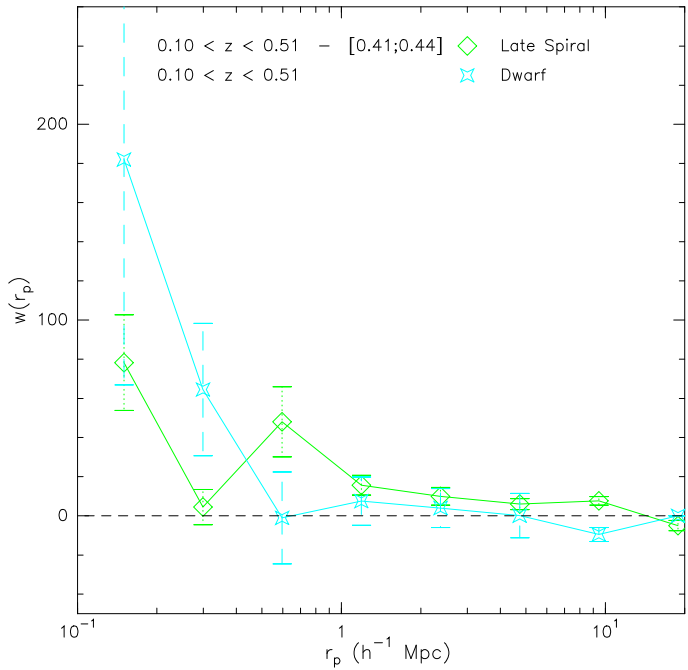}}
\caption{Projected correlation function $w(r_\mathrm{p})$ in linear
scale for the ESO-Sculptor late spiral galaxies (green open diamonds)
and dwarf galaxies (cyan open stars). The over-density in the
interval $0.41<z<0.44$ has been removed from the late spiral sample.}
\label{wrp-lum-lin}
\end{figure}

A remarkable result in \fg\ref{wrp-lum-wod} is that for
$r_\mathrm{p}>0.3$\hmpc, the correlation function for the dwarf
galaxies is consistent with null clustering within the error
bars. This is best seen in \fg\ref{wrp-lum-lin}, which shows
$w(r_\mathrm{p})$ in linear scale for the dwarf galaxies (with the
same color coding as in \fg\ref{wrp-lum-wod}). Significant clustering
of the dwarf galaxies is only detected for
$r_\mathrm{p}\le0.3$\hmpc~(with a $2\sigma$ significance level).

We emphasize the noteworthy differences in the correlation functions
for the three galaxy types in \fg\ref{wrp-lum-wod}.  First, pairs of
early-type galaxies tend to dominate over pairs of both other galaxy
types at all scales (except maybe at $r_\mathrm{p}\ge10$\hmpc), and
the effect is even stronger at small scales,
$r_\mathrm{p}\le0.3$\hmpc. Once the early-type galaxies are set aside,
the complementary clustering of the late spiral and dwarf galaxies is
worth attention: dwarf galaxies have a dominating clustering at
$r_\mathrm{p}\le0.3$\hmpc, and fall-off to null clustering at larger
scales; in contrast, \fg\ref{wrp-lum-lin} shows that $w(r_\mathrm{p})$
for the late spiral galaxies have moderate clustering at
$r_\mathrm{p}\simeq0.15$\hmpc, a factor 2 below the dwarf galaxies;
then $w(r_\mathrm{p})$ is consistent with null clustering at
$r_\mathrm{p}\simeq0.3$\hmpc, whereas significant signal is detected
at larger scales, from $r_\mathrm{p}\simeq0.6$\hmpc\ to
$r_\mathrm{p}\simeq10$\hmpc.  

These differences with galaxy types are quantified by the power-law
fits of $w(r_\mathrm{p})$. For the early-type galaxies, we obtain
\begin{equation}
r_0= 3.80\pm0.67\ h^{-1}{\rm Mpc,} \qquad \gamma=2.11\pm0.10
\label{eq:wrp-early-fit}
\end{equation}
(fit over the $0.15\le r_\mathrm{p}\le 5$\hmpc\ interval).  For the late
spiral galaxies, both the amplitude and the slope are smaller:
\begin{equation}
r_0= 2.72\pm0.64\ h^{-1}{\rm Mpc,} \qquad \gamma=1.60\pm0.08
\label{eq:wrp-late-spiral-fit}
\end{equation}
(fit over the $0.15\le r_\mathrm{p}\le 10$\hmpc\ interval). In contrast,
the dwarf galaxies have an even smaller amplitude and significantly
steeper slope than for the giant galaxies:
\begin{equation}
r_0= 1.85\pm0.83\ h^{-1}{\rm Mpc,} \qquad \gamma=2.46\pm0.38
\label{eq:wrp-dwarf-fit}
\end{equation}
(fit over the $0.15\le r_\mathrm{p}\le 2.5$\hmpc\ interval).

\citet{binggeli90} showed from a local wide-angle survey of low
surface brightness galaxies that although dwarf galaxies delineate the
same large-scale structures as the giant galaxies, there is a strong
segregation among dwarf galaxies: (1) dE lie preferentially in
concentrations of galaxies, whereas dI are more dispersed; (2) outside
clusters, dE also tend to be satellites of giant galaxies. Indeed,
studies of dense galaxy clusters show extensive populations of red
dwarf or dE galaxies \citep{andreon01c,trentham97}. In less dense
environments, one well-studied example being the Local Group, dE
galaxies and the even fainter dwarf spheroidal (dSph) galaxies are
preferentially found as satellites of the giant spiral galaxies, with
typical distances $\le0.1$\hmpc; in contrast, the dI galaxies are more
sparsely distributed, and in the Local Group, most of them populate
the outskirts at distances of $\le0.5$\hmpc\
(http://www.astro.washington.edu/mayer/LG/LG.html).  It is remarkable
that the 2 quoted scales are consistent with the interval over which
the ESS dwarf galaxies show significant clustering. In the picture of
the local Universe, the dE would be responsible for the dwarf
clustering at $r_{\mathrm p}\simeq0.15$\hmpc\ whereas the dI would
contribute to the signal at $r_\mathrm{p}\simeq0.3$\hmpc.



\subsection{Cross-correlation of the giant and dwarf galaxies      \label{wrp_cross}}

\begin{figure}
\resizebox{\hsize}{!}{\includegraphics{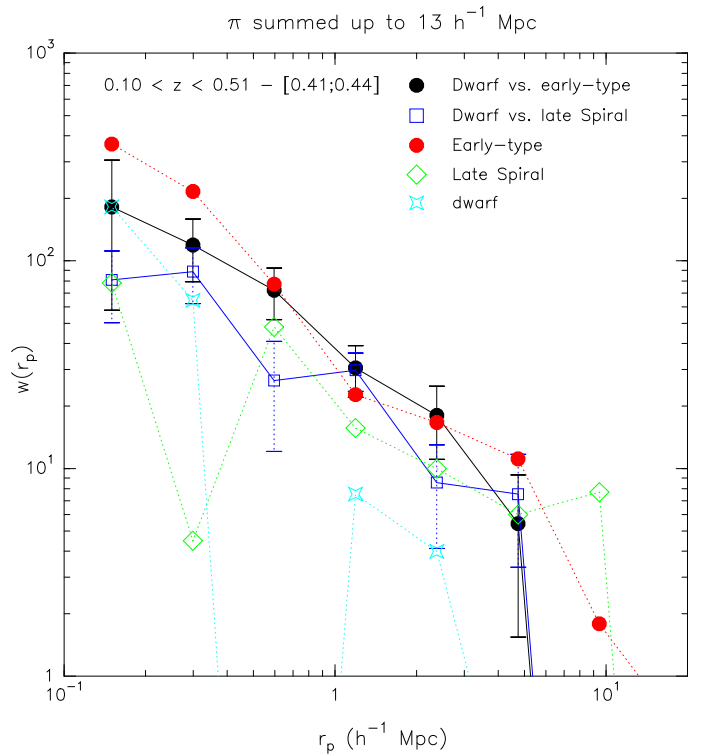}}
\caption{Projected cross-correlation function $w(r_\mathrm{p})$ of the
ESO-Sculptor dwarf galaxies with the early-type galaxies (black filled
circles) and the late spiral galaxies (blue open squares); both curves
are connected by solid lines. For comparison, the auto-correlation
function for the early-type (red filled circles), late spiral (green
open diamonds), and dwarf galaxies (cyan open stars) are over-plotted,
connected by dotted lines.  The over-density in the interval
$0.41<z<0.44$ has been removed from the early-type and late spiral
samples.}
\label{wrp-cross-dwarf-E-Sp}
\end{figure}

To directly measure how dwarf galaxies cluster around giant galaxies,
we plot in \fg\ref{wrp-cross-dwarf-E-Sp} the cross-correlation
$w(r_\mathrm{p})$ between the dwarf galaxies and either the early-type
galaxies or the late spiral galaxies. As for the auto-correlation
function, we adopt the LS estimator with $J3$ weighting (see
\sct\ref{estimators}).  Strikingly, the cross-correlation functions of
the dwarf galaxies versus both types of giant galaxies show
significant signal, with an amplitude comparable to or intermediate
between the auto-correlation functions of the early-type and late
spiral galaxies at $r_\mathrm{p}\le5$\hmpc\ (also shown in the graph).

First, we have also calculated the dwarf versus giant galaxy
cross-correlation functions after excluding the dE galaxies: the
resulting curves are identical to those shown in
\fg\ref{wrp-cross-dwarf-E-Sp}, indicating that both the dE galaxies
\emph{and} the dI galaxies contribute similarly to the
cross-correlation signal. The smaller number of dE galaxies in the
dwarf sub-sample may nevertheless indicate a weaker correlation for
these objects, compared to the dI galaxies.

In \fg\ref{wrp-cross-dwarf-E-Sp}, the cross-correlation of the dwarf
versus early-type galaxies has a $\sim1.5-2$ higher amplitude
($1\sigma$ effect) than that for the dwarf versus late spiral
galaxies, over the full scale range $r_\mathrm{p}\le5$\hmpc. Moreover,
both functions have a comparable amplitude as the auto correlation of
the dwarf galaxies at small scales ($r_\mathrm{p}\le0.3$\hmpc). At
these scales, the auto-correlation of the early-type galaxies is
stronger by a factor $\sim2$ ($1\sigma$ effect) than the
dwarf/early-type cross-correlation, and by a factor $\sim3-4$
($2\sigma$ effect) than the dwarf/late spiral cross-correlation.  This
indicates that at $r_\mathrm{p}\le0.3$\hmpc, pairs of early-type
galaxies may dominate over all the other types of pairs, namely
late-spiral/late-spiral, dwarf/dwarf, dwarf/early-type, and
dwarf/late-spiral pairs. It also suggests that the small-scale
clustering of the dwarf galaxies may be due to the combined effects of
them being satellites of early-type galaxies, and of the early-type
galaxy clustering.

In the intermediate scale range $0.6< r_\mathrm{p}\le5$\hmpc, mixed
pairs of dwarf and giant galaxies appear to contribute equally to the
clustering as pairs of giant galaxies, whereas pairs of dwarf galaxies
are significantly less frequent. At larger scale
($r_\mathrm{p}>5$\hmpc), the cross-correlation signal of dwarf
versus giant galaxies vanishes, indicating that the remaining
clustering signal is fully dominated by pairs of giant galaxies.

\begin{figure} 
\resizebox{\hsize}{!}{\includegraphics{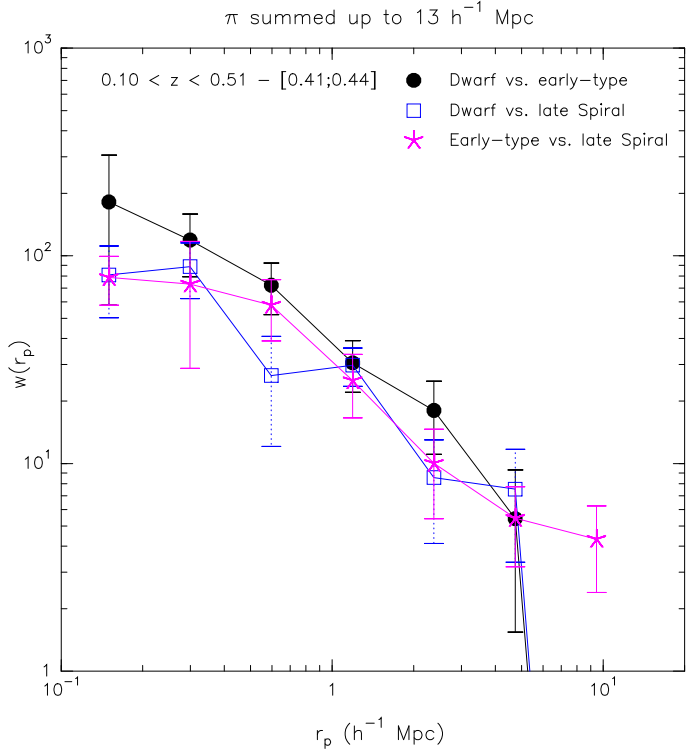}}
\caption{The same cross-correlation functions as in
\fg\ref{wrp-cross-dwarf-E-Sp}, namely the ESO-Sculptor dwarf versus
early-type galaxies (filled black circles) and the dwarf versus late
spiral galaxies (blue open squares), overlaid with the
cross-correlation function of the late spiral versus early-type
galaxies (magenta asterisks).  The over-density in the interval
$0.41<z<0.44$ has been removed from the early-type and late spiral
samples.}
\label{wrp-cross-dwarf-cross-E-Sp}
\end{figure}

The significant cross-correlation signal between the dwarf galaxies
and the late spiral galaxies in the full scale range
$r_\mathrm{p}\le5$\hmpc, seen in \fg\ref{wrp-cross-dwarf-E-Sp},
provides direct indication that the dE and dI galaxies are also
clustered in the vicinity of the late spiral galaxies.  However, the
different behavior of the auto-correlation functions for the two
galaxy populations, with the dominating clustering of the dwarf
galaxies at $r_\mathrm{p}\le0.3$\hmpc\ and that of the late spiral
galaxies at larger scales, may indicates that the two populations have
distinct distributions at these scales.

In \fg\ref{wrp-cross-dwarf-cross-E-Sp}, we compare the
cross-correlation function of the dwarf versus late spiral galaxies
with the cross-correlation of the late spiral versus early-type
galaxies. We observe that at all scales with $r_\mathrm{p}\le 5$\hmpc,
both functions are indistinguishable\footnote{The late spiral versus
early-type cross-correlation function might be smoother than the dwarf
versus late spiral cross-correlation because there are 72\% more
early-type galaxies than dwarf galaxies in the considered samples (see
Table~\ref{T1}).}, whereas the amplitude of the dwarf versus
early-type cross-correlation function (also plotted in
\fg\ref{wrp-cross-dwarf-cross-E-Sp}) is a factor of $1.5-2$ higher
($1\sigma$ effect).  This suggests the interesting property that the
clustering of the dwarf galaxies around late spiral galaxies might be
a consequence of how both galaxy types cluster in the environment of
early-type galaxies. This is further developed in the next
sub-section, where we interpret these observed clustering properties
in terms of the occupation of the dark matter halos by the different
galaxy types.


\section{Interpretation in terms of dark matter haloes   \label{halo}}

\begin{figure}
\resizebox{\hsize}{!}{\includegraphics{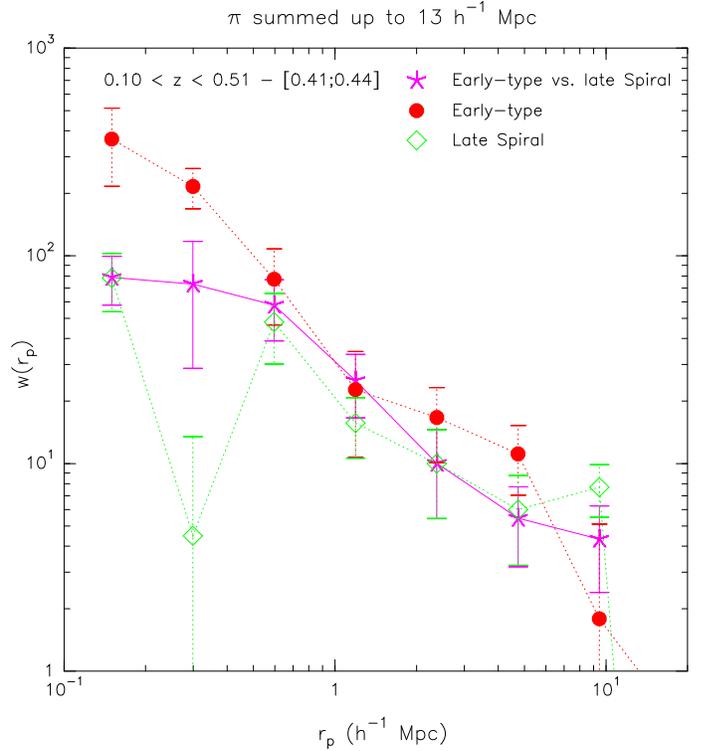}}
\caption{Projected cross-correlation function $w(r_\mathrm{p})$ of the
ESO-Sculptor late spiral galaxies with the early-type galaxies
(magenta asterisks). For comparison, the auto-correlation functions
for the early-type galaxies (red filled circles) and the late spiral
galaxies (green open diamonds) are also shown. The over-density in the
interval $0.41<z<0.44$ has been removed from the early-type and late
spiral samples.}
\label{wrp-cross-auto-E-Sp}
\end{figure}

Based on respectively the 2dF Galaxy Redshift Survey (2dFGRS
hereafter) and the Sloan Digital Sky Survey (SDSS hereafter),
\citet{magliocchetti03} and \citet{zehavi04} showed evidence for a
deviation of the projected correlation function $w(r_\mathrm{p})$ from
a power-law, with a change of slope at
$r_\mathrm{p}\simeq2$\hmpc. Both groups of authors interpret this
inflexion point as the transition from the small-scale regime where
pairs of galaxies located within the same dark matter halos dominate
(denoted hereafter ``1-halo component''), to the large-scale regime
where pairs of galaxies residing in separate halos overtake the
clustering signal (denoted hereafter ``2-halo component''), this
transition occurring near the virial diameter of the halos. This
interpretation is further confirmed by the excellent fit of the
observed deviations of $w(r_\mathrm{p})$ from a power-law using the
general formalism of the ``halo occupation distribution'' (HOD
hereafter; \citealp{magliocchetti03,zehavi04,zehavi05}).  This
approach has the advantage of providing an analytical description of
the clustering of biased galaxy populations
\citep{benson00,berlind02}, and comparison with observations provides
constraints on the HOD parameters.

The projected correlation function $w(r_\mathrm{p})$ for the full ESS
sample without the over-density at $0.41<z<0.44$, shown in
\fg\ref{wrp-fit}, displays a similar deviation from a power-law, with
an inflexion point at $\sim1$\hmpc. At scales smaller than the
inflexion point, the correlation function $w(r_\mathrm{p})$ is poorly
fitted by a power-law. This matches the theoretical expectation that
the 1-halo component follows the halo mass function, which flattens
off at small scales \citep{zehavi04,jenkins01}. At
$r_\mathrm{p}\sim1$\hmpc, the 2-halo regime takes over and is
determined by the matter correlation function and the halo bias
\citep{zehavi04}. At larger scales, the correlation function is also
expected to deviate from a power-law, but this is not visible in
\fg\ref{wrp-fit}, due to limited statistics.


\subsection{Dependence on galaxy type                                   \label{halo_type}}

We now turn to the analysis of the ESS correlation functions by galaxy
type, as they provide a new insight into the contribution of the
different galaxy populations to the halo components.
\fg\ref{wrp-lum-wod} shows that for the three ESS galaxy types,
$w(r_\mathrm{p})$ does deviate from a simple power-law fit. For the
early-type galaxies, the 1-halo to 2-halo transition is clearly
detected, and is located at $r_\mathrm{p}\simeq1$\hmpc. Moreover, both
the 1-halo and 2-halo components have the similar non power-law
behavior as that measured from the 2dFGRS early-type galaxies by
\citet{magliocchetti03}; the two components are modelled by a standard
mass profile \citep{navarro97}, and a prescription for the two-point
correlation function of dark matter halos respectively. Similar
results and modelling are derived from the SDSS by \citet{zehavi05}.

Nevertheless, there are indications of differences between the ESS
early-type clustering and those measured locally from the 2dFGRS and
SDSS: an apparently smaller transition scale of
$r_\mathrm{p}\simeq1$\hmpc, instead of $r_\mathrm{p}\simeq2$\hmpc; and
the fall-off of the ESS large-scale clustering power at
$r_\mathrm{p}\ge10$\hmpc, whereas both the 2dFGRS and SDSS have
significant clustering out to at least $r_\mathrm{p}\simeq20$\hmpc.
The cut-off at $r_\mathrm{p}\simeq10$\hmpc\ for all the ESS auto and
cross-correlation functions is likely due to the limited angular
extent of the survey: its angular size is $\sim1.0^{\circ}$, which
subtends $r_\mathrm{p}\simeq10$\hmpc\ at the median redshift of the
survey ($z=0.3$); as a result, any existing correlation signal beyond
$\sim10$\hmpc\ cannot be detected in $w(r_\mathrm{p})$.  In contrast,
the smaller transition scale in the ESS may be real and could be due
to evolution effects related to the higher redshift range of the ESS
($0.1\le z\le0.5$) compared to $z\le0.1$ for the 2dFGRS and SDSS.

\fg\ref{wrp-lum-wod} shows that the higher amplitude and slope of the
ESS early-type auto-correlation compared to that for the late spiral
galaxies can be decomposed into a 50\% higher amplitude but similar
slope for the 2-halo component, and a factor $\sim2-4$ higher
amplitude and significantly flatter slope for the 1-halo component.
This is in good agreement with the results obtained by
\citet{magliocchetti03} and \citet{zehavi05} from the 2dFGRS and SDSS
respectively. These authors successfully model the clustering of both
galaxy types (early/red, late/blue) using the HOD prescription, which
confirms that both galaxy types follow the dark matter distribution
within the halos.

One of the major predictions of hierarchical clustering is that the
most massive halos have the strongest two-point clustering
\citep{zheng02}.  In this context, the 50\% excess amplitude of the
early-type auto-correlation function over that for the late spiral
galaxies in the 2-halo regime indicates that early-type galaxies tend
to reside in more massive halos than the late spiral
galaxies. Moreover, the relative behavior in the 1-halo regime shows
that the excess of early-type clustering is even stronger within the
halos, and increases at higher clustering levels. Given that the
clustering is stronger in higher density regions, at all scales from
$0.1$ to $30$\hmpc\ \citep{abbas06}, and that the density increases
towards the center of the halos \citep{navarro97}, the excess small
scale clustering of the ESS early-type galaxies is consistent with
them being preferentially located at smaller radii from the halos
centers than the late spiral galaxies.

The auto-correlation function of the ESS dwarf galaxies can also be
conveniently interpreted in terms of halo membership.
\fg\ref{wrp-lum-wod} shows a clear 1-halo component which falls off at
$r_\mathrm{p}>0.3$\hmpc, thus indicating that the dwarf galaxies are
confined to the densest parts of the halos. This is in agreement with
the fact that these regions are dominated by early-type galaxies, and
that dwarf galaxies are preferentially satellites of early-type
galaxies (see \sct\ref{wrp_cross}). A possible weak 2-halo component
of the dwarf galaxies at $1\le r_\mathrm{p}\le 2$\hmpc\
(\fg\ref{wrp-lum-wod}) could be the replica of the early-type 2-halo
component with the appropriately scaled amplitude.

Information on the degree of mixing of the early-type and late spiral
galaxies within halos can be obtained from their cross-correlation
function.  It is displayed in \fg\ref{wrp-cross-auto-E-Sp}, where it
is compared with the auto-correlation functions $w(r_\mathrm{p})$ for
both galaxy types.  The cross-correlation is close to the mean of the
two auto-correlation functions at all scales but $r_{\mathrm
p}\sim0.15$\hmpc. In the 2-halo regime, this is naturally expected and
provides no additional information over the auto-correlation functions
\citep{zehavi05}. In contrast, in the 1-halo regime, it suggests that
both galaxy types are well mixed within halos at
$r_\mathrm{p}\ge0.3$\hmpc, that is that they \emph{do not} avoid
residing in the same halos.  Using the cross-correlation function,
\citet{zehavi05} also found a good level of mixing of the red and blue
SDSS galaxies, at all scales of the 1-halo regime.

Nevertheless, in the ESS analysis presented here, the identical
amplitude of the early-type versus late spiral cross-correlation
function and the late spiral auto-correlation function at $r_{\mathrm
p}\simeq0.15$\hmpc, together with a factor $\sim5$ higher amplitude
for the early-type auto-correlation function at this scale indicate a
lack of mixing of the two galaxy types: pairs of early-type galaxies
dominate over pairs of late spiral galaxies and cross-pairs of
early-type and late spiral galaxies.  We notice that
$r_\mathrm{p}\simeq0.15$\hmpc\ is also the smallest detected scale of
correlation signal in the 2dFGRS \citep{magliocchetti03} and SDSS
\citep{zehavi05}, which have higher statistics than the ESS. This
scale is likely to correspond to the highest density regions, hence to
the very centers of the most massive halos \citep{navarro97}. The
cross-correlation functions therefore bring the additional information
that the centers of the most massive halos are dominated by pairs of
early-type galaxies.  Conversely, these observations suggest that the
late spiral galaxies tend to lie either in the outer regions of the
densest halos or in the centers of less dense halos.

These results are consistent with the HOD-based model proposed by
\citet{zehavi05}, in which blue galaxies are the central galaxies of
the least massive halos, whereas red-type galaxies are the central
galaxies in all other halos, including the most massive.  Such a
segregation effect is also detected by \citet{magliocchetti03}, whose
HOD modelling using a common dark matter profile can successfully
predict $w(r_\mathrm{p})$ for both the early-type and late-type
galaxies, provided that the former populate the halos out to one
virial radius, and the latter are allowed to extend out to twice that
distance. The marked deficit of pairs of ESS late spiral galaxies at
$r_\mathrm{p}\simeq0.3$\hmpc\ (see \fg\ref{wrp-lum-wod}) could be
interpreted as further evidence that a significant part of these
objects tend to populate the outer regions of the halos.


\subsection{Dependence on galaxy luminosity                      \label{halo_lum}}

\begin{figure*}
\centerline{\psfig{figure=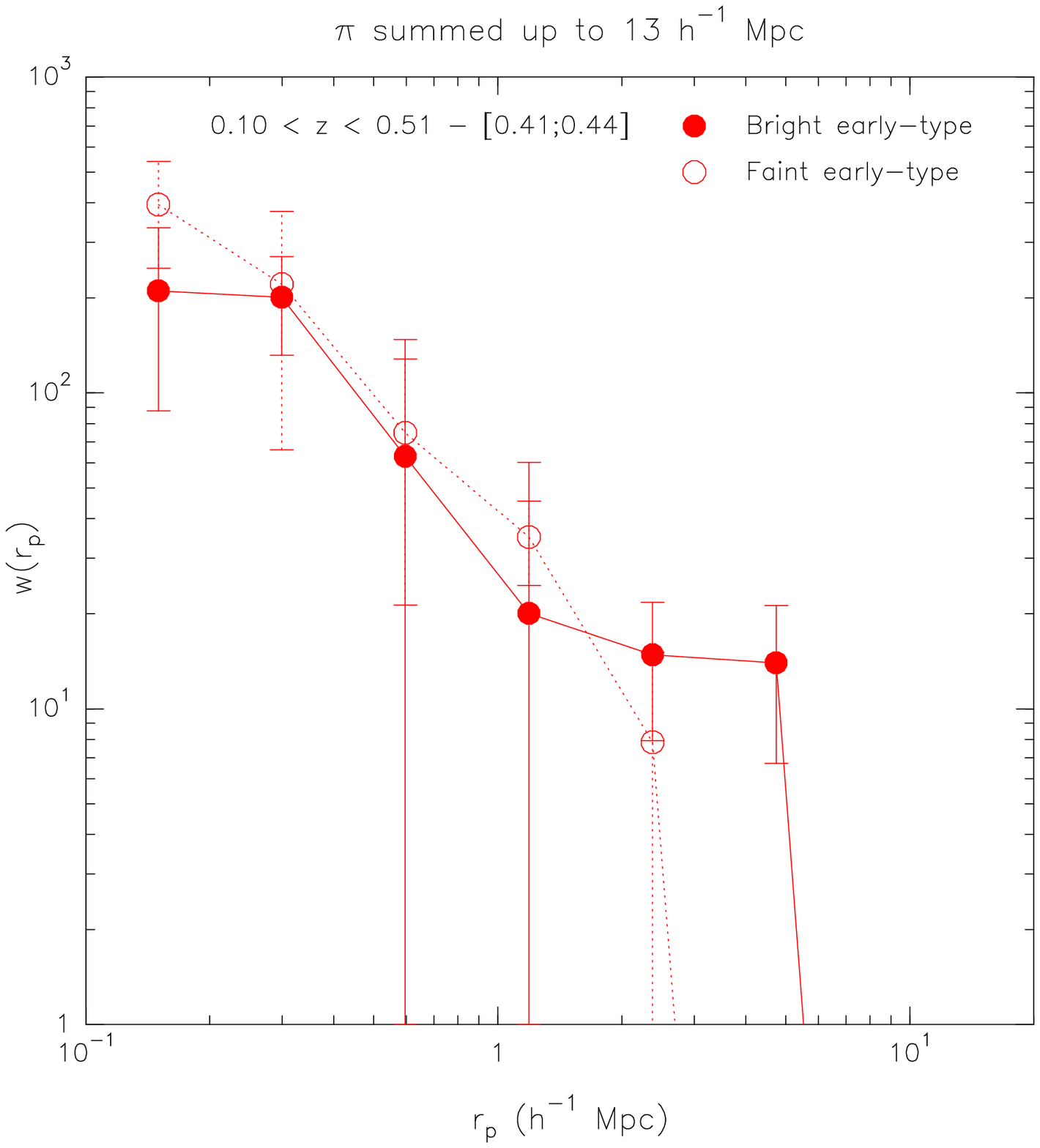,height=9cm,angle=0}
\qquad\qquad\psfig{figure=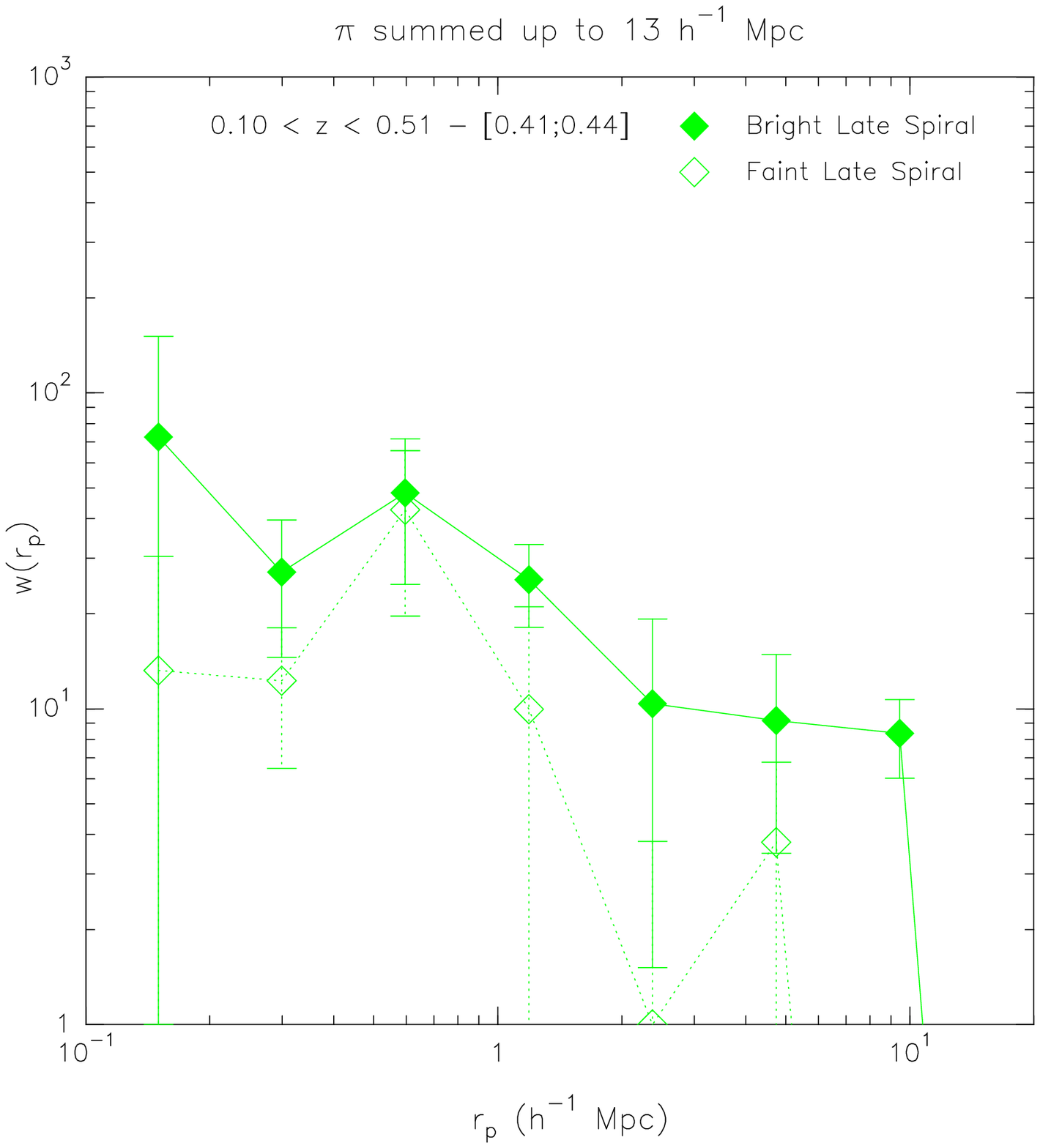,height=9cm,angle=0}}
\caption{Projected correlation function $w(r_\mathrm{p})$ for the
ESO-Sculptor sub-samples split at their median absolute magnitude: the
early-type, late spiral galaxies are shown in the left and right
panels respectively; the filled symbols and solid lines mark the
bright sub-samples, the open symbols and dotted lines the faint
sub-samples. In both panels, the over-density in the interval
$0.41<z<0.44$ has been removed from the samples.}
\label{wrp-bright-faint}
\end{figure*}

It has been widely observed that intrinsically luminous galaxies
cluster more strongly than faint ones \citep[e.g.][]{benoist96,
guzzo00, zehavi02}. To examine whether such systematic variations are
present in the ESS, we separate the early-type and late spiral
galaxies into bright and faint sub-samples using the median absolute
magnitude of each sample: $-21.14$ for the early-type galaxies,
$-20.56$ for the late spiral galaxies; the corresponding numbers of
galaxies for the sub-samples are listed in Table~\ref{T1}. The
resulting projected auto-correlation functions $w(r_\mathrm{p})$ for
both the early-type and late spiral sub-samples are displayed in
\fg\ref{wrp-bright-faint}, using filled, open symbols for the bright
and faint sub-samples respectively.

For the early-type galaxies (left panel of \fg\ref{wrp-bright-faint}),
the correlation function is unchanged when restricting to the bright
sub-sample, except at $r_\mathrm{p}\simeq0.15$\hmpc\ where the signal
decreases by a factor $\sim2$, and at $r_\mathrm{p}\simeq10$\hmpc\
where the signal vanishes. The faint sub-sample has nearly the same
clustering strength as the full early-type sample at
$r_\mathrm{p}\la1$\hmpc; then the correlation function decreases to
nearly half that for the bright galaxies at $r_{\mathrm
p}\simeq2.5$\hmpc, and the power vanishes beyond. Altogether, the
relative behavior of the early-type luminosity sub-samples indicates
that in the 1-halo regime, both sub-samples contribute equally, with
maybe a dominant contribution from the fainter galaxies at the very
small scale $r_{\mathrm p}\simeq0.15$\hmpc. Whereas in the 2-halo
regime, the bright early-type galaxies tend to dominate at
$r_\mathrm{p}\ge2.5$\hmpc.  This is in agreement with the results of
\citet{zehavi05} that bright red galaxies exhibit the strongest
clustering at large scale, whereas faint red galaxies exhibit the
strongest clustering at small scales. The authors reproduce this
behavior using an HOD model in which nearly all faint red galaxies are
satellite in high mass halos.

When compared to the early-type galaxies, the relative clustering of
the faint and bright late spiral galaxies (right panel of
\fg\ref{wrp-bright-faint}) shows a somewhat similar behavior at large
scales, and a difference at small scales. Indeed, the 2-halo component
of the bright late spiral sub-sample is consistent with a factor
$\sim2-5$ ($1\sigma$ deviation) stronger clustering than for the faint
sub-sample. In contrast to the early-type galaxies, the 1-halo
component of the bright late spiral sub-sample tends to have stronger
clustering than for the faint sub-sample. This is again in agreement
with the steadily decreasing amplitude of $w(r_\mathrm{p})$ for the
faint blue SDSS galaxies.  At variance with the early-type galaxies,
there is no indication of excess clustering of the faint late spiral
galaxies at $r_\mathrm{p}\simeq0.15$\hmpc; on the contrary, they might
be less clustered than their bright analogs.

Yet another prediction of hierarchical clustering is that luminous
galaxies are expected to be preferentially located within massive
halos, which in turn are more strongly clustered.  Here again,
although the clustering deviations in the luminosity sub-samples of
the early-type and late spiral galaxies are only significant at the
2$\sigma$ level at most, we take them at face value and derive an
interpretation in terms of dark matter halo membership. The excess
clustering in the 2-halo regime of the ESS bright sub-samples over the
faint sub-samples, which is detected for both the early-type and late
spiral galaxies in \fg\ref{wrp-bright-faint}, is consistent with this
expected property of hierarchical clustering.

In contrast, the difference in the relative behavior of the 1-halo
components for the early-type and late spiral galaxies suggests that
the two galaxy types trace the dark matter profiles of the halo in a
different way.  In the previous section, we suggested that early-type
galaxies tend to occupy the centers of the most massive halos, and
that late spiral galaxies tend to lie either in the centers of less
dense halos and/or in the outer regions of the densest halos.  The
additional information brought here is that this is nearly independent
of luminosity for the early-type galaxies, whereas faint late spiral
galaxies might tend to reside in even less dense regions than their
bright analogs. This implies a specific spatial segregation of the
early-type and late spiral galaxies inside the dark matter halos.


\section{Comparison with other surveys                      \label{comp}}

\begin{table*}
\caption{Parameters of power-law fits to the projected-separation correlation function for other redshift surveys.}
\label{T3}
\begin{center}
\begin{tabular}{llcrccl}
\hline
\hline
Survey & Redshift range & Absolute magnitude range & Numb. & $r_0$ (\hmpc) & $\gamma$ & Reference \\
Galaxy type & \\
\hline
ESS    & $0.1\le z\le0.51$\dag  & $M_{R_\mathrm{c}}-5\log h\le -16.0$ &     654 & $3.50\pm1.21$ & $1.93\pm0.09$ & present analysis\\
VVDS   & $0.1\le z\le0.5$       & $M_B-5\log h\le-17.0$               &   1 089 & $3.0 \pm0.5 $ & $1.7 \pm0.1 $ & \citet{pollo06} \\
SDSS   & $z\le0.04$             & $M_r-5\log h\le-18.0$~*             &   8 730 & $3.7 \pm0.3 $ & $1.87\pm0.05$ & \citet{zehavi05} \\
SDSS   & $z\le0.06$             & $M_r-5\log h\le-17.0$~*             &  23 560 & $4.6 \pm0.2 $ & $1.89\pm0.03$ & {\it id.} \\
2dFGRS & $0.01\le z\le0.20$     &                                     & 165 659 & $5.0 \pm0.3 $ & $1.70\pm0.03$ & \citet{hawkins03} \\
\hline
ESS: \\
- early-type galaxies    & $0.1\le z\le0.51$  &  $M_{R_\mathrm{c}}-5\log h\le -16.0$ &     218 & $3.80\pm0.67$ & $2.11\pm0.10$ & present analysis\\
- late spiral galaxies   & $0.1\le z\le0.51$  &  $M_{R_\mathrm{c}}-5\log h\le -16.0$ &     279 & $2.72\pm0.64$ & $1.60\pm0.08$ & {\it id.}\\
- dwarf galaxies         & $0.1\le z\le0.51$  &  $M_{R_\mathrm{c}}-5\log h\le -16.0$ &     159 & $1.85\pm0.83$ & $2.46\pm0.38$ & {\it id.}\\
VVDS: \\
- elliptical/S0 galaxies & $0.2\le z\le0.6$   & $M_{B_{AB}}-5\log h\le-15.0$         &     164 & $3.1 \pm0.8$    & $2.4\pm0.3$    & \citet{meneux06} \\
- Sb-Sc galaxies         & $0.2\le z\le0.6$   & $M_{B_{AB}}-5\log h\le-15.0$         &     736 & $2.8 \pm0.4$    & $1.9\pm0.2$    & {\it id.} \\
- Magellanic irregulars  & $0.2\le z\le0.6$   & $M_{B_{AB}}-5\log h\le-15.0$         &     507 & $1.7 \pm0.3$    & $1.8\pm0.1$    & {\it id.} \\
2dFGRS: \\
- passive galaxies       & $0.01\le z\le0.20$ &                                      &  36 318 & $6.05 \pm 0.35$ & $1.94 \pm 0.03$ & \citet{madgwick03} \\
- active galaxies        & $0.01\le z\le0.20$ &                                      &  60 473 & $3.89 \pm 0.31$ & $1.55\pm 0.04$  & {\it id.} \\
SDSS: \\
- red  galaxies          & $0.03\le z\le0.07$ & $-20\le M_r-5\log h\le-19.0$~*       &   5 804 & $5.7 \pm 0.3$   & $2.10 \pm 0.05$ & \citet{zehavi05} \\
- blue galaxies          & $0.03\le z\le0.07$ & $-20\le M_r-5\log h\le-19.0$~*       &   8 419 & $3.6 \pm 0.3$   & $1.70 \pm 0.05$ & {\it id.} \\
SDSS: \\
- red  $w(\theta)$       &                    &   $-21\le M_r-5\log h$               &     343 & $6.59\pm0.17$   & $1.96 \pm 0.05$ & \citet{budavari03} \\
- blue $w(\theta)$       &                    &   $-21\le M_r-5\log h$               &     316 & $4.51\pm0.19$   & $1.68 \pm 0.09$ & {\it id.} \\
\hline
\end{tabular}
\smallskip
\end{center}
\begin{list}{}{}
\item[\underline{Notes:}]
\item[\dag]The over-density in the interval $0.41<z<0.44$ is excluded from the listed redshift interval.
\item[*]The star symbol in the absolute magnitude range column indicates
that the corresponding sample is volume limited to the quoted absolute
magnitude limits. Elsewhere, the indicated absolute magnitude results
from the combination of the apparent magnitude and redshift limits of
the sample.
\end{list}
\end{table*}

Before the establishment of the currently standard cosmology
$\Omega_\mathrm{m}=0.3$ and $\Omega_\Lambda=0.7$
\citep{riess98,perlmutter99,phillips01,tonry03}, various surveys have
obtained measures of the galaxy two-point correlation function in
redshift space and/or projected separation assuming either a low or
high matter density Universe and a null cosmological constant (at
$z\simeq0.1-0.5$:
\citealt{cole94,lefevre96,small99,guzzo00,hogg00,carlberg00,shepherd01};
at $z\simeq0$:
\citealt{loveday92,park94,baugh96,tucker97,ratcliffe98c,giuricin01}).
To compare our results with those from the other surveys, we
thus consider only the more recent measurements, which use the new
standard cosmological parameters.

We specifically focus on the projected-separation correlation function
$w(r_\mathrm{p})$ and list in Table~\ref{T3} its amplitude and slope
obtained for the full ESS sample without the over-density in the
redshift interval $0.41\le z<0.44$. For comparison, we list the
measurements from the SDSS, the 2dFGRS, and the VIMOS-VLT Deep Survey
(VVDS).  The ESS values are in good agreement with those obtained from
the VVDS \citep{pollo06} in a similar redshift interval. The
comparatively higher ESS amplitude $r_0= 5.25\pm1.82$\hmpc, derived
when the over-density at $0.41\leq z<0.44$ is included, strengthens
our conclusion that this structure is a peculiar region of the survey.

The parameters of $w(r_\mathrm{p})$ derived from the ESS are also
consistent with those obtained from the 8 730 SDSS galaxies with
$M_r\le-18.0$ \citep{zehavi05}. Our value of $r_0$ is nevertheless
below that obtained from 23 560 SDSS galaxies with $M_r\le-17.0$
\citep{zehavi05}, and that from 165 659 2dFGRS galaxies (for
supposedly $M_{b_J}\le-17.5$, \citealp{hawkins03}). In a $\Lambda$CDM
universe undergoing hierarchical clustering, an evolution in $r_0$ is
expected between redshifts 0 and $0.5$, specifically a decrease by
$\sim0.5$\hmpc\ \citep{benson01,kauffmann99b}. Although such a
variation is compatible with the comparison of the ESS and the local
SDSS and 2dFGRS correlation measurements, the wide error bars of the
ESS correlation function do not allow us to draw any firm conclusion
on the evolution in $r_0$.

The second part of Table~\ref{T3} lists the parameters of the
power-law fits to $w(r_\mathrm{p})$ for the same surveys as quoted in
the top of the Table, split by galaxy type.  For the VVDS, we list the
measurements for the elliptical/S0 (type 1), Sb-Sc spiral (type 2) and
Magellanic irregular galaxies (type 4) \citep{meneux06}; note that we
do not consider the correlation function for galaxy types Sc-Sd (type
3) measured by \citet{meneux06}, because the results are nearly
identical to those for types Sb-Sc; moreover, among the ESS late
spiral galaxies, intermediate-type galaxies which correspond to Sb-Sc
type dominate in number over later type galaxies (see
Table~\ref{T1}). Both the ESS early-type and VVDS elliptical/S0 on the one
hand, and the ESS late spiral and the VVDS Sb-Sc galaxies on the other
hand, have values of the amplitude $r_0$ and the slope $\gamma$ which
are in 1$\sigma$ agreement.  The amplitude for the ESS dwarf galaxies
and the VVDS Magellanic irregulars are also in good agreement, whereas
the slope is significantly steeper in the ESS. The slope of
$w(r_\mathrm{p})$ for the ESS dwarf galaxies could be even steeper as
the signal at $r_\mathrm{p}\ge1$\hmpc\ may not be real in this
population (see \fg\ref{wrp-lum-wod}).

At the smaller redshifts covered by the SDSS and 2dFGRS, the
correlation function for the red galaxies has a higher amplitude and
steeper slope than for the blue galaxies \citep{zehavi05,madgwick03},
indicating similar segregation effects as in the ESS.  This type
effect was also detected from the SDSS angular correlation function
\citep{budavari03}, with good agreement in the power-law parameters.
Note however that the correlation functions by galaxy type for the
SDSS and 2dFGRS have higher amplitudes than for the ESS and VVDS,
which may also be the trace of clustering evolution.


\section{Summary of results                           \label{summary}}

We calculate the two-point correlation function for the ESO-Sculptor
redshift survey. The sample is limited to the 765 galaxies with
$R_\mathrm{c}\le21.5$ in the redshift interval $0.1\le z\le 0.51$.  We
use on the one hand the template free spectral classification of the
sample into early, intermediate, and late-type galaxies, which
correspond to the following mixes of morphological type: E + S0 + Sa,
Sb + Sc, and Sc + Sd/Sm, respectively \citep{galaz98}; and on the
other hand the results of the ESS luminosity function analysis, which
indicates that the three ESS spectral classes contain two additional
components, dwarf elliptical and dwarf irregular galaxies, mixed into
the intermediate and late-type classes respectively
\citep{lapparent03}. This leads us to separate the intermediate-type
and late-type spectral classes into their giant and dwarf galaxy
components, which we merge into two classes dominated by late spiral
(Sb + Sc + Sd/Sm), and dwarf (dE + dI) galaxies respectively. The
resulting three galaxy classes (early-type, late spiral, dwarf
galaxies) are therefore defined by spectral/morphological and
luminosity criteria, which are both relevant for studying segregation
effect in galaxy clustering.  We use the corresponding Schechter and
Gaussian luminosity functions for defining the selection function for
each of the three galaxy types.

We test the various estimators of the correlation function, and adopt
the \citet{landy93} estimator with $J3$ weighting, which combines
stability and minimum variance.  The redshift-space correlation
function $\xi(s)$ can be fitted by a power-law with amplitude
$s_0=7.49 \pm 3.18$\hmpc\ and slope $\gamma=0.90 \pm 0.13$ in the
interval $0.5<s< 5$\hmpc. At larger scales, $\xi(s)$ oscillates
between negative and positive low amplitude values, with a peak at
$\sim35$\hmpc\ and its multiples.  This is due to the combination of
the pencil-beam geometry of the ESS survey with the alternation of
walls and voids, as demonstrated by a pair separation analysis. The
ESS also contains an over-dense region located in the redshift
interval $0.41< z<0.44$, which affects the correlation function by
adding excess large-scale power in $\xi(s)$. When removing this
region, the power-law fit to $\xi(s)$ yields $s_0=4.22 \pm 1.15$\hmpc\
and $\gamma=1.22 \pm 0.15$ (in $0.5\le s\le 5$\hmpc), in better
agreement with the other existing surveys.

We then calculate the redshift-space correlation function $\xi(s)$ for
the three galaxy types. These show marked differences, with a dominant
signal originating from the early-type galaxies at nearly all scales.
The late spiral galaxies show a weaker correlation amplitude at small
and large scales, in agreement with the type-density relationship
\citep{blanton05}.  The new result is that the dwarf galaxies show a
very steep correlation function over a narrow range of scales: $\xi(s)$
decreases from the clustering amplitude of the early-type galaxies at
very small scale, to more than one order of magnitude weaker at
$r_\mathrm{p}\simeq4$\hmpc. These segregation effects in the two-point
clustering quantify the visual impression drawn from the redshift cone
of the ESS.

To free the correlation function measurements from the effect of
peculiar velocities, as they decrease the clustering amplitude at
small scales due to random motions and increase its amplitude at large
scales due to coherent bulk flows, we then calculate the real-space
correlation function as a function of projected separation
$w(r_\mathrm{p})$. This is done by integrating the 2-dimensional
correlation function $\xi(r_\mathrm{p},\pi)$ along the line-of-sight
separation $\pi$. The resulting projected-separation correlation
function $w(r_\mathrm{p})$ can be adjusted by a power-law over a
larger range of scales than $\xi(s)$, from $0.15$\hmpc\ to $10$\hmpc.
In this scale range, and after removing the over-density at $0.41<
z<0.44$, we obtain a best fit amplitude $r_0=3.50\pm1.21$\hmpc\ and a
slope $\gamma=1.93\pm0.09$ which, as expected, is significantly
steeper than that measured from $\xi(s)$.

When splitting the ESS by galaxy type, the projected-separation
correlation function $w(r_\mathrm{p})$ shows similarities with
$\xi(s)$, with again the early-type galaxies dominating over the other
types at all scales. At variance with $\xi(s)$, the dwarf galaxies
clustering dominates over the late spiral galaxies at
$r_\mathrm{p}\simeq0.15$\hmpc.  At larger scales, the dwarf galaxies
have a spatial correlation function consistent with null clustering,
whereas the late spiral galaxies take over at
$r_\mathrm{p}\ge0.6$\hmpc, with a similar shape of the correlation
function as for the early-type galaxies and a 50\% lower amplitude.

Comparison of $\xi(s)$ and $w(r_\mathrm{p})$ with and without
inclusion of the over-dense region at $0.41<z<0.44$ provides useful
clues on the nature of this region: the excess clustering appears
entirely due to the fact that it contains richer and more densely
clustered groups of galaxies than in the rest of the survey; and both
the early-type and late spiral galaxies contribute to the excess
clustering in the region. We then consider that the correlation
functions for the ESS survey without the over-density are more
representative of the overall galaxy distribution.

A subsequent analysis of the cross-correlation of the dwarf galaxies
with the early-type and late spiral galaxies provides direct evidence
that the dwarf galaxies are satellites of the giant galaxies.  At
$r_\mathrm{p}\le0.3$\hmpc, pairs of early-type galaxies dominate over
all the other types of pairs. Pairs of dwarf/early-type galaxies,
dwarf galaxies, dwarf/late spiral galaxies, and late spiral galaxies
are the next contributors to the small-scale two-point clustering, in
decreasing order of contribution to the correlation function. In the
intermediate scale range $0.3<r_\mathrm{p}\le5$\hmpc, mixed pairs of
dwarf and giant galaxies contribute equally to the clustering as pairs
of giant galaxies. Then at $r_\mathrm{p}\simeq10$\hmpc, the clustering
signal is dominated by pairs of giant galaxies.  Moreover, the
cross-correlation analysis indicates that dwarf and late spiral
galaxies are \emph{not} well mixed at
$r_\mathrm{p}\le0.3$\hmpc. Altogether, this suggests that the
clustering of the dwarf galaxies around late spiral galaxies at
$r_\mathrm{p}\le 5$\hmpc\ may be an indirect consequence of how both
galaxy types cluster in the environment of early-type galaxies.

We then interpret the variations in the correlation function with
galaxy type in terms of membership to the underlying dark matter
halos. This approach is eased by the separation into the giant and
dwarf galaxies, as they exhibit a clear dichotomy in their halo
components: the correlation function for the early-type galaxies shows
a dip at $r_\mathrm{p}\simeq 1$\hmpc, which is interpreted as the
transition between the regimes in which the 1-halo and 2-halo pairs
dominate \resp, and both components show a significant contribution;
in contrast, the dwarf and late spiral galaxy correlation functions
are dominated by their 1-halo and 2-halo components, at small and
large scales respectively.  Altogether, this indicates that early-type
galaxies tend to lie predominantly at the centers of the massive
halos, whereas late spiral galaxies tend to lie either in the centers
of less dense halos and/or in the outer regions of the densest halos.
The small scale clustering is then not only determined by the dominant
galaxies in the massive halos, but also by their dwarf satellites.

We also examine the two-point clustering for the bright and faint
sub-samples of the early-type and late spiral galaxies.  For both the
early-type and late spiral galaxies, we detect a 1$\sigma$ excess
clustering in the 2-halo regime of the bright sub-samples over the
faint sub-samples, which is consistent with the expected properties of
hierarchical clustering: the most massive halos have the strongest
2-halo clustering, and luminous galaxies are preferentially located
within massive halos. Comparison of the 1-halo component brings the
additional information that the relationship between halo mass and
giant galaxy type is nearly independent of luminosity for the
early-type galaxies, whereas faint late spiral galaxies might tend to
reside in even less dense regions than their bright analogs.

At last, we compare our results with those from the other published
analyses. Our power-law fits to $w(r_\mathrm{p})$ for the full ESS
sample, and for the sub-samples by galaxy type are consistent with
those measured at comparable and lower redshifts from the other
surveys.


\section{Discussion and perspectives                              \label{discussion}}

\subsection{The halo components of the correlation function}

One of the major results obtained here from the ESO-Sculptor redshift
survey is that the projected-separation correlation function $w(r_\mathrm{p})$
for each of the three galaxy types (early-type, late spiral and dwarf galaxies)
presents marked deviations from a power-law, which can be interpreted
as the transition between galaxies belonging to a same dark matter
halo, and galaxies belonging to two different halos. This provides
confirmation that the results obtained at low redshift ($z\la0.1$) by
\citet{zehavi05} and \citet{magliocchetti03} extend to higher redshift
($z\la0.5$). A similar result was recently obtained at even higher
redshifts ($z\la1.2$) based on the COMBO17 survey with broad and
medium band photometric redshifts ($\sigma(z)/z\sim0.01$;
\citealt{phleps06}). The ESS brings a useful confirmation based on
spectroscopic redshifts, with $\sigma(z)\sim0.00055$ at $0.1\le
z\le0.51$.

In this context, the ESS results provide evidence in favor of the
gravitational instability scenario for the formation of structure, in
which the evolution of galaxy clustering is driven by the hierarchical
merging of halos.  Most recently, \citet{conroy06} have directly
demonstrated that high-resolution dissipationless $\Lambda$CDM
simulations can reproduce the observed bimodal behavior of the
correlation function for absolute magnitude-limited samples at various
redshift limits: these results are entirely based on combining the
spatial clustering of the halos with a prescription that relates the
galaxy luminosities to the maximum circular velocity of the sub-halos
at the time of accretion.  The hierarchical merging scenario is
further validated by the detailed shape of the correlation functions
for ESS early-type and late spiral galaxies, which allow a
straightforward identification of the 1-halo and 2-halo components.

Direct modelling of the projected-separation correlation function
using the ``Halo Occupation Distribution'' (HOD)
\citep{benson00,berlind02}, or the more refined ``Conditional
Luminosity Function'' approach (which takes into account the
luminosity and colour distribution of galaxies within dark matter
halos of varying mass; \citealp{yang03}), provides constraints on the
halo parameters describing the central and satellite galaxies
parameters
\citep{magliocchetti03,phleps03,abazajian05,zehavi05,cooray06}.
Nevertheless, several measurements of the correlation function at
small scale also challenge the current version of the halo model for
galaxy clustering: the very small scale clustering of luminous red
galaxies from the SDSS is too steep and would require either a steeper
dark halo profile or a galaxy distribution which is steeper than the
dark matter at scales $0.01\le r_{\mathrm p}\le0.1$\hmpc\
\citep{masjedi06}. In contrast, \citet{diaz05} obtained projected
density profiles of galaxy groups which are too flat compared to the
standard \citet{navarro97} profile.

\subsection{Early-type versus late spiral segregation}

The second major result obtained from the ESS is the markedly
different clustering properties of the two giant galaxy types,
early-type and late spiral.  Both types have comparable 2-halo
components with a 50\% higher amplitude for the early-type galaxies,
whereas the 1-halo component of the early-type galaxies largely
dominates over that for the late spiral galaxies. These results are
remarkably similar to the predicted correlation functions calculated
by \citet{kauffmann99} in a $\Lambda$CDM semi-analytical simulation,
which exhibit a higher amplitude and steeper slope for the
early-type/red galaxies, and a lower amplitude and a marked
small-scale flattening at $r_{\mathrm p}\le1.0$\hmpc\ for the
starforming galaxies.

In the framework of hierarchical clustering of the dark matter halos,
our results imply a specific spatial segregation of the early-type and
late spiral galaxies inside the dark matter halos, with the early-type
galaxies residing in the center of the most massive halos, whereas the
late spiral reside in their outskirts or in less dense halos.  These
segregation effects are consistent with the 50\% higher pairwise
velocity dispersion measured by \citet{madgwick03} for the 2dFGRS
passive galaxies compared to the starforming galaxies, as it indicates that
the passive galaxies inhabit preferentially the cores of high-mass
virialized regions.

The link between the ESS type-segregation effects and the dark matter
halos also finds direct confirmation from other recent analyses based
on group catalogs.  \citet{zandivarez03} and \citet{yang05b} show that
the clustering properties of galaxy groups in the 2dFGRS match those
of the dark matter halos in $\Lambda$CDM N-body simulations.
\citet{yang05} thus derive a ``halo-based'' group finder algorithm
which is optimized to associate a group to those galaxies which belong
to the same dark matter halo. This allows one to directly examine the
link between galaxies and their dark matter halos. Most interestingly,
\citet{yang05b} separate the galaxy correlation function into the
1-group and 2-group components, and thus directly measure the
individual 1-halo and 2-halo components.

The detected different distribution of ESS early-type and late spiral
galaxies inside the dark matter halos raises the issue of whether the
effect is due to each population belonging to different types of
halos, or whether both galaxy types coexist within the same halos, but
with a different spatial distribution. In the former case, the
segregation effect would be related to the global halo properties like
mass, whereas in the latter case, it would be related to the local
properties such as dark matter density. The existence of both a 1-halo
and 2-halo components in the cross-correlation function of the two ESS
giant galaxy types (see \fg\ref{wrp-cross-auto-E-Sp}) indicates that
both effects may be at play. This is confirmed from the analyses
performed with the 2dFGRS group catalogue of \citet{yang05}: on the
one hand, \citet{yang05c} measure that central galaxies in high-mass,
low-mass halos are mostly early-type, respectively late-type galaxies;
on the other hand, \citet{yang05d} obtain direct evidence that
early-type galaxies are closer to the luminosity-weighted group center
than the late-type galaxies.

Another analysis of the three-dimensional density profiles of the
2dFGRS and SDSS groups found by a ``friend-of-friend'' algorithm
indicates a type segregation, namely a decrease of the early-type
galaxy fraction at larger group-centric distance and a corresponding
increase of the late-type fraction \citep{diaz05}.  At higher
redshifts ($0.7\le z\le1.5$), \citet{coil06} show that red galaxies
are more centrally concentrated than blue galaxies in the galaxy
groups extracted from the DEEP2 survey; this work uses yet another
algorithm for group selection, based on the search for galaxy
over-densities in redshift space which accounts for redshift-space
distortions.

These various results find confirmation in the thorough analysis of
\citet{weinmann06}, based on the SDSS ``halo-based'' catalogue: the
authors show that the fractions of early and late galaxy types not
only vary with distance from the halo center, but also with halo mass
over the full mass range probed, with more massive halos having
higher/lower fractions of early-type/late-type galaxies
\citep{weinmann06}. Interestingly, \citet{weinmann06} find a flat
distribution of intermediate-type galaxies as a function of mass and
distance to the halo center, with types based on both color and
specific star formation rate.

We also note that the detected segregation effects in the distribution
of early-type and late-type galaxies within the dark matter halos are
consistent with the type/density relationship, namely the trend for
early-type galaxies to preferentially inhabit high-density regions.
Although dense regions of a survey contain more galaxies that in the
other regions, it is not obvious that they undergo stronger
clustering.  However, \citet{abbas06} have shown that in high density
regions of the SDSS, galaxies are more clustered than in low density
regions, and this is valid at all scales from $0.1$ to $30$\hmpc.
This remarkable property is interpreted by \citet{abbas06} as
providing strong support to the hierarchical models, as it is well
reproduced by numerical and analytical models in which the entire
effect is due to the correlation of galaxy properties with the mass of
the parent halo, and to the fact that more massive halos populate
dense regions.

The type-density relationship was clarified by the detailed study of
\citet{blanton05}, who found that color (present star formation rate)
and luminosity (hence stellar mass, resulting from the history of past
star formation) are the two properties most predictive of local
density. Therefore, the detected segregation effects in the clustering
of ESS galaxies for different spectral type (early-type versus late
spiral) and luminosity (giant versus dwarf) are naturally
expected. The uniqueness of the results presented here is that we
identify for the first time the joint type/luminosity clustering
segregation effect in terms of the very galaxy types which correspond
to the locally well known morphological types.

\subsection{The dwarf galaxy correlation function}

The third major result obtained by the present ESS analysis is the
correlation function for the dwarf galaxies. This function is measured
for the first time, thanks to the separation of the dwarf galaxy
component which the ESS allows, based on the type specific luminosity
functions \citep{lapparent03}.  The projected-separation correlation
function of the ESS dwarf galaxies can also be interpreted in terms of
halo membership, as it displays a clear 1-halo component which falls
off at $r_\mathrm{p}>0.3$\hmpc.  By their stronger clustering at all
scales, the early-type galaxies appear as a key component of galaxy
clustering. The auto-correlation function of the dwarf galaxies and
their cross-correlation function with the early-type galaxies indicate
that they are the next contributor to galaxy clustering at small
scales. The additional evidence, based on the cross-correlation
analysis, that dwarf and late spiral galaxies are \emph{not} well
mixed at scales $\le0.3$\hmpc\ leads to a picture in which dwarf
galaxies are confined to the densest, hence central parts of the
halos, and are preferentially satellites of early-type galaxies.  This
is in agreement with the observation that in local groups and clusters
of galaxies, the dwarf-to-giant galaxy ratio is an increasing function
of the richness of the galaxy concentration
\citep{ferguson91,trentham02a,trentham02b}.

Most halo models parameterize the halo content in terms of a dominant
galaxy and it satellites \citep{berlind03,cooray06}. In their analysis
of SDSS data, \citet{zehavi05} perform a two-component HOD modelling
based on either red and blue central galaxies surrounded by red and
blue satellite respectively. However, in the ESS, the dwarf galaxy
sample is largely dominated by dwarf irregular hence blue galaxies,
and these appear as satellites of the early-type hence red
galaxies. Moreover, the late spiral galaxies may play the role of
central galaxies in the less massive halos, whereas in the more
massive halos, they may be considered as satellite galaxies.  Another
subtle effect is that detected by \citet{weinmann06} in the relative
distribution of central and satellite galaxies in halos, which they
name ``galaxy conformity'': for a halo of a given mass, the early-type
fraction of satellites is significantly higher when the central galaxy
is early-type rather than late-type. Altogether, these various results
indicate that reality may be more complex than the simple
two-component HOD models.

To illustrate the variety in galaxy types among a given halo, and
their specific spatial distributions, let us consider our local group,
which is typical of an intermediate-mass halo. It is dominated by the
Milky Way, an Sb galaxy, and Andromeda, an Sab galaxy. In the ESS
classification, the Milky Way would be classified as a late spiral,
and Andromeda would be at the limit between early-type and late
spiral. The third giant, although smaller galaxy, M33, is an Sdm galaxy,
and would be classified as a late spiral. The blue Sm/Irr satellites
of the Milky Way, the Magellanic Clouds, and the red dE satellites of
Andromeda, M32 and M110, would all be classified as dwarf
galaxies. There are in addition many dSph and dI galaxies in the local
group, distributed at typically $0.05-0.1$\hmpc\ and $0.5$\hmpc\ from
the giant galaxies (http://www.astro.washington.edu/mayer/LG/LG.html).

\subsection{Perspectives}

The present analysis emphasizes the need for further studies of galaxy
clustering as a function of galaxy type. This requires statistical
analyses of large galaxy samples effectively containing halos within a
large mass range, and a detailed knowledge of their galaxy content as
a function of galaxy mass, luminosity and type. The specific
clustering of the dwarf galaxies evidenced in the present ESS analysis
suggests that probing the dark halo content in terms of the full
sequence of giant and dwarf galaxy types would significantly enrich
our understanding of galaxy clustering. Higher signal-to-noise
measurement of the galaxy correlation functions for the various giant
and dwarf galaxy types, and interpretation using the ``Halo Occupation
Distribution'' or ``Conditional Luminosity Function'' would bring new
insight into their distribution within the dark matter halos, and the
relative role of the central and satellite galaxies in the halos.

The ultimate goal when identifying the different segregation effects
which galaxies undergo in a given halo, is to make a link with the
past history of star formation and mass accumulation of each system.
In such studies, the definition of the galaxy types will be important,
and the choice of the classification method will have an decisive
impact. The innovative approach of programme EFIGI at IAP (see
http://terapix.iap.fr/), aimed at obtaining a quantitative
morphological classification (Baillard \etal in preparation), should
allow one to reliably classify large samples of galaxies, hence to
better understand how each morphological type contributes to galaxy
clustering.


\begin{acknowledgements}
We thank C. Benoist (OCA), A. Cappi (INAF, Osservatorio Astronomico di Bologna)
and S. Maurogordato (OCA/CNRS) for useful discussions about subtleties of the
two-point correlation function estimators. We also wish to express our thanks
to F. Mignard (OCA/CNRS) who generously supplied the frequency mapping FAMOUS
software (ftp://ftp.obs-nice.fr/pub/mignard/Famous) which has been used to
produce the periodogram displayed in \fg\ref{pair-sep}.
\end{acknowledgements}


\begin{appendix}


\section{Distances  \label{distances}}

Redshifts $z$ are converted into comoving distances $r$
from the observer by using $\tau$ the dimensionless radial comoving
coordinate of the Robertson-Walker line element \citep{weinberg72}:
\begin{equation}
\begin{array}{ll}
r &= {c \over H_0}\ \tau \\
\mathrm{with}&\\
\tau &= \int_0^z[\;\Omega_\mathrm{m}(1+\upsilon)^3+\Omega_\Lambda\ ]^{-{1\over 2}}\;\mathrm{d}\upsilon\,
\end{array}
\label{eq:r-z}
\end{equation}

Given the chosen flat geometry, the comoving separation between
any two objects $i$ and $j$ with angular separation $\theta$ on the
sky is expressed from the usual law of cosines as:
\begin{equation}
s \equiv s_{ij} =
{c\over H_0}\ \sqrt{\tau_i^2 + \tau_j^2 - 2\tau_i\tau_j\cos\theta}\ .
\label{eq:r-sep}
\end{equation}


\section{Estimators                 \label{estimators}}

The galaxy-galaxy correlation function in redshift space $\xi(s)$ is
defined as the probability in excess of a homogeneous Poisson
distribution of finding in any direction two galaxies at distance $s$
from each other:
\begin{equation}
\delta P = \rho\ [ 1+\xi(s) ]\ \delta V,
\end{equation}
where $\rho$ is the mean space number density of galaxies and $\delta
V$ is the volume element \citep[see][]{peebles80}.  When the
distribution is homogeneous, $\xi(s)\equiv0$, but any uncertainty in
the mean density of the galaxy sample $\rho$ under study may result in
an error in the correlation amplitude, especially at large spatial
scales where the signal is below the fractional uncertainty in the
density.  To overcome this difficulty, together with the problem of
selection and boundaries effects in the data sample, several
estimators have been introduced which allow one to measure $\xi(s)$
from a finite set of objects with minimum bias and variance.  They are
generally defined as suitably normalized ratios of counts of galaxy
pairs separated by distance $s$ in a narrow interval of distances
$\delta s$ centered on $s$. The various considered pair counts are:
(i) the weighted number of pairs of observed galaxies
\begin{equation}
DD(s)=\Sigma_{i,j>i}\ w_i^{\rm (d)} w_j^{\rm (d)};
\label{eq:DD}
\end{equation}
(ii) the weighted number of pairs for a computer-generated random
distribution with the same selection criteria as the galaxy sample
(see \sct\ref{norm})
\begin{equation}
RR(s)=\Sigma_{i,j>i}\ w_i^{\rm (r)} w_j^{\rm (r)};
\label{eq:RR}
\end{equation}
and (iii) the weighted number of pairs between the set of random
objects and the observed galaxies
\begin{equation}
DR(s)=\Sigma_{i,j>i}\ w_i^{\rm (d)} w_j^{\rm (r)}.
\label{eq:DR}
\end{equation}
Note that a given pair of objects $(i,j)$ is only counted once in
\eqs\ref{eq:DD}, \ref{eq:RR} and \ref{eq:DR}.  The adopted functions for the
weights $w_i^{\mathrm{d}}$ and $w_i^{\mathrm{r}}$ are discussed in
\sct\ref{weights}.

In the following, we also denote $N_\mathrm{d}$ the number of observed
galaxies in the data sample and $N_\mathrm{r}$ the number of points in
the corresponding random set (see \sct\ref{norm}).

For investigating the correlation properties of the ESO-Sculptor
redshift survey, we consider three of the demonstrated best estimators
of $\xi(s)$.  The first estimator is that by \citet[][ denoted DP
estimator hereafter]{davis83}. If we denote $C^\mathrm{(d)}$ and
$C^\mathrm{(r)}$ the weighted object counts in the galaxy and random
samples \resp~(see \sct\ref{norm} for the definition of these
quantities), this estimator may be defined for large enough samples
($N_\mathrm{d},N_\mathrm{r} \gg$~1) as:
\begin{equation}
1+\hat{\xi}_\mathrm{DP}(s)\simeq 2\ {C^\mathrm{(r)} \over C^\mathrm{(d)}}\ {DD(s) \over
DR(s)},
\label{eq:xi-DP} 
\end{equation}
with the sums in the pair counts extending over all independent pairs
with redshift-space separations between $s-\delta s/2$ and $s+\delta
s/2$; $\hat{x}$ is the standard notation to refer to an estimator of
quantity $x$. The DP estimator is poorly sensitive to the adopted
edge correction but its variance varies as $1/\rho$.  With
its quadratic dependence on the uncertainty in the mean density
$\rho$, the \citet{hamilton93} estimator (denoted H estimator
hereafter) performs better than the DP estimator for sparse samples
with a poorly determined mean density. The H estimator takes into
account the pair count within the random sample according to:
\begin{equation}
1+\hat{\xi}_\mathrm{H}(s)\simeq 4\ {DD(s) \times RR(s) \over [\ DR(s)\ ]^2},
\label{eq:xi-H}
\end{equation}
which includes thereby a measure of the relative densities of the two
catalogues at any separation, via the pair counts (independent pairs
only); this allows one to bypass the density normalization factor
present in \eq \ref{eq:xi-DP}.

To minimize the effects of the finite solid angle on the sky,
\citet{landy93} introduced yet another quadratic estimator, denoted LS
estimator hereafter:
\begin{equation}
1+\hat{\xi}_\mathrm{LS}(s)\simeq 2\ + \left[{C^\mathrm{(r)} \over
C^\mathrm{(d)}}\right]^2\ {DD(s)
\over RR(s)} - {C^\mathrm{(r)} \over C^\mathrm{(d)}}\ {DR(s) \over RR(s)};
\label{eq:xi-LS}
\end{equation}
again, only independent pairs are counted. The authors show that this
estimator performs very well with a nearly Poisson variance for
uncorrelated data (for other clustering regimes, see
\citealt{bernstein94}), and is less sensitive to the number of points
in the random distribution than the H estimator \citep{kerscher00}.
Each estimator has its own theoretical advantages and weak points
which depend on the scale range under study \citep{pons-borderia99}.
Even if recent analyses have shown that the 3 estimators agree within
the error bars (\citealp{tucker97,guzzo00,zehavi02}; but see
\citealp{loveday95}), here we choose to calculate the 3 estimators for
each ESS sub-sample and to compare the estimates. This allows us to
secure our conclusions on the behavior of the correlation function.

Finally, we define the cross-correlation function between two
different sub-samples. It measures the excess probability over random
of finding a galaxy belonging to sample \#2 at a separation $s$ from a
galaxy belonging to sample \#1.  The same estimators as for the
auto-correlation function, but with slightly modified expressions and
normalization factors can be used.  The DP, H and LS estimators for
the two-point cross-correlation can be written respectively as~:
\begin{equation}
1+\hat{\xi}_\mathrm{DP}(s)\simeq \sqrt{ {C^\mathrm{(r_1)}C^\mathrm{(r_2)} \over 
C^\mathrm{(d_1)}C^\mathrm{(d_2)} }}\ 
{D_1D_2(s) \over \sqrt{D_1R_2(s)\ D_2R_1(s)} },
\label{eq:xi-DP-cross}
\end{equation}
\begin{equation}
1+\hat{\xi}_\mathrm{H}(s)\simeq {D_1D_2(s) \times R_1R_2(s) \over D_1R_2(s)\ D_2R_1(s) },
\label{eq:xi-H-cross}
\end{equation}
\begin{eqnarray}
1+\hat{\xi}_\mathrm{LS}(s)\simeq 2\ &+& {C^\mathrm{(r_1)}C^\mathrm{(r_2)} \over
C^\mathrm{(d_1)}C^\mathrm{(d_2)} } \ {D_1D_2(s) \over R_1R_2(s)}\nonumber \\
&-& {C^\mathrm{(r_1)} \over C^\mathrm{(d_1)}}\ {D_1R_2(s) \over R_1R_2(s)}
- {C^\mathrm{(r_2)} \over C^\mathrm{(d_2)}}\ {D_2R_1(s) \over R_2R_1(s)},
\label{eq:xi-LS-cross}
\end{eqnarray}
where the $D_1D_2(s)$, $R_1R_2(s)$ sums are the weighted numbers of
{\sl all} pairs with separations $s\pm \delta s / 2$, and the
$D_1R_2(s)$, $D_2R_1(s)$ sums are the weighted numbers of all
cross-reference data-random and random-data pairs. We have checked
that the three estimators in \eqs\ref{eq:xi-DP-cross} to
\ref{eq:xi-LS-cross} yield consistent measures for the various
cross-correlation functions considered in \sct\ref{wrp_cross}.


\section{Selection functions          \label{selfunc}}

In magnitude-limited surveys, the observed galaxy density varies
strongly with distance $r$ from the origin, because such surveys do
not include all the galaxies within a limiting redshift distance, but
only those bright enough to be detected. To calculate the correlation
functions, one must account for this selection effect. The
corresponding selection function is derived from the galaxy luminosity
function, denoted $\phi(M)$. The probability that a galaxy at comoving
distance $r$ (see \sct\ref{distances}) with absolute magnitude $M$ is
detected in a sample can be written as:
\begin{equation}
p(M) = {\phi(M)\over
\int_{M_\mathrm{bright}(r)}^{M_\mathrm{faint}(r)}\phi(M)\ \mathrm{d}M },
\end{equation}
where $M_\mathrm{bright}(r)$ and $M_\mathrm{faint}(r)$ are the
brightest and faintest absolute magnitudes observable at distance
$r$. The selection function $\psi(r)$ is then defined as the ratio
between the number of the detectable objects at $r$ and the total
number of galaxies which would be observed in a homogeneous sample
between absolute magnitudes $M_1$ to $M_2$:
\begin{equation}
\psi(r) = {\int_{\max(M_\mathrm{bright}(r),M_1)}^{\min(M_\mathrm{faint}(r),M_2)}
\phi(M)\ \mathrm{d}M\ \over \int_{M_1}^{M_2} \phi(M)\ \mathrm{d}M }.
\label{eq:selfunc}
\end{equation}
Note that $\psi(r)$ takes values in the interval $[0,1]$, with
$\psi(0)=1$ and $\psi(r)\rightarrow 0$ when $r\rightarrow\infty$. 

Here, we use for $M_1$ and $M_2$ the effective boundaries (rounded to
the first decimal place) of each considered sub-sample, in order to
have the same distribution in absolute magnitude for the observed
sample and the comparison random set (see \sct\ref{norm}). The values
of $M_1$ and $M_2$ adopted for each considered ESS sub-sample are
listed in the third column of Table~\ref{T1}. In the case of a
sub-sample with a cut in absolute magnitude, $M_1$ or $M_2$ in
\eq\ref{eq:selfunc} are replaced with the appropriate bound.

Because only 92\% of the ESS galaxies with $R_\mathrm{c}\le20.5$, and
52\% with $R_\mathrm{c}\le21.5$ have a measured redshift, we also
include the redshift incompleteness in the calculation of the
selection function. As the redshift incompleteness is uncorrelated
with the position on the sky and only depends on the apparent
magnitude \citep[see][]{lapparent03}, we proceed as follows. We bin
the redshift incompleteness in fixed intervals of 0.5 mag{.} in
apparent magnitude.  At each comoving distance $r$, we calculate the
corresponding intervals in absolute magnitude and split the numerator
integral in \eq\ref{eq:selfunc} into sub-integrals using these
intervals; then in each sub-integral, the incompleteness is accounted
for as a constant factor $\le1$.

Combined sub-samples including more than one class have their
selection function defined as follows.  The expected distance
distribution for a homogeneous distribution with a single spectral
class is
\begin{equation}
N(r)=\phi_0\psi(r)\int_{M_1}^{M_2}\varphi(M)\;\mathrm{d}M
\end{equation}
where $\varphi(M)$ is the shape of the luminosity function, 
defined as
\begin{equation}
\phi(M)\; \mathrm{d}M = \phi_0 \varphi(M)\; \mathrm{d}M\\
\label{eq:lf-shape}
\end{equation}
for a Gaussian parameterization. We also use \eq\ref{eq:lf-shape} for
the composite luminosity functions of the ESS intermediate-type and
late-type samples (see \sct\ref{lf}); then, the shape of the additive
Schechter component contributed to $\varphi(M)$ is scaled by
$\phi^*/\phi_0$.

By equating the total expected number of galaxies to the sum of the
expected numbers for each sub-sample, we obtain
\begin{equation}
\psi(r)={\Sigma_{k=1}^{K}\ \psi_k(r)\;\phi_{0k}\;\int_{M_1}^{M_2}
\varphi_k(M)\;\mathrm{d}M \over
\phi_0\;\int_{M_1}^{M_2}\varphi(M)\;\mathrm{d}M}
\label{eq:n-sum}
\end{equation}
The integral in the denominator is unknown, as the parametric form of
the luminosity function corresponding to the total sample is a priori
unknown. It can however be determined using the boundary condition
that \eq\ref{eq:n-sum} must also be valid at $r=0$, where all selection
functions $\psi(r)$ and $\psi_k(r)$ are equal to unity (see
\eq\ref{eq:selfunc}). This yields
\begin{equation}
\psi(r)={\Sigma_{k=1}^{K}\ \psi_k(r)\;\phi_{0k}\;\int_{M_1}^{M_2}
\varphi_k(M)\;\mathrm{d}M \over
\Sigma_{k=1}^{K}\ \phi_{0k}\;\int_{M_1}^{M_2}\varphi_k(M)\;\mathrm{d}M}
\label{eq:selfunc-comb}
\end{equation}
For ESS sub-samples with a cut in absolute magnitude, the selection
function for a single spectral class (\eq\ref{eq:selfunc}) is
calculated with the modified values of the bounds $M_1$ and $M_2$. For
ESS sub-samples from which is extracted a redshift interval, the
selection function is set to zero in that interval. For both types of
cuts, the selection function for a combined sample is derived using
\eq\ref{eq:selfunc-comb}.


\section{Weights                \label{weights}}

The selection functions described in \sct\ref{selfunc} can be
accounted for in the calculation of correlation functions by weighting
each pair of galaxies in the various estimators of
\sct\ref{estimators} according to three different schemes.  When the
weighting function is constant
\begin{equation}
w(r)\equiv1
\label{eq:w1}
\end{equation}
(denoted ``equal pair'' weighting), pairwise estimates of the
correlation function are biased against the few distant galaxies. In
contrast, weighting the galaxies in proportion to the inverse radial
selection function $\psi(r)$
\begin{equation}
w(r)\equiv1/\psi(r)
\label{eq:w-psi}
\end{equation}
(denoted ``equal volume'' weighting) gives too small a weight to the
well-sampled nearby regions where clustering dominates the galaxy shot
noise. This leads to the introduction of the minimum-variance
weighting scheme in which each object at distance $r$ in a pair with
separation $s$ is applied a weight
\begin{equation}
w(r,s) = 1\ /\ [\ 1 + 4\pi\ \rho\ \psi(r)\ J_3(s)\ ],
\label{eq:w-J3}
\end{equation}
where $4\pi J_3(s)$ is the volume integral of the two-point
correlation function $\xi(s)$ out to a separation $s$; note that this
approach can only be used if $\xi(s)$ vanishes on scales larger than
some scale $s_\mathrm{c}$.  The minimum-variance weighting (also
denoted $J3$ weighting) is intermediate between the two other
weighting schemes: although $\psi(r)$ increases at small values of
$r$, the decrease of $J3$ with $r$ dominates and $4\pi\ \rho\ \psi(r)\
J_3(s)\ll1$ at small $r$, so that $w(r,s)\sim1$; in contrast, at large
values of $r$, $4\pi\ \rho\ \psi(r)\ J_3(s)\gg1$ and $w(r,s)$ behaves
as $1/\psi(r)$.

It was also shown that the $J3$ weighting scheme gives the minimum
uncertainty in the clustering amplitude on scales where $\xi(s)\leq$~1
\citep{efstathiou88c,saunders92}.  \eq\ref{eq:w-J3} results from a
separable approximation of the true minimum-variance pair weighting
which is valid in the linear regime applicable at large separations,
when higher-order statistics can be neglected \citep{hamilton93}.  The
integral $J_3(s)$ can be calculated without involving any iterative
technique by modeling the required correlation function with a
power-law model and still yield accurate estimates of $\xi(s)$,
especially if one uses unbiased estimators for $\xi(s)$ (see
\sct\ref{estimators}; see also \citealt{ratcliffe98c,guzzo00}).

Here, we estimate $J_3(s)$ using the power-law model which provides a
good fit to most observed samples
\begin{equation}
\xi(s)=(s/s_0)^{-\gamma}
\label{eq:xi-power}
\end{equation}
with 
\begin{equation}
\begin{array}{ll}
\gamma&=1.6 \\
   s_0&=6\;h^{-1}\mathrm{Mpc}
\end{array}
\label{eq:xi-power-param}
\end{equation}
as measured from the existing redshift surveys
\citep{lapparent88,maurogordato92,loveday95,hermit96,tucker97,willmer98,
ratcliffe98b,zehavi02,hawkins03}. A posteriori, this is also in
acceptable agreement with the results for the ESS (see
\eq\ref{eq:xi-fit-wod}). The power-law parameterization of
\eq\ref{eq:xi-power} is used only for separations smaller than
$s_\mathrm{c}=30$\hmpc; we use $\xi(s)=0$ otherwise. This yields
\begin{equation}
\begin{array}{lll}
J_3(s)&=12.6\;s^{1.4}\ h^{-3}\mathrm{Mpc}^3 &\quad\mathrm{for}\quad s\le s_\mathrm{c},\\ 
J_3(s)&=1468\ h^{-3}\mathrm{Mpc}^3 &\quad\mathrm{for}\quad s>s_\mathrm{c}.
\end{array}
\label{eq:J3-power-1}
\end{equation}

Note that the $J_3$ weighting favors low-luminosity pairs at small
separations while luminous objects dominate the estimate on large
scales \citep{guzzo00}. The overall shape of the correlation function
may then change in case of any luminosity dependence of the galaxy
clustering. The $J_3$ weighting results must therefore be compared
with those obtained in the ``equal pair'' weighting, in particular at
small scales.


\section{Normalization                     \label{norm}}

Because the two-point correlation measures the excess number of pairs
over a homogeneous distribution, it requires a normalization which is
obtained by comparison of the number of pairs in a given ESS sample,
with that derived from a mock homogeneous distribution occupying the
same volume as the ESS sample, and having the selection function
$\psi(r)$ derived from the luminosity function of that sample. Because
the redshift incompleteness is accounted for in the selection function
$\psi(r)$ (see \sct\ref{selfunc}), it is automatically accounted
for in the random distributions, and does not need to be included into
the weighting functions of \eqs\ref{eq:w1}, \ref{eq:w-psi} and
\ref{eq:w-J3}.

The Monte-Carlo set containing $N_\mathrm{r}$ points, and
corresponding to each data sub-sample defined in Table~\ref{T2}
includes at least fifty times as many objects as the observational
catalogue:
\begin{equation}
\begin{array}{llll}
N_\mathrm{r}&=50\ N_\mathrm{d} &\quad\mathrm{for}\quad N_\mathrm{d}&\ge100, \\
N_\mathrm{r}&=5000             &\quad\mathrm{for}\quad N_\mathrm{d}&<100.
\end{array}
\label{eq:n-random}
\end{equation}
The number of random points $N_\mathrm{r}$ is then large enough to
ensure that the fluctuation in $C^\mathrm{(r)}$ (see below) and the
related uncertainty in $\xi(s)$ are negligible, so that the
uncertainties in the correlation function are dominated by those in
the pair count $DD(s)$ (see \eq\ref{eq:DD}).  The normalizing factor
${C^\mathrm{(r)} /C^\mathrm{(d)}}$ (defined below) then allows one to
normalize the density of the random distribution to that of the data
sample for the DP and LS estimators (\eqs\ref{eq:xi-DP} and
\ref{eq:xi-LS}; in the H estimator of \eq\ref{eq:xi-H}, the
normalizing factor cancels out).  This normalization of the observed
number of pairs is equivalent to adopting the mean density for each
observed sample as the reference density.

Each random distribution is then generated by randomly drawing points
with a redshift probability distribution defined by the selection
function $\psi(r)$ corresponding to the data sub-sample
(\eq\ref{eq:selfunc}). The RA and Dec coordinates of each random point
are also drawn randomly between the ESS extreme values while
accounting for the small excluded RA and Dec regions due to saturated
stars.  For a sub-sample in Table~\ref{T1} which is based on one
spectral type \emph{and} has cuts in either redshift or absolute
magnitude, the number of observed objects $N_\mathrm{d}$ after
applying the redshift or magnitude cut is listed in Table~\ref{T1} and
used in \eq\ref{eq:n-random}, and the random distributions is
generated using the cut-updated selection function for that spectral
type (see \sct\ref{selfunc}).  For the combined samples (for example
the sample containing ``all galaxies''), we use the reunion of the
random sets corresponding to each spectral-type sub-sample and each
satisfying \eq\ref{eq:n-random}, which ensures that the selection
functions and relative proportions for each spectral-type are taken
into account.

Then, from each observed sample and its corresponding random set, we
calculate the data pairs counts $DD(s)$ (\eq\ref{eq:DD}) and the
comparison random pair counts $DR(s)$ and $RR(s)$ (\eqs\ref{eq:DR} and
\ref{eq:RR}).  In the case of ``equal pair'' or ``equal volume''
weighting, the normalizing ratio of weighted counts of objects which
appear in the DP and LS estimators (see \eqs\ref{eq:xi-DP} and
\ref{eq:xi-LS}) is defined as
\begin{equation}
{C^\mathrm{(r)} \over C^\mathrm{(d)}} = { 
\Sigma_{l=1}^{N_\mathrm{r}}\ w(r_l)
\over
\Sigma_{k=1}^{N_\mathrm{d}}\ w(r_k) }
\label{eq:cr-cd}
\end{equation}
where the sums run over the $N_\mathrm{r}$, $N_\mathrm{d}$ objects of
the random, \resp~observed distributions, and the weights are defined
in \eqs\ref{eq:w-psi} and \ref{eq:w-J3}.  In the case of $J_3$
weighting, the ratio of weighted pair counts in the DP and LS
estimators is computed as
\begin{equation}
{C^\mathrm{(r)} \over C^\mathrm{(d)}} = { 
\Sigma_{l=1}^{N_{\rm r}}\ w(r_l,s_\mathrm{c})
\over
\Sigma_{k=1}^{N_\mathrm{d}}\ w(r_k,s_\mathrm{c}) }
\quad\mathrm{with}\quad s_\mathrm{c} \equiv 30\ h^{-1}~\mathrm{Mpc}.
\end{equation}
By fixing the value of the $J3$ weights at the large pair separation
$s_\mathrm{c}=30$\hmpc, we ensure that the weighted pair counts
$C^\mathrm{(d)}$ and $C^\mathrm{(r)}$ are not affected by galaxy
clustering.


\section{Mean density        \label{density}}

In the case of the $J_3$ pair-weighting function (\eq\ref{eq:w-J3}),
one must define an estimator of the mean number density $\rho$. Given
a magnitude-limited sample of $N$ galaxies, we denote
$\rho(M_1<M<M_2)$ the mean density of galaxies with absolute magnitude
$M$ in the interval $M_1<M<M_2$ (corresponding to the bounds used in
the selection function, in \eq\ref{eq:selfunc}). An estimator of
$\rho(M_1<M<M_2)$ which is unbiased by the selection function
$\psi(r)$ can be obtained using
\begin{equation}
\hat\rho(M_1<M<M_2) = {\sum_{i=1}^N w(r_i)
\over 
\int_{r_{\min}}^{r_{\max}} w(r)\ \psi(r)\ {\mathrm{d}V \over \mathrm{d}r}\;\mathrm{d}r},
\label{eq:mean-dens}
\end{equation}
where $w(r)$ is a weighting function (see \eqs\ref{eq:w1},
\ref{eq:w-psi} and \ref{eq:w-J3}). The comoving distances $r_{\min}$
and $r_{\max}$ correspond to the redshift boundaries of the sample
(\eq\ref{eq:rlim}).

\citet{davis82b} showed that in the case of ``equal pair'' weighting
(\eq\ref{eq:w1}), this estimator is the most stable, but heavily weights
galaxies near the peak of the redshift distribution.  In the case of
``equal volume'' weighting (\eq\ref{eq:w-psi}), \citet{davis82b} also
showed that \eq\ref{eq:mean-dens} is close to the minimum variance
estimator, but that it heavily weights distant structures.  The $J_3$
weighting defined in \eq\ref{eq:w-J3} provides an intermediate estimate
of the mean density.

To estimate the mean density required for a $J_3$ pair-weighting
(\eq\ref{eq:w-psi}), we iterate over \eqs\ref{eq:mean-dens} and
\ref{eq:w-J3}. The ``equal volume'' weighting (\eq\ref{eq:w1}) is
used for calculating a first value of $\rho$.  Then, in each
calculation of $w(r)$, we use \eq\ref{eq:w-J3} with $J_3(s)\equiv
J_3(30\ h^{-1}~\mathrm{Mpc})=1468$\hmpcc\ (see
\eq\ref{eq:J3-power-1}), because the estimate of $\rho$ in
\eq\ref{eq:mean-dens} requires a weighting function with the comoving
distance $r$ as the only variable, whereas the $J_3$ weighting
function $w(r,s)$ varies with both $r$ and separation $s$; we thus
ensure that the weights, and therefore the mean density estimate, are
not affected by galaxy clustering.  This yields for the 3
spectral-type sub-samples:
\begin{eqnarray}
\hat{\rho}=&\ 8.47\ 10^{-3}\ h^3~\mathrm{Mpc}^{-3}&\quad\mathrm{for}\ \rm{early-type}\nonumber\\ 
\hat{\rho}=&33.58\ 10^{-3}\ h^3~\mathrm{Mpc}^{-3}&\quad\mathrm{for}\ \rm{intermediate-type}\\ 
\hat{\rho}=&52.41\ 10^{-3}\ h^3~\mathrm{Mpc}^{-3}&\quad\mathrm{for}\ \rm{late-type}\nonumber
\end{eqnarray}
With these values of $\rho$, $4\pi\ \rho\ \psi(r)\ J_3(30)\ge10$ for
$r\le1500$\hmpc. As a result, $w(r,s)\propto1/\psi(r)$, which
corresponds to the ``equal volume'' weighting (\eq\ref{eq:w-psi}).


\section{Projected correlation functions        \label{proj}}

Peculiar velocities distort the redshift-space correlation function
$\xi(s)$, which then differs from the real-space correlation function
$\xi(r)$.  In redshift-space, internal random motions within bound
structures create the so-called ``finger-of-god'' structures
(elongated along the line-of-sight), while coherent motions on large
scales tend to flatten the over-densities along the observer's line of
sight \citep{hawkins03}. Moreover, the rms uncertainty of
$\sim1.6$\hmpc\ on the line-of-sight separation caused by the
redshift measurement uncertainties in the ESS, also contributes to
smooth out any clustering in redshift space on scales comparable
$s\la3$\hmpc\ \citep[see $\S$3.1 and \fg~4 in][]{cole94}.

Because the redshift-space distortions are
only radial, one can compute the correlation function as a function of
separation parallel ($\pi$) and perpendicular ($r_\mathrm{p}$) to the
line-of-sight, which allows one to disentangle the effects of peculiar
velocities from the genuine spatial correlations. Following the
formalism of \citet{fisher94a}, for any two galaxies with redshift
positions $\mathbf{P_1}$ and $\mathbf{P_2}$, the redshift separation
and line-of-sight vectors are defined as $\mathbf{S}\equiv
\mathbf{P_2} - \mathbf{P_1}$ and $\mathbf{L}\equiv
0.5\times(\mathbf{P_1}+\mathbf{P_2)}$, respectively. Therefore, the
parallel and perpendicular separations are:
\begin{equation}
\begin{array}{ll}
\pi = { \mathbf{S}\cdot\mathbf{L} / \ \vert L \vert}, \\
 r^2_\mathrm{p}=\mathbf{S}\cdot\mathbf{S} -\ \pi^2.
\end{array}
\label{eq:rp-pi}
\end{equation}
The redshift-space correlation function $\xi(r_\mathrm{p},\pi)$ can
then be derived for each estimators by replacing $DD(s)$, $RR(s)$ and
$DR(s)$ in \eqs\ref{eq:xi-DP}, \ref{eq:xi-H} and \ref{eq:xi-LS} with
$DD(r_\mathrm{p},\pi)$ $RR(r_\mathrm{p},\pi)$ $DR(r_\mathrm{p},\pi)$,
which refer to the data--data, random--random and data--random pair
counts \resp\ at each value of $(r_\mathrm{p},\pi)$.

In a second stage, $\xi(r_\mathrm{p},\pi)$ allows one to derive the
correlation function $w(r_\mathrm{p})$ as a function of projected
separation $r_\mathrm{p}$, which is unaffected by redshift distortions
\citep{davis83}, and is obtained by integrating
$\xi(r_\mathrm{p},\pi)$ over $\pi$:
\begin{equation}
w(r_\mathrm{p}) = 2\int_0^\infty \xi(r_\mathrm{p},\pi)\ {\mathrm d}\pi
=2\ \sum_i \xi(r_\mathrm{p},\pi_i) \Delta_i,
\label{eq:wrp}
\end{equation}
The summation yields an unbiased estimate of $w(r_\mathrm{p})$
\citep{jing98}, which is related to the real-space correlation
function $\xi(r)$ by:
\begin{equation}
w(r_\mathrm{p}) = 2\int_0^\infty \xi(\sqrt{r^2_\mathrm{p}+y^2})\ {\mathrm d}y.
\label{eq:wrp-xi}
\end{equation}
If $\xi(r)$ is modelled as a power-law $\xi(r)=(r/r_0)^{-\gamma}$,
the integral in \eq\ref{eq:wrp-xi} can be calculated analytically, and yields:
\begin{equation}
w(r_\mathrm{p}) = r_\mathrm{p}^{1-\gamma}\ \ r_0^\gamma\
{ \Gamma(1/2)\Gamma(\gamma/2-1/2) \over \Gamma(\gamma/2)},
\label{eq:wrp-gamma}
\end{equation}
where $\Gamma(x)$ is the Gamma function.

The model parameters $r_0$ and $\gamma$ in \eq\ref{eq:wrp-gamma} are
derived by minimizing the value of $\chi^2$ defined as:
\begin{equation}
\chi^2=\sum_i{ [\xi(r_i)-(r_i/r_0)^{-\gamma}]^2 \over \sigma_i^2},
\label{eq:chi2}
\end{equation}
where $\xi(r_i)$ and $\sigma_i$ are the measured values of the
correlation function and its rms fluctuation at a separation $r_i$,
assuming thereby that the correlation between $\xi(r_i)$ values
leads to a small enough bias on the final result.

\end{appendix}



\bibliography{aamnem99,7150_final2}

\begin{thebibliography}{115}
\expandafter\ifx\csname natexlab\endcsname\relax\def\natexlab#1{#1}\fi

\bibitem[{{Abazajian} {et~al.}(2005){Abazajian}, {Zheng}, {Zehavi}, {Weinberg},
  {Frieman}, {Berlind}, {Blanton}, {Bahcall}, {Brinkmann}, {Schneider}, \&
  {Tegmark}}]{abazajian05}
{Abazajian}, K., {Zheng}, Z., {Zehavi}, I., {et~al.} 2005, \apj, 625, 613

\bibitem[{{Abbas} \& {Sheth}(2006)}]{abbas06}
{Abbas}, U. \& {Sheth}, R.~K. 2006, \mnras, 372, 1749

\bibitem[{{Andreon} \& {Cuillandre}(2002)}]{andreon01c}
{Andreon}, S. \& {Cuillandre}, J.-C. 2002, ApJ, 569, 144

\bibitem[{{Arnouts} {et~al.}(1997){Arnouts}, {de Lapparent}, {Mathez},
  {Mazure}, {Mellier}, {Bertin}, \& {Kruszewski}}]{arnouts97}
{Arnouts}, S., {de Lapparent}, V., {Mathez}, G., {et~al.} 1997, A\&A, 124, 163

\bibitem[{{Baldi} {et~al.}(2001){Baldi}, {Bardelli}, \& {Zucca}}]{baldi01}
{Baldi}, A., {Bardelli}, S., \& {Zucca}, E. 2001, MNRAS, 324, 509

\bibitem[{{Baugh}(1996)}]{baugh96}
{Baugh}, C.~M. 1996, MNRAS, 280, 267

\bibitem[{{Bellanger} \& {de Lapparent}(1995)}]{bellanger95b}
{Bellanger}, C. \& {de Lapparent}, V. 1995, ApJ Lett., 455, L103

\bibitem[{{Bellanger} {et~al.}(1995){Bellanger}, {de Lapparent}, {Arnouts},
  {Mathez}, {Mazure}, \& {Mellier}}]{bellanger95a}
{Bellanger}, C., {de Lapparent}, V., {Arnouts}, S., {et~al.} 1995, A\&AS, 110,
  159

\bibitem[{{Benoist} {et~al.}(1996){Benoist}, {Maurogordato}, {da Costa},
  {Cappi}, \& {Schaeffer}}]{benoist96}
{Benoist}, C., {Maurogordato}, S., {da Costa}, L.~N., {Cappi}, A., \&
  {Schaeffer}, R. 1996, ApJ, 472, 452

\bibitem[{{Benson} {et~al.}(2000){Benson}, {Cole}, {Frenk}, {Baugh}, \&
  {Lacey}}]{benson00}
{Benson}, A.~J., {Cole}, S., {Frenk}, C.~S., {Baugh}, C.~M., \& {Lacey}, C.~G.
  2000, \mnras, 311, 793

\bibitem[{{Benson} {et~al.}(2001){Benson}, {Frenk}, {Baugh}, {Cole}, \&
  {Lacey}}]{benson01}
{Benson}, A.~J., {Frenk}, C.~S., {Baugh}, C.~M., {Cole}, S., \& {Lacey}, C.~G.
  2001, \mnras, 327, 1041

\bibitem[{{Berlind} \& {Weinberg}(2002)}]{berlind02}
{Berlind}, A.~A. \& {Weinberg}, D.~H. 2002, ApJ, 575, 587

\bibitem[{{Berlind} {et~al.}(2003){Berlind}, {Weinberg}, {Benson}, {Baugh},
  {Cole}, {Dav{\'e}}, {Frenk}, {Jenkins}, {Katz}, \& {Lacey}}]{berlind03}
{Berlind}, A.~A., {Weinberg}, D.~H., {Benson}, A.~J., {et~al.} 2003, ApJ, 593,
  1

\bibitem[{{Bernstein}(1994)}]{bernstein94}
{Bernstein}, G.~M. 1994, ApJ, 424, 569

\bibitem[{{Binggeli} {et~al.}(1990){Binggeli}, {Tarenghi}, \&
  {Sandage}}]{binggeli90}
{Binggeli}, B., {Tarenghi}, M., \& {Sandage}, A. 1990, A\&A, 228, 42

\bibitem[{{Blanton} {et~al.}(2005){Blanton}, {Eisenstein}, {Hogg}, {Schlegel},
  \& {Brinkmann}}]{blanton05}
{Blanton}, M.~R., {Eisenstein}, D., {Hogg}, D.~W., {Schlegel}, D.~J., \&
  {Brinkmann}, J. 2005, ApJ, 629, 143

\bibitem[{{Brainerd} {et~al.}(1995){Brainerd}, {Smail}, \&
  {Mould}}]{brainerd95}
{Brainerd}, T.~G., {Smail}, I., \& {Mould}, J. 1995, MNRAS, 275, 781

\bibitem[{{Bromley} {et~al.}(1998){Bromley}, {Press}, {Lin}, \&
  {Kirshner}}]{bromley98}
{Bromley}, B.~C., {Press}, W.~H., {Lin}, H., \& {Kirshner}, R.~P. 1998, ApJ,
  505, 25

\bibitem[{{Budav{\'a}ri} {et~al.}(2003){Budav{\'a}ri}, {Connolly}, {Szalay},
  {Szapudi}, {Csabai}, {Scranton}, {Bahcall}, {Brinkmann}, {Eisenstein},
  {Frieman}, {Fukugita}, {Gunn}, {Johnston}, {Kent}, {Loveday}, {Lupton},
  {Tegmark}, {Thakar}, {Yanny}, {York}, \& {Zehavi}}]{budavari03}
{Budav{\'a}ri}, T., {Connolly}, A.~J., {Szalay}, A.~S., {et~al.} 2003, \apj,
  595, 59

\bibitem[{{Carlberg} {et~al.}(2000){Carlberg}, {Yee}, {Morris}, {Lin}, {Hall},
  {Patton}, {Sawicki}, \& {Shepherd}}]{carlberg00}
{Carlberg}, R.~G., {Yee}, H.~K.~C., {Morris}, S.~L., {et~al.} 2000, \apj, 542,
  57

\bibitem[{{Coil} {et~al.}(2004){Coil}, {Davis}, {Madgwick}, {Newman},
  {Conselice}, {Cooper}, {Ellis}, {Faber}, {Finkbeiner}, {Guhathakurta},
  {Kaiser}, {Koo}, {Phillips}, {Steidel}, {Weiner}, {Willmer}, \&
  {Yan}}]{coil04}
{Coil}, A.~L., {Davis}, M., {Madgwick}, D.~S., {et~al.} 2004, \apj, 609, 525

\bibitem[{{Coil} {et~al.}(2006){Coil}, {Gerke}, {Newman}, {Ma}, {Yan},
  {Cooper}, {Davis}, {Faber}, {Guhathakurta}, \& {Koo}}]{coil06}
{Coil}, A.~L., {Gerke}, B.~F., {Newman}, J.~A., {et~al.} 2006, \apj, 638, 668

\bibitem[{{Cole} {et~al.}(1994){Cole}, {Ellis}, {Broadhurst}, \&
  {Colless}}]{cole94}
{Cole}, S., {Ellis}, R.~S., {Broadhurst}, T.~J., \& {Colless}, M.~M. 1994,
  MNRAS, 267, 541

\bibitem[{{Colless} {et~al.}(2001){Colless}, {Dalton}, {Maddox}, {Sutherland},
  {Norberg}, {Cole}, {Bland-Hawthorn}, {Bridges}, {Cannon}, {Collins}, {Couch},
  {Cross}, {Deeley}, {De Propris}, {Driver}, {Efstathiou}, {Ellis}, {Frenk},
  {Glazebrook}, {Jackson}, {Lahav}, {Lewis}, {Lumsden}, {Madgwick}, {Peacock},
  {Peterson}, {Price}, {Seaborne}, \& {Taylor}}]{colless01}
{Colless}, M., {Dalton}, G., {Maddox}, S., {et~al.} 2001, MNRAS, 328, 1039

\bibitem[{{Conroy} {et~al.}(2006){Conroy}, {Wechsler}, \&
  {Kravtsov}}]{conroy06}
{Conroy}, C., {Wechsler}, R.~H., \& {Kravtsov}, A.~V. 2006, \apj, 647, 201

\bibitem[{{Cooray}(2006)}]{cooray06}
{Cooray}, A. 2006, \mnras, 365, 842

\bibitem[{{Courteau} \& {van den Bergh}(1999)}]{courteau99}
{Courteau}, S. \& {van den Bergh}, S. 1999, AJ, 118, 337

\bibitem[{{Davis} \& {Huchra}(1982)}]{davis82b}
{Davis}, M. \& {Huchra}, J. 1982, ApJ, 254, 437

\bibitem[{{Davis} \& {Peebles}(1983)}]{davis83}
{Davis}, M. \& {Peebles}, P.~J.~E. 1983, ApJ, 267, 465

\bibitem[{{de Lapparent} {et~al.}(2004){de Lapparent}, {Arnouts}, {Galaz}, \&
  {Bardelli}}]{lapparent04}
{de Lapparent}, V., {Arnouts}, S., {Galaz}, G., \& {Bardelli}, S. 2004, A\&A,
  422, 841

\bibitem[{{de Lapparent} {et~al.}(2003){de Lapparent}, {Galaz}, {Bardelli}, \&
  {Arnouts}}]{lapparent03}
{de Lapparent}, V., {Galaz}, G., {Bardelli}, S., \& {Arnouts}, S. 2003, A\&A,
  404, 831

\bibitem[{{de Lapparent} {et~al.}(1986){de Lapparent}, {Geller}, \&
  {Huchra}}]{lapparent86}
{de Lapparent}, V., {Geller}, M.~J., \& {Huchra}, J.~P. 1986, ApJ Lett., 302,
  L1

\bibitem[{{de Lapparent} {et~al.}(1988){de Lapparent}, {Geller}, \&
  {Huchra}}]{lapparent88}
---. 1988, ApJ, 332, 44

\bibitem[{{D{\'{\i}}az} {et~al.}(2005){D{\'{\i}}az}, {Zandivarez},
  {Merch{\'a}n}, \& {Muriel}}]{diaz05}
{D{\'{\i}}az}, E., {Zandivarez}, A., {Merch{\'a}n}, M.~E., \& {Muriel}, H.
  2005, \apj, 629, 158

\bibitem[{{Dressler}(1980)}]{dressler80}
{Dressler}, A. 1980, ApJ, 236, 351

\bibitem[{{Efstathiou}(1988)}]{efstathiou88c}
{Efstathiou}, G. 1988, in LNP Vol. 297: Comets to Cosmology, 312--+

\bibitem[{{Ekholm} {et~al.}(2001){Ekholm}, {Baryshev}, {Teerikorpi}, {Hanski},
  \& {Paturel}}]{ekholm01}
{Ekholm}, T., {Baryshev}, Y., {Teerikorpi}, P., {Hanski}, M.~O., \& {Paturel},
  G. 2001, A\&A, 368, L17

\bibitem[{{Ferguson} \& {Sandage}(1991)}]{ferguson91}
{Ferguson}, H.~C. \& {Sandage}, A. 1991, AJ, 101, 765

\bibitem[{{Fioc} \& {Rocca-Volmerange}(1997)}]{fioc97}
{Fioc}, M. \& {Rocca-Volmerange}, B. 1997, A\&A, 326, 950

\bibitem[{{Fisher} {et~al.}(1994){Fisher}, {Davis}, {Strauss}, {Yahil}, \&
  {Huchra}}]{fisher94a}
{Fisher}, K.~B., {Davis}, M., {Strauss}, M.~A., {Yahil}, A., \& {Huchra}, J.
  1994, MNRAS, 266, 50

\bibitem[{{Folkes} {et~al.}(1996){Folkes}, {Lahav}, \& {Maddox}}]{folkes96}
{Folkes}, S.~R., {Lahav}, O., \& {Maddox}, S.~J. 1996, MNRAS, 283, 651

\bibitem[{{Galaz} \& {de Lapparent}(1998)}]{galaz98}
{Galaz}, G. \& {de Lapparent}, V. 1998, A\&A, 332, 459

\bibitem[{{Giuricin} {et~al.}(2001){Giuricin}, {Samurovi{\' c}}, {Girardi},
  {Mezzetti}, \& {Marinoni}}]{giuricin01}
{Giuricin}, G., {Samurovi{\' c}}, S., {Girardi}, M., {Mezzetti}, M., \&
  {Marinoni}, C. 2001, ApJ, 554, 857

\bibitem[{{Grant} {et~al.}(2005){Grant}, {Kuipers}, \& {Phillipps}}]{grant05}
{Grant}, N.~I., {Kuipers}, J.~A., \& {Phillipps}, S. 2005, \mnras, 363, 1019

\bibitem[{{Guzzo} {et~al.}(2000){Guzzo}, {Bartlett}, {Cappi}, {Maurogordato},
  {Zucca}, {Zamorani}, {Balkowski}, {Blanchard}, {Cayatte}, {Chincarini},
  {Collins}, {Maccagni}, {MacGillivray}, {Merighi}, {Mignoli}, {Proust},
  {Ramella}, {Scaramella}, {Stirpe}, \& {Vettolani}}]{guzzo00}
{Guzzo}, L., {Bartlett}, J.~G., {Cappi}, A., {et~al.} 2000, A\&A, 355, 1

\bibitem[{{Hamilton}(1993)}]{hamilton93}
{Hamilton}, A.~J.~S. 1993, ApJ, 417, 19

\bibitem[{{Hawkins} {et~al.}(2003){Hawkins}, {Maddox}, {Cole}, {Lahav},
  {Madgwick}, {Norberg}, {Peacock}, {Baldry}, {Baugh}, {Bland-Hawthorn},
  {Bridges}, {Cannon}, {Colless}, {Collins}, {Couch}, {Dalton}, {De Propris},
  {Driver}, {Efstathiou}, {Ellis}, {Frenk}, {Glazebrook}, {Jackson}, {Jones},
  {Lewis}, {Lumsden}, {Percival}, {Peterson}, {Sutherland}, \&
  {Taylor}}]{hawkins03}
{Hawkins}, E., {Maddox}, S., {Cole}, S., {et~al.} 2003, MNRAS, 346, 78

\bibitem[{{Hermit} {et~al.}(1996){Hermit}, {Santiago}, {Lahav}, {Strauss},
  {Davis}, {Dressler}, \& {Huchra}}]{hermit96}
{Hermit}, S., {Santiago}, B.~X., {Lahav}, O., {et~al.} 1996, MNRAS, 283, 709

\bibitem[{{Hogg} {et~al.}(2000){Hogg}, {Cohen}, \& {Blandford}}]{hogg00}
{Hogg}, D.~W., {Cohen}, J.~G., \& {Blandford}, R. 2000, \apj, 545, 32

\bibitem[{{Jenkins} {et~al.}(2001){Jenkins}, {Frenk}, {White}, {Colberg},
  {Cole}, {Evrard}, {Couchman}, \& {Yoshida}}]{jenkins01}
{Jenkins}, A., {Frenk}, C.~S., {White}, S.~D.~M., {et~al.} 2001, MNRAS, 321,
  372

\bibitem[{{Jerjen} \& {Tammann}(1997)}]{jerjen97b}
{Jerjen}, H. \& {Tammann}, G.~A. 1997, A\&A, 321, 713

\bibitem[{{Jing} {et~al.}(1998){Jing}, {Mo}, \& {Boerner}}]{jing98}
{Jing}, Y.~P., {Mo}, H.~J., \& {Boerner}, G. 1998, ApJ, 494, 1

\bibitem[{{Kauffmann} {et~al.}(1999{\natexlab{a}}){Kauffmann}, {Colberg},
  {Diaferio}, \& {White}}]{kauffmann99}
{Kauffmann}, G., {Colberg}, J.~M., {Diaferio}, A., \& {White}, S.~D.~M.
  1999{\natexlab{a}}, \mnras, 303, 188

\bibitem[{{Kauffmann} {et~al.}(1999{\natexlab{b}}){Kauffmann}, {Colberg},
  {Diaferio}, \& {White}}]{kauffmann99b}
---. 1999{\natexlab{b}}, \mnras, 307, 529

\bibitem[{{Kennicutt}(1992)}]{kennicutt92}
{Kennicutt}, R.~C. 1992, 79, 255

\bibitem[{{Kerscher} {et~al.}(2000){Kerscher}, {Szapudi}, \&
  {Szalay}}]{kerscher00}
{Kerscher}, M., {Szapudi}, I., \& {Szalay}, A.~S. 2000, ApJ Lett., 535, L13

\bibitem[{{Landy} \& {Szalay}(1993)}]{landy93}
{Landy}, S.~D. \& {Szalay}, A.~S. 1993, ApJ, 412, 64

\bibitem[{{Le Fevre} {et~al.}(1996){Le Fevre}, {Hudon}, {Lilly}, {Crampton},
  {Hammer}, \& {Tresse}}]{lefevre96}
{Le Fevre}, O., {Hudon}, D., {Lilly}, S.~J., {et~al.} 1996, \apj, 461, 534

\bibitem[{{Li} {et~al.}(2006){Li}, {Kauffmann}, {Jing}, {White}, {B{\"o}rner},
  \& {Cheng}}]{li06}
{Li}, C., {Kauffmann}, G., {Jing}, Y.~P., {et~al.} 2006, \mnras, 368, 21

\bibitem[{{Lilly} {et~al.}(1995){Lilly}, {Tresse}, {Hammer}, {Crampton}, \& {Le
  Fevre}}]{lilly95}
{Lilly}, S.~J., {Tresse}, L., {Hammer}, F., {Crampton}, D., \& {Le Fevre}, O.
  1995, ApJ, 455, 108

\bibitem[{{Lin} {et~al.}(1999){Lin}, {Yee}, {Carlberg}, {Morris}, {Sawicki},
  {Patton}, {Wirth}, \& {Shepherd}}]{lin99}
{Lin}, H., {Yee}, H. K.~C., {Carlberg}, R.~G., {et~al.} 1999, ApJ, 518, 533

\bibitem[{{Ling} {et~al.}(1986){Ling}, {Barrow}, \& {Frenk}}]{ling86}
{Ling}, E.~N., {Barrow}, J.~D., \& {Frenk}, C.~S. 1986, MNRAS, 223, 21P

\bibitem[{{Loveday} {et~al.}(1992){Loveday}, {Efstathiou}, {Peterson}, \&
  {Maddox}}]{loveday92}
{Loveday}, J., {Efstathiou}, G., {Peterson}, B.~A., \& {Maddox}, S.~J. 1992,
  \apjl, 400, L43

\bibitem[{{Loveday} {et~al.}(1995){Loveday}, {Maddox}, {Efstathiou}, \&
  {Peterson}}]{loveday95}
{Loveday}, J., {Maddox}, S.~J., {Efstathiou}, G., \& {Peterson}, B.~A. 1995,
  ApJ, 442, 457

\bibitem[{{Loveday} {et~al.}(1999){Loveday}, {Tresse}, \& {Maddox}}]{loveday99}
{Loveday}, J., {Tresse}, L., \& {Maddox}, S. 1999, MNRAS, 310, 281

\bibitem[{{Madgwick} {et~al.}(2003){Madgwick}, {Hawkins}, {Lahav}, {Maddox},
  {Norberg}, {Peacock}, {Baldry}, {Baugh}, {Bland-Hawthorn}, {Bridges},
  {Cannon}, {Cole}, {Colless}, {Collins}, {Couch}, {Dalton}, {De Propris},
  {Driver}, {Efstathiou}, {Ellis}, {Frenk}, {Glazebrook}, {Jackson}, {Lewis},
  {Lumsden}, {Peterson}, {Sutherland}, \& {Taylor}}]{madgwick03}
{Madgwick}, D.~S., {Hawkins}, E., {Lahav}, O., {et~al.} 2003, \mnras, 344, 847

\bibitem[{{Magliocchetti} \& {Porciani}(2003)}]{magliocchetti03}
{Magliocchetti}, M. \& {Porciani}, C. 2003, \mnras, 346, 186

\bibitem[{{Masjedi} {et~al.}(2006){Masjedi}, {Hogg}, {Cool}, {Eisenstein},
  {Blanton}, {Zehavi}, {Berlind}, {Bell}, {Schneider}, {Warren}, \&
  {Brinkmann}}]{masjedi06}
{Masjedi}, M., {Hogg}, D.~W., {Cool}, R.~J., {et~al.} 2006, \apj, 644, 54

\bibitem[{{Maurogordato} {et~al.}(1992){Maurogordato}, {Schaeffer}, \& {da
  Costa}}]{maurogordato92}
{Maurogordato}, S., {Schaeffer}, R., \& {da Costa}, L.~N. 1992, ApJ, 390, 17

\bibitem[{{Meneux} {et~al.}(2006){Meneux}, {Le F{\`e}vre}, {Guzzo}, {Pollo},
  {Cappi}, {Ilbert}, {Iovino}, {Marinoni}, {McCracken}, {Bottini}, {Garilli},
  {Le Brun}, {Maccagni}, {Picat}, {Scaramella}, {Scodeggio}, {Tresse},
  {Vettolani}, {Zanichelli}, {Adami}, {Arnouts}, {Arnaboldi}, {Bardelli},
  {Bolzonella}, {Charlot}, {Ciliegi}, {Contini}, {Foucaud}, {Franzetti},
  {Gavignaud}, {Marano}, {Mazure}, {Merighi}, {Paltani}, {Pell{\`o}},
  {Pozzetti}, {Radovich}, {Zamorani}, {Zucca}, {Bondi}, {Bongiorno},
  {Busarello}, {Cucciati}, {Gregorini}, {Lamareille}, {Mathez}, {Mellier},
  {Merluzzi}, {Ripepi}, \& {Rizzo}}]{meneux06}
{Meneux}, B., {Le F{\`e}vre}, O., {Guzzo}, L., {et~al.} 2006, \aap, 452, 387

\bibitem[{{Navarro} {et~al.}(1997){Navarro}, {Frenk}, \& {White}}]{navarro97}
{Navarro}, J.~F., {Frenk}, C.~S., \& {White}, S.~D.~M. 1997, \apj, 490, 493

\bibitem[{{Norberg} {et~al.}(2002){Norberg}, {Baugh}, {Hawkins}, {Maddox},
  {Madgwick}, {Lahav}, {Cole}, {Frenk}, {Baldry}, {Bland-Hawthorn}, {Bridges},
  {Cannon}, {Colless}, {Collins}, {Couch}, {Dalton}, {De Propris}, {Driver},
  {Efstathiou}, {Ellis}, {Glazebrook}, {Jackson}, {Lewis}, {Lumsden},
  {Peacock}, {Peterson}, {Sutherland}, \& {Taylor}}]{norberg02}
{Norberg}, P., {Baugh}, C.~M., {Hawkins}, E., {et~al.} 2002, \mnras, 332, 827

\bibitem[{{Park} {et~al.}(1994){Park}, {Vogeley}, {Geller}, \&
  {Huchra}}]{park94}
{Park}, C., {Vogeley}, M.~S., {Geller}, M.~J., \& {Huchra}, J.~P. 1994, \apj,
  431, 569

\bibitem[{{Pearce} {et~al.}(2001){Pearce}, {Jenkins}, {Frenk}, {White},
  {Thomas}, {Couchman}, {Peacock}, \& {Efstathiou}}]{pierce01}
{Pearce}, F.~R., {Jenkins}, A., {Frenk}, C.~S., {et~al.} 2001, \mnras, 326, 649

\bibitem[{{Peebles}(1980)}]{peebles80}
{Peebles}, P.~J.~E. 1980, {The large-scale structure of the universe} (Research
  supported by the National Science Foundation.~Princeton, N.J., Princeton
  University Press, 1980.~435 p.)

\bibitem[{{Perlmutter} {et~al.}(1999){Perlmutter}, {Aldering}, {Goldhaber},
  {Knop}, {Nugent}, {Castro}, {Deustua}, {Fabbro}, {Goobar}, {Groom}, {Hook},
  {Kim}, {Kim}, {Lee}, {Nunes}, {Pain}, {Pennypacker}, {Quimby}, {Lidman},
  {Ellis}, {Irwin}, {McMahon}, {Ruiz-Lapuente}, {Walton}, {Schaefer}, {Boyle},
  {Filippenko}, {Matheson}, {Fruchter}, {Panagia}, {Newberg}, {Couch}, \& {The
  Supernova Cosmology Project}}]{perlmutter99}
{Perlmutter}, S., {Aldering}, G., {Goldhaber}, G., {et~al.} 1999, ApJ, 517, 565

\bibitem[{{Phillips} {et~al.}(2001){Phillips}, {Weinberg}, {Croft},
  {Hernquist}, {Katz}, \& {Pettini}}]{phillips01}
{Phillips}, J., {Weinberg}, D.~H., {Croft}, R.~A.~C., {et~al.} 2001, ApJ, 560,
  15

\bibitem[{{Phleps} \& {Meisenheimer}(2003)}]{phleps03}
{Phleps}, S. \& {Meisenheimer}, K. 2003, \aap, 407, 855

\bibitem[{{Phleps} {et~al.}(2006){Phleps}, {Peacock}, {Meisenheimer}, \&
  {Wolf}}]{phleps06}
{Phleps}, S., {Peacock}, J.~A., {Meisenheimer}, K., \& {Wolf}, C. 2006, \aap,
  457, 145

\bibitem[{{Pollo} {et~al.}(2006){Pollo}, {Guzzo}, {Le F{\`e}vre}, {Meneux},
  {Cappi}, {Franzetti}, {Iovino}, {McCracken}, {Marinoni}, {Zamorani},
  {Bottini}, {Garilli}, {Le Brun}, {Maccagni}, {Picat}, {Scaramella},
  {Scodeggio}, {Tresse}, {Vettolani}, {Zanichelli}, {Adami}, {Arnouts},
  {Bardelli}, {Bolzonella}, {Charlot}, {Ciliegi}, {Contini}, {Foucaud},
  {Gavignaud}, {Ilbert}, {Marano}, {Mazure}, {Merighi}, {Paltani}, {Pell{\`o}},
  {Pozzetti}, {Radovich}, {Zucca}, {Bondi}, {Bongiorno}, {Busarello},
  {Cucciati}, {Gregorini}, {Lamareille}, {Mathez}, {Mellier}, {Merluzzi},
  {Ripepi}, \& {Rizzo}}]{pollo06}
{Pollo}, A., {Guzzo}, L., {Le F{\`e}vre}, O., {et~al.} 2006, \aap, 451, 409

\bibitem[{{Pons-Border{\'{\i}}a} {et~al.}(1999){Pons-Border{\'{\i}}a},
  {Mart{\'{\i}}nez}, {Stoyan}, {Stoyan}, \& {Saar}}]{pons-borderia99}
{Pons-Border{\'{\i}}a}, M., {Mart{\'{\i}}nez}, V.~J., {Stoyan}, D., {Stoyan},
  H., \& {Saar}, E. 1999, ApJ, 523, 480

\bibitem[{{Postman} \& {Geller}(1984)}]{postman84}
{Postman}, M. \& {Geller}, M.~J. 1984, \apj, 281, 95

\bibitem[{{Ramella} {et~al.}(1997){Ramella}, {Pisani}, \& {Geller}}]{ramella97}
{Ramella}, M., {Pisani}, A., \& {Geller}, M.~J. 1997, AJ, 113, 483

\bibitem[{{Ratcliffe} {et~al.}(1996){Ratcliffe}, {Shanks}, {Broadbent},
  {Parker}, {Watson}, {Oates}, {Fong}, \& {Collins}}]{ratcliffe96}
{Ratcliffe}, A., {Shanks}, T., {Broadbent}, A., {et~al.} 1996, MNRAS, 281, L47+

\bibitem[{{Ratcliffe} {et~al.}(1998{\natexlab{a}}){Ratcliffe}, {Shanks},
  {Parker}, {Broadbent}, {Watson}, {Oates}, {Collins}, \&
  {Fong}}]{ratcliffe98b}
{Ratcliffe}, A., {Shanks}, T., {Parker}, Q.~A., {et~al.} 1998{\natexlab{a}},
  MNRAS, 300, 417

\bibitem[{{Ratcliffe} {et~al.}(1998{\natexlab{b}}){Ratcliffe}, {Shanks},
  {Parker}, \& {Fong}}]{ratcliffe98c}
{Ratcliffe}, A., {Shanks}, T., {Parker}, Q.~A., \& {Fong}, R.
  1998{\natexlab{b}}, MNRAS, 296, 173

\bibitem[{{Riess} {et~al.}(1998){Riess}, {Filippenko}, {Challis},
  {Clocchiatti}, {Diercks}, {Garnavich}, {Gilliland}, {Hogan}, {Jha},
  {Kirshner}, {Leibundgut}, {Phillips}, {Reiss}, {Schmidt}, {Schommer},
  {Smith}, {Spyromilio}, {Stubbs}, {Suntzeff}, \& {Tonry}}]{riess98}
{Riess}, A.~G., {Filippenko}, A.~V., {Challis}, P., {et~al.} 1998, AJ, 116,
  1009

\bibitem[{{Saunders} {et~al.}(1992){Saunders}, {Rowan-Robinson}, \&
  {Lawrence}}]{saunders92}
{Saunders}, W., {Rowan-Robinson}, M., \& {Lawrence}, A. 1992, MNRAS, 258, 134

\bibitem[{{Schechter}(1976)}]{schechter76}
{Schechter}, P. 1976, ApJ, 203, 297

\bibitem[{{Shectman} {et~al.}(1996){Shectman}, {Landy}, {Oemler}, {Tucker},
  {Lin}, {Kirshner}, \& {Schechter}}]{shectman96}
{Shectman}, S.~A., {Landy}, S.~D., {Oemler}, A., {et~al.} 1996, ApJ, 470, 172

\bibitem[{{Shepherd} {et~al.}(2001){Shepherd}, {Carlberg}, {Yee}, {Morris},
  {Lin}, {Sawicki}, {Hall}, \& {Patton}}]{shepherd01}
{Shepherd}, C.~W., {Carlberg}, R.~G., {Yee}, H.~K.~C., {et~al.} 2001, ApJ, 560,
  72

\bibitem[{{Small} {et~al.}(1999){Small}, {Ma}, {Sargent}, \&
  {Hamilton}}]{small99}
{Small}, T.~A., {Ma}, C., {Sargent}, W.~L.~W., \& {Hamilton}, D. 1999, ApJ,
  524, 31

\bibitem[{{Small} {et~al.}(1997){Small}, {Sargent}, \& {Hamilton}}]{small97a}
{Small}, T.~A., {Sargent}, W.~L.~W., \& {Hamilton}, D. 1997, ApJS, 111, 1

\bibitem[{{Smoot} {et~al.}(1991){Smoot}, {Bennett}, {Kogut}, {Aymon}, {Backus},
  {de Amici}, {Galuk}, {Jackson}, {Keegstra}, {Rokke}, {Tenorio}, {Torres},
  {Gulkis}, {Hauser}, {Janssen}, {Mather}, {Weiss}, {Wilkinson}, {Wright},
  {Boggess}, {Cheng}, {Kelsall}, {Lubin}, {Meyer}, {Moseley}, {Murdock},
  {Shafer}, \& {Silverberg}}]{smoot91}
{Smoot}, G.~F., {Bennett}, C.~L., {Kogut}, A., {et~al.} 1991, ApJ Lett., 371,
  L1

\bibitem[{{Tonry} {et~al.}(2003){Tonry}, {Schmidt}, {Barris}, {Candia},
  {Challis}, {Clocchiatti}, {Coil}, {Filippenko}, {Garnavich}, {Hogan},
  {Holland}, {Jha}, {Kirshner}, {Krisciunas}, {Leibundgut}, {Li}, {Matheson},
  {Phillips}, {Riess}, {Schommer}, {Smith}, {Sollerman}, {Spyromilio},
  {Stubbs}, \& {Suntzeff}}]{tonry03}
{Tonry}, J.~L., {Schmidt}, B.~P., {Barris}, B., {et~al.} 2003, ApJ, 594, 1

\bibitem[{{Trentham}(1997)}]{trentham97}
{Trentham}, N. 1997, MNRAS, 286, 133

\bibitem[{{Trentham} \& {Hodgkin}(2002)}]{trentham02a}
{Trentham}, N. \& {Hodgkin}, S. 2002, MNRAS, 333, 423

\bibitem[{{Trentham} \& {Tully}(2002)}]{trentham02b}
{Trentham}, N. \& {Tully}, R.~B. 2002, MNRAS, 335, 712

\bibitem[{{Tucker} {et~al.}(1997){Tucker}, {Oemler}, {Kirshner}, {Lin},
  {Shectman}, {Landy}, {Schechter}, {Muller}, {Gottlober}, \&
  {Einasto}}]{tucker97}
{Tucker}, D.~L., {Oemler}, A., {Kirshner}, R.~P., {et~al.} 1997, MNRAS, 285, L5

\bibitem[{{Weinberg}(1972)}]{weinberg72}
{Weinberg}, S. 1972, {Gravitation and cosmology: Principles and applications of
  the general theory of relativity} (New York: Wiley, |c1972)

\bibitem[{{Weinmann} {et~al.}(2006){Weinmann}, {van den Bosch}, {Yang}, \&
  {Mo}}]{weinmann06}
{Weinmann}, S.~M., {van den Bosch}, F.~C., {Yang}, X., \& {Mo}, H.~J. 2006,
  \mnras, 366, 2

\bibitem[{{Willmer} {et~al.}(1998){Willmer}, {da Costa}, \&
  {Pellegrini}}]{willmer98}
{Willmer}, C.~N.~A., {da Costa}, L.~N., \& {Pellegrini}, P.~S. 1998, AJ, 115,
  869

\bibitem[{{Yahil} {et~al.}(1977){Yahil}, {Tammann}, \& {Sandage}}]{yahil77}
{Yahil}, A., {Tammann}, G.~A., \& {Sandage}, A. 1977, ApJ, 217, 903

\bibitem[{{Yang} {et~al.}(2005{\natexlab{a}}){Yang}, {Mo}, {Jing}, \& {van den
  Bosch}}]{yang05c}
{Yang}, X., {Mo}, H.~J., {Jing}, Y.~P., \& {van den Bosch}, F.~C.
  2005{\natexlab{a}}, \mnras, 358, 217

\bibitem[{{Yang} {et~al.}(2004){Yang}, {Mo}, {Jing}, {van den Bosch}, \&
  {Chu}}]{yang04}
{Yang}, X., {Mo}, H.~J., {Jing}, Y.~P., {van den Bosch}, F.~C., \& {Chu}, Y.
  2004, \mnras, 350, 1153

\bibitem[{{Yang} {et~al.}(2003){Yang}, {Mo}, \& {van den Bosch}}]{yang03}
{Yang}, X., {Mo}, H.~J., \& {van den Bosch}, F.~C. 2003, \mnras, 339, 1057

\bibitem[{{Yang} {et~al.}(2005{\natexlab{b}}){Yang}, {Mo}, {van den Bosch}, \&
  {Jing}}]{yang05}
{Yang}, X., {Mo}, H.~J., {van den Bosch}, F.~C., \& {Jing}, Y.~P.
  2005{\natexlab{b}}, \mnras, 356, 1293

\bibitem[{{Yang} {et~al.}(2005{\natexlab{c}}){Yang}, {Mo}, {van den Bosch}, \&
  {Jing}}]{yang05b}
---. 2005{\natexlab{c}}, \mnras, 357, 608

\bibitem[{{Yang} {et~al.}(2005{\natexlab{d}}){Yang}, {Mo}, {van den Bosch},
  {Weinmann}, {Li}, \& {Jing}}]{yang05d}
{Yang}, X., {Mo}, H.~J., {van den Bosch}, F.~C., {et~al.} 2005{\natexlab{d}},
  \mnras, 362, 711

\bibitem[{{Yoshida} {et~al.}(2001){Yoshida}, {Colberg}, {White}, {Evrard},
  {MacFarland}, {Couchman}, {Jenkins}, {Frenk}, {Pearce}, {Efstathiou},
  {Peacock}, \& {Thomas}}]{yoshida01}
{Yoshida}, N., {Colberg}, J., {White}, S.~D.~M., {et~al.} 2001, \mnras, 325,
  803

\bibitem[{{Zandivarez} {et~al.}(2003){Zandivarez}, {Merch{\'a}n}, \&
  {Padilla}}]{zandivarez03}
{Zandivarez}, A., {Merch{\'a}n}, M.~E., \& {Padilla}, N.~D. 2003, \mnras, 344,
  247

\bibitem[{{Zehavi} {et~al.}(2002){Zehavi}, {Blanton}, {Frieman}, {Weinberg},
  {Mo}, {Strauss}, {Anderson}, {Annis}, {Bahcall}, {Bernardi}, {Briggs},
  {Brinkmann}, {Burles}, {Carey}, {Castander}, {Connolly}, {Csabai},
  {Dalcanton}, {Dodelson}, {Doi}, {Eisenstein}, {Evans}, {Finkbeiner},
  {Friedman}, {Fukugita}, {Gunn}, {Hennessy}, {Hindsley}, {Ivezi{\' c}},
  {Kent}, {Knapp}, {Kron}, {Kunszt}, {Lamb}, {Leger}, {Long}, {Loveday},
  {Lupton}, {McKay}, {Meiksin}, {Merrelli}, {Munn}, {Narayanan}, {Newcomb},
  {Nichol}, {Owen}, {Peoples}, {Pope}, {Rockosi}, {Schlegel}, {Schneider},
  {Scoccimarro}, {Sheth}, {Siegmund}, {Smee}, {Snir}, {Stebbins}, {Stoughton},
  {SubbaRao}, {Szalay}, {Szapudi}, {Tegmark}, {Tucker}, {Uomoto}, {Vanden
  Berk}, {Vogeley}, {Waddell}, {Yanny}, \& {York}}]{zehavi02}
{Zehavi}, I., {Blanton}, M.~R., {Frieman}, J.~A., {et~al.} 2002, ApJ, 571, 172

\bibitem[{{Zehavi} {et~al.}(2004){Zehavi}, {Weinberg}, {Zheng}, {Berlind},
  {Frieman}, {Scoccimarro}, {Sheth}, {Blanton}, {Tegmark}, {Mo}, {Bahcall},
  {Brinkmann}, {Burles}, {Csabai}, {Fukugita}, {Gunn}, {Lamb}, {Loveday},
  {Lupton}, {Meiksin}, {Munn}, {Nichol}, {Schlegel}, {Schneider}, {SubbaRao},
  {Szalay}, {Uomoto}, \& {York}}]{zehavi04}
{Zehavi}, I., {Weinberg}, D.~H., {Zheng}, Z., {et~al.} 2004, ApJ, 608, 16

\bibitem[{{Zehavi} {et~al.}(2005){Zehavi}, {Zheng}, {Weinberg}, {Frieman},
  {Berlind}, {Blanton}, {Scoccimarro}, {Sheth}, {Strauss}, {Kayo}, {Suto},
  {Fukugita}, {Nakamura}, {Bahcall}, {Brinkmann}, {Gunn}, {Hennessy},
  {Ivezi{\'c}}, {Knapp}, {Loveday}, {Meiksin}, {Schlegel}, {Schneider},
  {Szapudi}, {Tegmark}, {Vogeley}, \& {York}}]{zehavi05}
{Zehavi}, I., {Zheng}, Z., {Weinberg}, D.~H., {et~al.} 2005, ApJ, 630, 1

\bibitem[{{Zheng} {et~al.}(2002){Zheng}, {Tinker}, {Weinberg}, \&
  {Berlind}}]{zheng02}
{Zheng}, Z., {Tinker}, J.~L., {Weinberg}, D.~H., \& {Berlind}, A.~A. 2002,
  \apj, 575, 617

\end{thebibliography}
\bibliographystyle{aa}


\end{document}